\documentclass[12pt,a4paper]{article}
\usepackage{ifthen} % for conditional statements
\newboolean{pdflatex}
\setboolean{pdflatex}{true} % False for eps figures 
\newboolean{articletitles}
\setboolean{articletitles}{true} % False removes titles in references

\newboolean{uprightparticles}
\setboolean{uprightparticles}{false} %True for upright particle symbols
\usepackage{booktabs} % ADD TO MAKE THE TABLES THAT FELICIA LIKES
\usepackage{multirow}
\def\paperauthors{LHCb collaboration} % Leave as is for PAPER, CONF and FIGURE
\def\paperasciititle{Angular analysis of the decay Lb -> L(1520) mu+mu- } % Set ASCII title here !! MAKE sure it's only ASCII characters !! 
\def\papertitle{Angular analysis of the decay \LbToLresmm} % Latex formatted title
\def\paperkeywords{{High Energy Physics}, {LHCb}} % Comma separated list
\def\papercopyright{\the\year\ CERN for the benefit of the LHCb collaboration} % new since 9/Apr/2018
\def\paperlicence{CC BY 4.0 licence}
\def\paperlicenceurl{https://creativecommons.org/licenses/by/4.0/}

\newif\ifEnableSectionTOCLinks
\EnableSectionTOCLinksfalse % deactivated

\usepackage[top=1in, bottom=1.25in, left=1in, right=1in]{geometry}

\usepackage{microtype}
\usepackage{lineno}  % for line numbering during review
\usepackage{xspace} % To avoid problems with missing or double spaces after
\usepackage{caption} %these three command get the figure and table captions automatically small

\usepackage{graphicx}  % to include figures (can also use other packages)
\usepackage{color}
\usepackage{colortbl}
\graphicspath{{./figs/}} % Make Latex search fig subdir for figures
\usepackage{amsmath} % Adds a large collection of math symbols
\usepackage{amssymb}
\usepackage{amsfonts}
\usepackage{upgreek} % Adds in support for greek letters in roman typeset

\newcommand*\patchAmsMathEnvironmentForLineno[1]{%
\expandafter\let\csname old#1\expandafter\endcsname\csname #1\endcsname
\expandafter\let\csname oldend#1\expandafter\endcsname\csname
end#1\endcsname
 \renewenvironment{#1}%
   {\linenomath\csname old#1\endcsname}%
   {\csname oldend#1\endcsname\endlinenomath}%
}
\newcommand*\patchBothAmsMathEnvironmentsForLineno[1]{%
  \patchAmsMathEnvironmentForLineno{#1}%
  \patchAmsMathEnvironmentForLineno{#1*}%
}
\AtBeginDocument{%
\patchBothAmsMathEnvironmentsForLineno{equation}%
\patchBothAmsMathEnvironmentsForLineno{align}%
\patchBothAmsMathEnvironmentsForLineno{flalign}%
\patchBothAmsMathEnvironmentsForLineno{alignat}%
\patchBothAmsMathEnvironmentsForLineno{gather}%
\patchBothAmsMathEnvironmentsForLineno{multline}%
\patchBothAmsMathEnvironmentsForLineno{eqnarray}%
}

\usepackage[pdftex,
            pdfauthor={\paperauthors},
            pdftitle={\paperasciititle},
            pdfkeywords={\paperkeywords}]{hyperref}
\usepackage{hyperxmp}
\hypersetup{
    pdfcopyright={Copyright (C) \papercopyright},
    pdflicenseurl={\paperlicenceurl}
}
\usepackage[bottom,flushmargin,hang,multiple]{footmisc}

\usepackage[all]{hypcap} % Internal hyperlinks to floats.

\usepackage{xspace} 
\usepackage{upgreek}

\def\lhcb   {\mbox{LHCb}\xspace}

\def\MagUp {\mbox{\em Mag\kern -0.05em Up}\xspace}

\ifdefined\ifuprightparticles
\else
\newboolean{uprightparticles}
\setboolean{uprightparticles}{false} %True for upright particle symbols
\fi

\ifthenelse{\boolean{uprightparticles}}%
{

 \def\Pmu         {\ensuremath{\upmu}\xspace}

 \def\Ppi         {\ensuremath{\uppi}\xspace}

 \def\Ppsi        {\ensuremath{\uppsi}\xspace}

 \def\PDelta      {\ensuremath{\Delta}\xspace}                 
 \def\PXi         {\ensuremath{\Xi}\xspace}                 
 \def\PLambda     {\ensuremath{\Lambda}\xspace}                 
 \def\PSigma      {\ensuremath{\Sigma}\xspace}                 
 \def\POmega      {\ensuremath{\Omega}\xspace}                 
 \def\PUpsilon    {\ensuremath{\Upsilon}\xspace}
 \let\oldPi\Pi
 \def\PPi         {\ensuremath{\oldPi}\xspace}

 \def\PB      {\ensuremath{\mathrm{B}}\xspace}                 
 \def\PD      {\ensuremath{\mathrm{D}}\xspace}                 
 \def\PJ      {\ensuremath{\mathrm{J}}\xspace}                 
 \def\PK      {\ensuremath{\mathrm{K}}\xspace}                 
 \def\Pb      {\ensuremath{\mathrm{b}}\xspace}                 
 \def\Pc      {\ensuremath{\mathrm{c}}\xspace}

 \def\Pp      {\ensuremath{\mathrm{p}}\xspace}                 

 \def\Ps      {\ensuremath{\mathrm{s}}\xspace}

 \def\thebaroffset{0.0em}
}
{

 \def\Pmu         {\ensuremath{\mu}\xspace}

 \def\Ppi         {\ensuremath{\pi}\xspace}

 \def\Ppsi        {\ensuremath{\psi}\xspace}                 
                  
 \mathchardef\PDelta="7101
 \mathchardef\PXi="7104
 \mathchardef\PLambda="7103
 \mathchardef\PSigma="7106
 \mathchardef\POmega="710A
 \mathchardef\PUpsilon="7107
 \mathchardef\PPi="7105
 \def\PB      {\ensuremath{B}\xspace}                 
 \def\PD      {\ensuremath{D}\xspace}                 
 \def\PJ      {\ensuremath{J}\xspace}                 
 \def\PK      {\ensuremath{K}\xspace}                 
 \def\Pb      {\ensuremath{b}\xspace}                 
 \def\Pc      {\ensuremath{c}\xspace}

 \def\Pp      {\ensuremath{p}\xspace}                 

 \def\Ps      {\ensuremath{s}\xspace}

 \def\thebaroffset{0.18em}
}
\newcommand{\offsetoverline}[2][\thebaroffset]{\kern #1\overline{\kern -#1 #2}}%

\makeatletter
\ifcase \@ptsize \relax% 10pt
  \newcommand{\miniscule}{\@setfontsize\miniscule{4}{5}}% \tiny: 5/6
\or% 11pt
  \newcommand{\miniscule}{\@setfontsize\miniscule{5}{6}}% \tiny: 6/7
\or% 12pt
  \newcommand{\miniscule}{\@setfontsize\miniscule{5}{6}}% \tiny: 6/7
\fi
\makeatother

\DeclareRobustCommand{\optbar}[1]{\shortstack{{\miniscule (\rule[.5ex]{1.25em}{.18mm})}
  \\ [-.7ex] $#1$}}

\def\mup        {{\ensuremath{\Pmu^+}}\xspace}
\def\mun        {{\ensuremath{\Pmu^-}}\xspace} % muon negative (\mum is taken)

\def\mumu       {{\ensuremath{\Pmu^+\Pmu^-}}\xspace}

\def\ellell     {\ensuremath{\ell^+ \ell^-}\xspace}

\def\squark    {{\ensuremath{\Ps}}\xspace}

\def\cquark    {{\ensuremath{\Pc}}\xspace}

\def\bquark    {{\ensuremath{\Pb}}\xspace}

\def\pion   {{\ensuremath{\Ppi}}\xspace}

\def\pip    {{\ensuremath{\pion^+}}\xspace}
\def\pim    {{\ensuremath{\pion^-}}\xspace}

\def\kaon    {{\ensuremath{\PK}}\xspace}
\def\KorKbar {\kern \thebaroffset\optbar{\kern -\thebaroffset \PK}{}\xspace}

\def\Kp      {{\ensuremath{\kaon^+}}\xspace}
\def\Km      {{\ensuremath{\kaon^-}}\xspace}

\def\Kstarz  {{\ensuremath{\kaon^{*0}}}\xspace}

\def\D       {{\ensuremath{\PD}}\xspace}

\def\DorDbar {\kern \thebaroffset\optbar{\kern -\thebaroffset \PD}\xspace}
\def\Dz      {{\ensuremath{\D^0}}\xspace}

\def\Dp      {{\ensuremath{\D^+}}\xspace}
\def\Dm      {{\ensuremath{\D^-}}\xspace}

\def\DpDm    {\ensuremath{\Dp {\kern -0.16em \Dm}}\xspace}

\def\B       {{\ensuremath{\PB}}\xspace}

\def\BorBbar {\kern \thebaroffset\optbar{\kern -\thebaroffset \PB}\xspace}

\def\Bd      {{\ensuremath{\B^0}}\xspace}

\def\BdorBdbar {\kern \thebaroffset\optbar{\kern -\thebaroffset \Bd}\xspace}
\def\Bu      {{\ensuremath{\B^+}}\xspace}

\def\Bp      {{\ensuremath{\Bu}}\xspace}

\def\Bs      {{\ensuremath{\B^0_\squark}}\xspace}

\def\BsorBsbar {\kern \thebaroffset\optbar{\kern -\thebaroffset \Bs}\xspace}

\def\jpsi     {{\ensuremath{{\PJ\mskip -3mu/\mskip -2mu\Ppsi}}}\xspace}
\def\psitwos  {{\ensuremath{\Ppsi{(2S)}}}\xspace}

\def\Y#1S{\ensuremath{\PUpsilon{(#1S)}}\xspace}

\def\proton      {{\ensuremath{\Pp}}\xspace}

\def\Lz          {{\ensuremath{\PLambda}}\xspace}
\def\Lbar        {{\ensuremath{\offsetoverline{\PLambda}}}\xspace}
\def\LorLbar     {\kern \thebaroffset\optbar{\kern -\thebaroffset \PLambda}\xspace}
\def\Lambdares   {{\ensuremath{\PLambda}}\xspace}

\def\Lc          {{\ensuremath{\Lz^+_\cquark}}\xspace}

\def\Lb           {{\ensuremath{\Lz^0_\bquark}}\xspace}
\def\Lbbar        {{\ensuremath{\Lbar{}^0_\bquark}}\xspace}

\newcommand{\decay}[2]{\ensuremath{\mathinner{#1\!\to #2}}\xspace}

\def\to                 {\ensuremath{\rightarrow}\xspace}

\def\qsq       {{\ensuremath{q^2}}\xspace}

\def\CP                {{\ensuremath{C\!P}}\xspace}

\def\BdToKstmm    {\decay{\Bd}{\Kstarz\mup\mun}}

\def\AT#1     {\ensuremath{A_{\mathrm{T}}^{#1}}\xspace}           % 2

\def\C#1      {\ensuremath{\mathcal{C}_{#1}}\xspace}                       % 9
\def\Cp#1     {\ensuremath{\mathcal{C}_{#1}^{'}}\xspace}                    % 7
\def\Ceff#1   {\ensuremath{\mathcal{C}_{#1}^{\mathrm{(eff)}}}\xspace}        % 9  
\def\Cpeff#1  {\ensuremath{\mathcal{C}_{#1}^{'\mathrm{(eff)}}}\xspace}       % 7
\def\Ope#1    {\ensuremath{\mathcal{O}_{#1}}\xspace}                       % 2
\def\Opep#1   {\ensuremath{\mathcal{O}_{#1}^{'}}\xspace}                    % 7

\newcommand{\aunit}[1]{\ensuremath{\text{\,#1}}}       
\newcommand{\tev}{\aunit{Te\kern -0.1em V}\xspace}
\newcommand{\gev}{\aunit{Ge\kern -0.1em V}\xspace}
\newcommand{\mev}{\aunit{Me\kern -0.1em V}\xspace}
\newcommand{\kev}{\aunit{ke\kern -0.1em V}\xspace}
\newcommand{\ev}{\aunit{e\kern -0.1em V}\xspace}
 
\newcommand{\mevc}{\ensuremath{\aunit{Me\kern -0.1em V\!/}c}\xspace}
\newcommand{\gevc}{\ensuremath{\aunit{Ge\kern -0.1em V\!/}c}\xspace}
\newcommand{\mevcc}{\ensuremath{\aunit{Me\kern -0.1em V\!/}c^2}\xspace}
\newcommand{\gevcc}{\ensuremath{\aunit{Ge\kern -0.1em V\!/}c^2}\xspace}
\newcommand{\gevgevcccc}{\ensuremath{\gev^2\!/c^4}\xspace} % for q^2

\def\fb   {\ensuremath{\aunit{fb}}\xspace}
\def\invfb   {\ensuremath{\fb^{-1}}\xspace}

\def\deriv {\ensuremath{\mathrm{d}}}

\def\gsim{{~\raise.15em\hbox{$>$}\kern-.85em
          \lower.35em\hbox{$\sim$}~}\xspace}
\def\lsim{{~\raise.15em\hbox{$<$}\kern-.85em
          \lower.35em\hbox{$\sim$}~}\xspace}

\def\sqs   {\ensuremath{\protect\sqrt{s}}\xspace}

\def\evtgen     {\mbox{\textsc{EvtGen}}\xspace}

\def\photos     {\mbox{\textsc{Photos}}\xspace}

\def\pythia     {\mbox{\textsc{Pythia}}\xspace}

\def\tell1  {TELL1\xspace}
\def\ukl1   {UKL1\xspace}

\newcommand{\ie}{\mbox{\itshape i.e.}\xspace}

\newcommand{\phm}{\phantom{-}}
\newcommand{\lhcborcid}[1]{\href{https://orcid.org/#1}{\hspace*{0.1em}\raisebox{-0.45ex}{\includegraphics[width=1em]{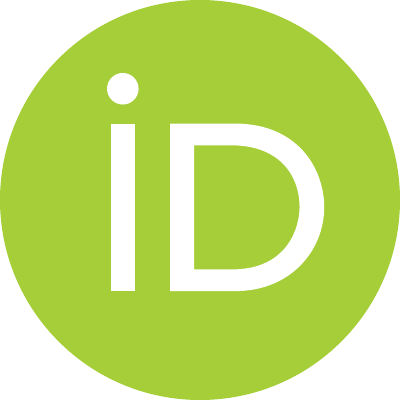}}}}

\hypersetup{
  colorlinks   = true, %Colours links instead of ugly boxes
  urlcolor     = blue, %Colour for external hyperlinks
  linkcolor    = blue, %Colour of internal links
  citecolor    = red   %Colour of citations
}

\ifEnableSectionTOCLinks
    \usepackage[explicit]{titlesec} % to change headings
    
    \let\oldcontentsline\contentsline
    \renewcommand

    \titleformat{\section}{\normalfont\Large\bf}{\hyperlink{tocsection.\thesection}{{\thesection} \parbox[t]{\dimexpr\textwidth-1pc}{#1}}}{1pc}{}

    \titleformat{\subsection}{\normalfont\bf}{\hyperlink{tocsubsection.\thesubsection}{{\thesubsection} \parbox[t]{\dimexpr\textwidth-1pc}{#1}}}{1pc}{}

    \titleformat{name=\section,numberless}[display]{}{}{0pt}{\normalfont\Huge\bfseries #1}
\fi

\usepackage{cite} % Allows for ranges in citations
\usepackage{mciteplus}
\def\pp{\ensuremath{\proton\proton}\xspace}

\def\Lst{\ensuremath{\PLambda^*}\xspace}
\def\Lstar{\ensuremath{\PLambda^*}\xspace}

\def\Lres{\ensuremath{\PLambda(1520)}\xspace}

\def\LbToLresmm{\decay{\Lb}{\Lres \mumu}}
\def\LbToLstmm{\decay{\Lb}{\Lst\mumu}}

\usepackage{longtable} % only for template; not usually to be used in PAPERs

\begin{document}

%%%%%%%%%%%%%%%%%%%%%%%%%
%%%%% Title     %%%%%%%%%
%%%%%%%%%%%%%%%%%%%%%%%%%
\renewcommand{\thefootnote}{\fnsymbol{footnote}}
\setcounter{footnote}{1}

% %%%%%%% CHOOSE TITLE PAGE--------
%\onecolumn
% ===============================================================================
% Purpose: LHCb-PAPER journal paper title page template
% Author: 
% Created on: 2010-09-25
% ===============================================================================

%%%%%%%%%%%%%%%%%%%%%%%%%
%%%%%  TITLE PAGE  %%%%%%
%%%%%%%%%%%%%%%%%%%%%%%%%
\begin{titlepage}
\pagenumbering{roman}

% Header ---------------------------------------------------
\vspace*{-1.5cm}
\centerline{\large EUROPEAN ORGANIZATION FOR NUCLEAR RESEARCH (CERN)}
\vspace*{1.5cm}
\noindent
\begin{tabular*}{\linewidth}{lc@{\extracolsep{\fill}}r@{\extracolsep{0pt}}}
\ifthenelse{\boolean{pdflatex}}% Logo format choice
{\vspace*{-1.5cm}\mbox{\!\!\!\includegraphics[width=.14\textwidth]{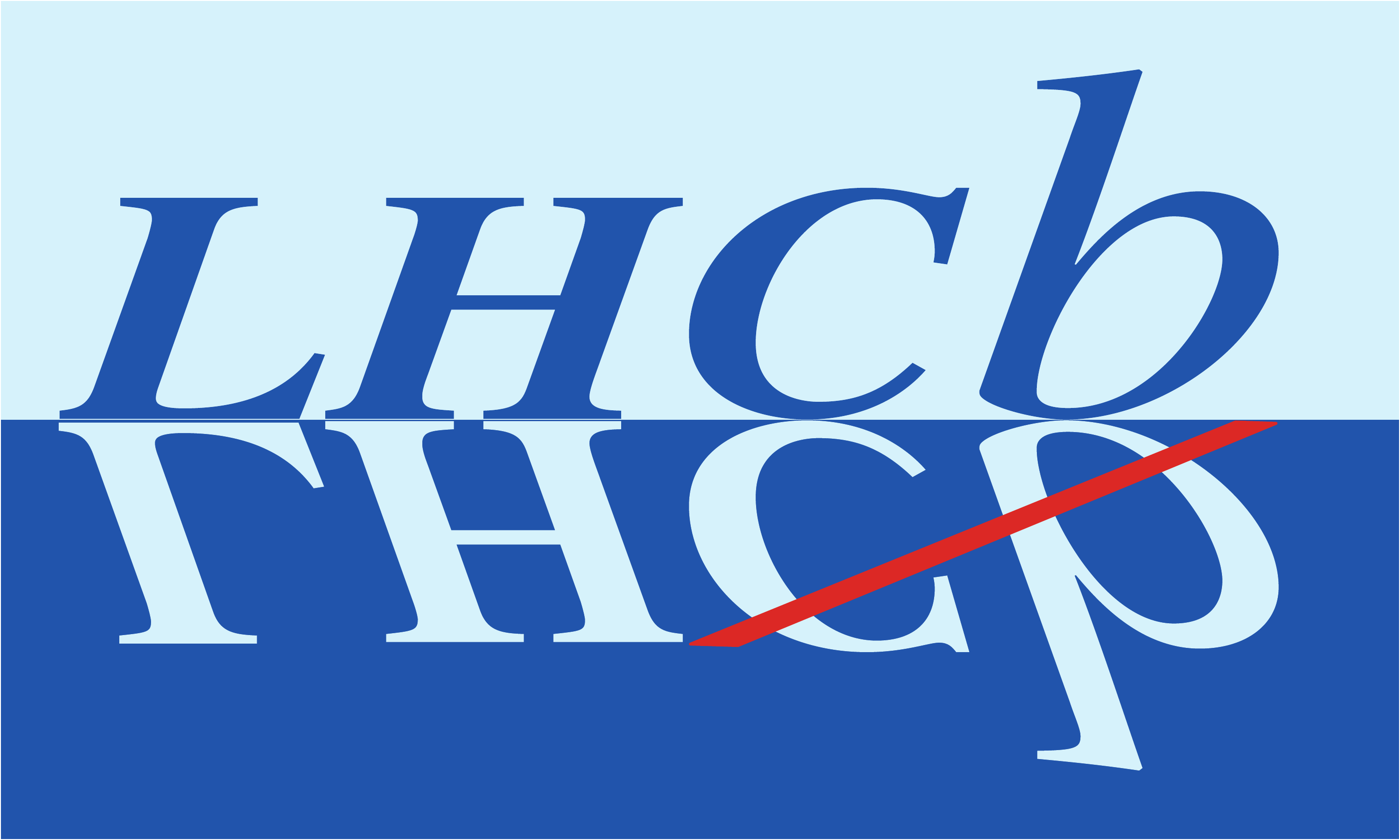}} & &}%
{\vspace*{-1.2cm}\mbox{\!\!\!\includegraphics[width=.12\textwidth]{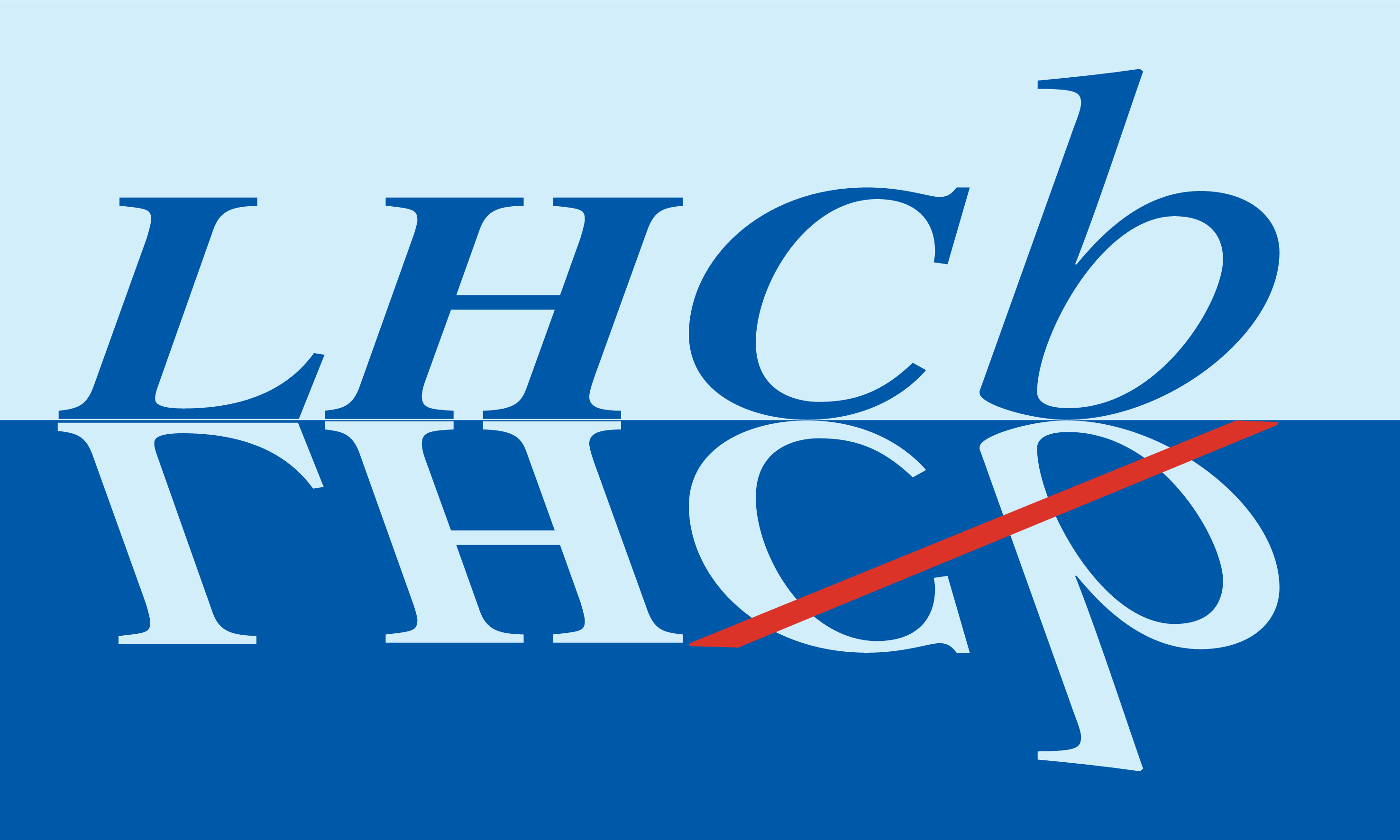}} & &}%
\\
 & & CERN-EP-2026-240 \\  % ID 
 & & LHCb-PAPER-2026-037 \\  % ID 
 & & 21 August 2026 \\ %\today \\ % Date - Can also hardwire e.g.: 23 March 2010
 & & \\
% not in paper \hline
\end{tabular*}

\vspace*{4.0cm}

% Title --------------------------------------------------
{\normalfont\bfseries\boldmath\huge
\begin{center}
% DO NOT EDIT HERE. Instead edit macro in main.tex to keep metadata correct
  \papertitle 
\end{center}
}

\vspace*{2.0cm}

% Authors -------------------------------------------------
\begin{center}
%In the footnote, replace 'paper' by 'Letter' in case of submission to PRL or PLB 
% Edit macro in main.tex to keep metadata correct
\paperauthors\footnote{Authors are listed at the end of this paper.}
\end{center}

\vspace{\fill}

% Abstract -----------------------------------------------
\begin{abstract}
  \noindent
  The first angular analysis of ${\it \Lambda}_{\it b}^{0} \to {\it \Lambda}(1520)\mu^{+}\mu^{-}$ decays is presented, using proton-proton collision data collected with the LHCb detector between 2011 and 2018, corresponding to an integrated luminosity of 9 fb$^{-1}$. The leptonic forward-backward asymmetry, $A_\text{FB,\,3/2}^\ell$, and the $C\!P$-averaged angular observable, $S_{1cc}$, are determined by fitting projections of the angular distributions in four intervals of the square of the dimuon invariant mass between 0.1 and 12.5 GeV$^2/c^4$. The results are in good agreement with predictions based on the Standard Model of particle physics.
\end{abstract}

\vspace*{2.0cm}

\begin{center}
  Submitted to
  JHEP 

\end{center}

\vspace{\fill}

{\footnotesize 
% Edit macro in main.tex to keep metadata correct
\centerline{\copyright~\papercopyright. \href{\paperlicenceurl}{\paperlicence}.}}
\vspace*{2mm}

\end{titlepage}

%%%%%%%%%%%%%%%%%%%%%%%%%%%%%%%%
%%%%%  EOD OF TITLE PAGE  %%%%%%
%%%%%%%%%%%%%%%%%%%%%%%%%%%%%%%%

%  empty page follows the title page ----
\newpage
\setcounter{page}{2}
\mbox{~}
%\newpage
%
%% Author List ----------------------------
%%  You need to get a new author list!
%\input{LHCb_authorlist.tex}
%
%The author list for journal publications is provided by the Membership Committee shortly after 'approval to go to paper' has been given.
%%It will be made available on the page
%%\verb!http://www.physik.uzh.ch/~strauman/forMemCo/LHCb-PAPER-XXXX-XXX/! .
%It will be sent to you by email shortly after a paper number has beens assigned.
%The author list should be included already at first circulation, 
%to allow new members of the collaboration to verify whether they have been included correctly.
%Occasionally a misspelled name is corrected or associated institutions become full members.
%In that case, a new author list will be sent to you.
%In case line numbering doesn't work well after including the authorlist, try moving the \verb!\bigskip! after the last author to a separate line.
%
%
%The authorship for Conference Reports should be ``The LHCb
%  collaboration'', with a footnote giving the name(s) of the contact
%  author(s), but without the full list of collaboration names.

%\twocolumn
% %%%%%%%%%%%%% ---------

\renewcommand{\thefootnote}{\arabic{footnote}}
\setcounter{footnote}{0}

%%%%%%%%%%%%%%%%%%%%%%%%%%%%%%%%
%%%%%  Table of Content   %%%%%%
%%%%%%%%%%%%%%%%%%%%%%%%%%%%%%%%
%%%% Uncomment if desired
%\tableofcontents

\cleardoublepage

%%%%%%%%%%%%%%%%%%%%%%%%%
%%%%% Main text %%%%%%%%%
%%%%%%%%%%%%%%%%%%%%%%%%%

\pagestyle{plain} % restore page numbers for the main text
\setcounter{page}{1}
\pagenumbering{arabic}

%% Uncomment during review phase. 
%% Comment before a final submission.
%\linenumbers

%% This is the main body
%% It is useful to have a single file so comments are not missed in overleaf.

\clearpage 
%-------%----------------%--------------- ^ ^ ---------------%
\section{Introduction}
\label{sec:introduction}
The flavour-changing neutral current decay \LbToLresmm involves a \decay{\bquark}{\squark} quark transition and is Cabibbo- and loop-suppressed in the Standard Model (SM) of particle physics.\footnote{Charge conjugation is implied throughout this paper unless stated otherwise.} Contributions of processes beyond the SM (BSM) to the decay amplitudes could have a size comparable to the SM rate. 
Observables such as decay rates and angular distributions are thus sensitive to the presence of BSM effects. Interesting tensions have been observed between predictions of the SM and measurements for \decay{\B}{\kaon\mumu}~\cite{LHCb-PAPER-2014-006,CMS:2024syx,Belle:2019xld} and \decay{\Bs}{\phi\mumu} \cite{LHCb-PAPER-2021-014} decays, as well as in the rate and angular distribution of \BdToKstmm decays~\cite{LHCb-PAPER-2020-002,LHCb-PAPER-2024-011, LHCb-PAPER-2023-032, LHCb-PAPER-2023-033,LHCb-PAPER-2025-041,LHCb-PAPER-2026-023,CMS:2024atz,ATLAS:2018gqc,Belle:2019oag,Belle:2016fev}. 
The decays of \Lb baryons are complementary in both theoretical and experimental aspects to their mesonic counterparts \Bu, \Bd and \Bs, which have been the subject of more extensive investigations. Thanks to the presence of an additional spectator quark, the theoretical understanding of the hadronic part of the interaction, which is associated with the largest theoretical uncertainties, is probed. The nonzero spins of the \Lb baryon and the \Lres resonance enable the investigation of the spin structure of the interaction complementary to the ones in $B$ mesons.

The decay \LbToLstmm was first observed at \lhcb \cite{LHCb-PAPER-2019-040}. The expression of $\Lstar$ refers to excited $\Lambdares$ resonances which decay via the strong interaction to $p\Km$. However, a full amplitude analysis has only been performed in the resonant $\Lb \to p\Km \jpsi$~\cite{LHCb-PAPER-2015-029} and $\Lb \to p\Km \gamma$~\cite{LHCb-PAPER-2023-036} decay modes. These analyses suggest that although various excited \Lst resonances contribute to the $p\Km$ final state, the \Lres state dominates with a relatively narrow width of 16\,\mevcc~\cite{PDG2026} and spin parity of $J^{P} = (3/2)^-$.

Standard Model predictions are available for the \decay{\Lb}{\Lz (1520)\mumu} angular distributions based on both Lattice QCD \cite{Meinel:2020owd, Meinel:2021mdj} and  dispersive bounds\cite{Amhis:2022vcd}. 
The first measurement of the differential decay rate in \decay{\Lb}{p\Km\mumu} decays with $m(p\Km)<2600$\,\mevcc is described in Ref.~\cite{LHCb-PAPER-2022-050} and angular coefficients using the full $p\Km$ mass spectrum are determined in Ref.~\cite{LHCb-PAPER-2024-024}. 

This paper presents the first angular analysis of \decay{\Lb}{\Lz(1520)\mumu} decays in intervals of the dimuon invariant mass squared, \qsq. 
The $\Lz(1520)$ resonance is reconstructed in the $p\Km$ final state and studied in narrow ([0.1, 3.0], [3.0, 6.0], [6.0, 8.0] and \mbox{[10.0, 12.5]}\,\gevgevcccc) and in one wide ([1.1, 6.0]\,\gevgevcccc) \qsq intervals typically used for measurements of $b \to s \ell\ell$ transitions. The \qsq regions of the tree-level decays via the \jpsi and the \psitwos resonances are removed. The measurements are performed using proton-proton ($pp$) collision data, corresponding to an integrated luminosity of 9\,\invfb, recorded by the LHCb experiment at centre-of-mass energies of \mbox{$\sqs=7$}, 8 and 13\tev between 2011 and 2018.
 
This paper is organised as follows: Sec.~\ref{sec:angular_distribution} introduces the formalism used to characterise the differential angular distribution for \decay{\Lb}{\Lz(1520)\mumu} decays; Sec.~\ref{sec:detector} describes the \lhcb detector and its simulation with the corresponding running conditions, while  Sec.~\ref{sec:selection} summarises the selection of \decay{\Lb}{\Lz(1520)\mumu} candidates. The modelling of the impact of the candidate selection criteria on the angular observables is presented in Sec.~\ref{sec:angular_eff}; fits made to the mass and angular distributions are described in Sec.~\ref{sec:mass_angular_fits}; with the discussion of potential sources of systematic uncertainties set out in Sec.~\ref{sec:systematics}. Results and conclusions of the analysis are given in Sec.~\ref{sec:results}.

%-------%----------------%--------------- ^ ^ ---------------%
\section{Angular distribution}
\label{sec:angular_distribution}
The full differential decay width of the \Lb baryon decaying to a lepton-antilepton pair and a \Lstar resonance with spin $J=3/2$, such as the \Lres, has been calculated in Ref.~\cite{Descotes-Genon:2019dbw} neglecting lepton masses, and in Ref.~\cite{Das:2020cpv} taking the lepton masses into account. Given the expected sensitivity of this analysis, the simplified expression in Ref.~\cite{Descotes-Genon:2019dbw} is adopted.\footnote{Due to the small sample size of the analysis, the lepton masses can be safely neglected.}

\begin{figure}[t]
	\begin{center}
		\includegraphics[width=0.65 \linewidth]{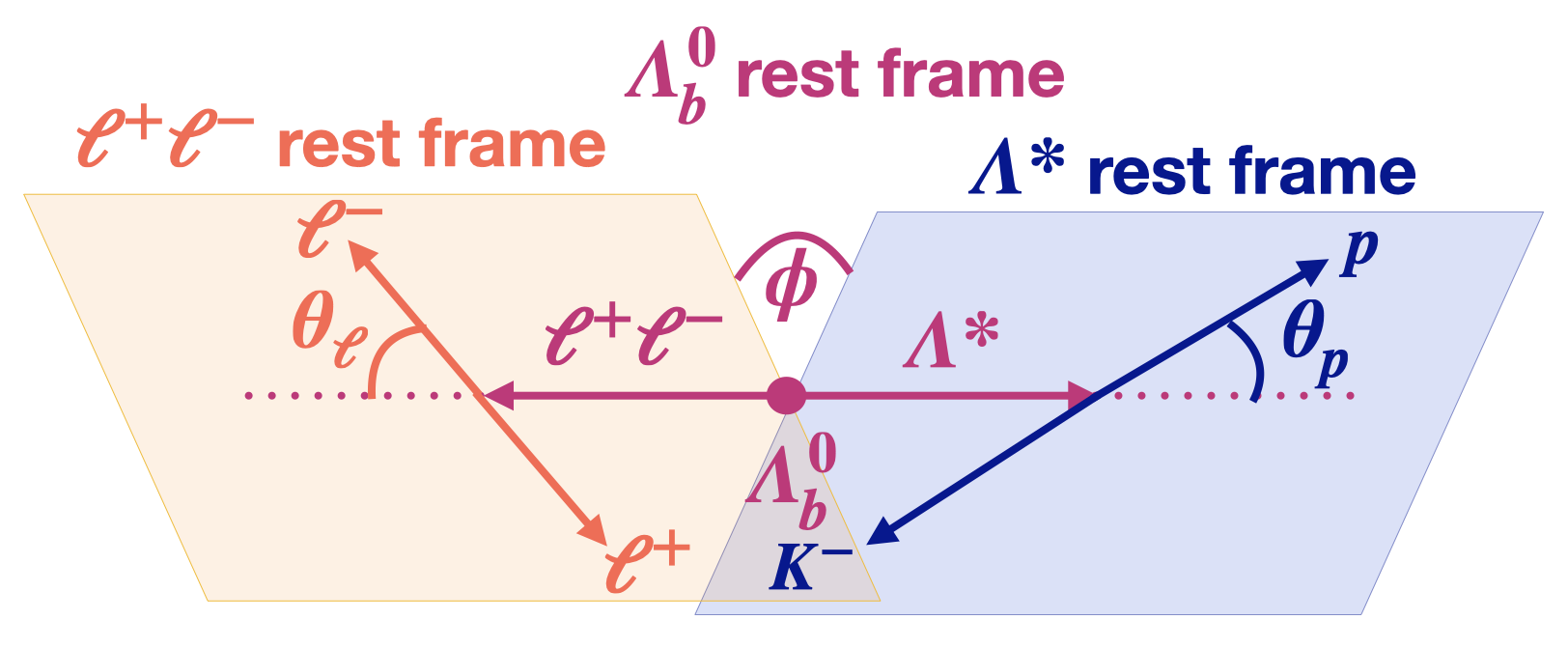}
	\caption{Definition of the ${\it\Lambda}_{\it b}^0\to pK^{-}\ell^{+}\ell^{-}$ decay angles in the helicity frame. The $\ell$ stands for leptons and are muons in this analysis.}
	\label{fig:AngleDef}
	\end{center}
\end{figure}
The distribution can be characterised in terms of \qsq and three decay angles $(\theta_\ell, \theta_p, \phi)$, following the definition in Ref.~\cite{Blake:2017une,LHCb-PAPER-2015-009,LHCb-PAPER-2018-029} and illustrated in Fig.~\ref{fig:AngleDef}, as 
\begin{align} 
	\frac{8\pi}{3}\frac{\deriv^4\Gamma}{\deriv q^2\,\deriv\cos{\theta_\ell}\,\deriv\cos{\theta_p}\,\deriv\phi}\nonumber
	 =&\cos ^2\theta_p \left(L_{1c} \cos \theta_\ell+L_{1cc} \cos ^2\theta_\ell+L_{1ss} \sin ^2\theta_\ell\right)\nonumber\\[-1.5ex]
	+& \sin ^2\theta_p \left(L_{2c} \cos
   \theta_\ell+L_{2cc} \cos ^2\theta_\ell+L_{2ss} \sin ^2\theta_\ell\right)\nonumber\\
	+& \sin ^2\theta_p \left(L_{3ss} \sin ^2\theta_\ell \cos^2
   \phi+L_{4ss} \sin ^2\theta_\ell \sin \phi \cos
   \phi\right)\nonumber\\
	+&\sin \theta_p \cos \theta_p \cos \phi (L_{5s} \sin \theta_\ell+L_{5sc} \sin \theta_\ell \cos \theta_\ell)\nonumber\\
	+&\sin \theta_p \cos \theta_p\sin \phi (L_{6s} \sin
   \theta_\ell+L_{6sc} \sin \theta_\ell \cos \theta_\ell), \label{eq:FullDiffWidth}
\end{align}
where the $L_i$ terms with $i \in \{1c, 1cc, 1ss, 2c, ... \}$ are angular coefficients. The differential decay width
\begin{align}
	\frac{\deriv\Gamma}{\deriv q^2}=\frac{1}{3}(L_{1cc}+2L_{1ss}+2L_{2cc}+4L_{2ss}+2L_{3ss})\label{eq:dGammaOverdq2}
\end{align}
is used as normalisation of the angular observables.

An optimised angular observable, which represents a leptonic forward-backward  asymmetry, which is particularly 
sensitive to physics beyond the SM, is defined~\cite{Amhis:2020phx} as
 \begin{equation}\label{eq:AFB}
 	A^\ell_{\mathrm{FB,\,3/2}}=\frac{3}{2}\frac{L_{1c}+2 L_{2c}}{L_{1cc}+2 L_{1ss}+2L_{2cc}+4 L_{2ss}+2L_{3ss}}
 \end{equation}
and measured in the angular analysis presented in this paper. 

In the $p\Km$ mass window around the \Lres resonance there are several other resonant $\Lambdares^*$ contributions, primarily the $\Lambdares(1405)$ and $\Lambdares(1600)$ resonances~\cite{LHCb-PAPER-2015-029,LHCb-PAPER-2023-036} with the $\Lambdares(1800)$ resonance observed at a lower level. As the $\Lambdares(1405)$ and $\Lambdares(1600)$ resonances are spin $J=1/2$ and their contribution in the signal region is small compared to the $\Lres$ resonance, they are considered collectively. Therefore, a single fit component models the spin $J = 1/2$ contribution in the $m(p\Km)$ and angular fit. In the rest of the paper, they are referred to as $\Lambdares^*_{J=1/2}$.

As the \Lb and \Lbbar decays are measured together, the differential decay width is \CP averaged. 
This paper adopts the differential decay width in the heavy-quark limit\footnote{In the heavy-quark limit, the angular coefficients $L_{1j}$ with $j \in \{ c, cc, ss\}$ become $4L_{2j}$ and all $L_i$ with $i \in \{ 3ss, 4ss, 5s, 5sc, 6s, 6sc\}$ vanish.} proposed in Ref.~\cite{Descotes-Genon:2019dbw,Amhis:2020phx}, which effectively approximates the Eq.~\ref{eq:FullDiffWidth} and may be expressed as
\begin{align}\label{eq:angobs_simpel_norm}
	\frac{8\pi}{3}\frac{\deriv^4 (\Gamma + \bar{\Gamma})}{\deriv q^2\, \deriv\cos\theta_\ell\, \deriv\cos \theta_{p}\, \deriv\phi} &\simeq \frac{1}{4} \biggl( S_{1ss}  \sin^2\theta_\ell + S_{1cc}\cos^2\theta_\ell + \frac{4}{3}A_{\mathrm{FB},3/2}^\ell \cos\theta_\ell \biggr) \notag \\
	&\quad \times  \left( 1+3\cos^2\theta_p \right), 
\end{align}
where the \CP-averaged angular observables 
are defined as 
\begin{align}
	S_i &= \frac{L_i + \bar{L}_i}{\deriv(\Gamma+\bar{\Gamma})/\deriv q^2}. \label{eq:Si} 
\end{align}
The heavy-quark limit impacts Eq.~\ref{eq:dGammaOverdq2} in the same way. In order to get a normalised probability density function (PDF), the resulting Eq.~\ref{eq:dGammaOverdq2} leads to the relationship
\begin{align}
    S_{1ss} = 1 - \frac{S_{1cc}}{2},
\end{align}
where $S_{1cc}$ is chosen to be measured.

Since at the time of the start of the analysis no dedicated theoretical model for strongly decaying, spin $J = 1/2$ \Lstar resonances had been available in the literature, the decomposition of the differential decay width is derived from Ref.~\cite{Boer:2014kda} for the \decay{\Lb}{\Lambdares\ellell} decay, which is written as 
\begin{align}
	\frac{8\pi}{3}\frac{\deriv^4\Gamma}{\deriv q^2\, \deriv\cos{\theta_\ell}\,\deriv\cos{\theta_p}\,\deriv\phi}\nonumber
	=&\left(K_{1c} \cos \theta_\ell+K_{1cc} \cos ^2\theta_\ell+K_{1ss} \sin ^2\theta_\ell\right)\nonumber\\[-1.5ex]
	+& \left(K_{2c} \cos
   \theta_\ell+K_{2cc} \cos ^2\theta_\ell+K_{2ss} \sin ^2\theta_\ell\right) \cos\theta_p \nonumber\\
	+& \left(K_{3sc} \sin\theta_\ell \cos
   \theta_\ell + K_{3s} \sin \theta_\ell \right) \sin\theta_p\sin\phi  \nonumber\\
	+& \left(K_{4sc} \sin\theta_\ell \cos
   \theta_\ell + K_{4s} \sin \theta_\ell \right)  \sin\theta_p\cos\phi, \label{eq:FullDecayWidth-L12}
\end{align}
where the $K_i$ terms represent the angular coefficients. 
Since the strongly decaying \Lstar resonances are studied instead of the weakly decaying ground state \Lambdares, the differential decay width can be simplified with all angular coefficients dependent on the parity-violating weak decay parameter $\alpha$~\cite{PDG2026}\footnote{The parameter $\alpha$ is defined in Eq.~79.8 of Ref.~\cite{PDG2026} and its value listed in the particle listings of the \Lambdares baryon under the name $\alpha_\mp$. } vanishing, leaving only terms in  $K_{1ss}$, $K_{1cc}$ and $K_{1c}$. The differential decay width is \CP averaged and normalised, which introduces parameters as in Eq.~\ref{eq:Si}. For convenience, even after \CP averaging, the observable name $K_{1cc}$ is kept, in order to distinguish it from $S_{1cc}$. In addition, the forward-backward asymmetry of the spin $J = 1/2$ \Lstar resonances, {\ensuremath{A_{\mathrm{FB}, 1/2}^{\ell}}\xspace}, is introduced analogously to what is described in Eq.~\ref{eq:AFB} for the spin $J = 3/2$ resonance.

Although the $\cos\theta_\ell$ distribution is expected to have a very similar shape for the \Lres and $\Lstar_{J=1/2}$ resonances, their distribution in $\cos\theta_p$ is expected to differ. 
Therefore, modelling the $p\Km$ invariant mass distribution is necessary both to estimate the fraction of the \Lres resonance, $f_{3/2}$, and to separate the \Lres and the $\Lstar_{J=1/2}$ angular contributions. The \Lres resonance is described by a relativistic Breit--Wigner (BW$_{\rm rel}$) with parameters from Ref.~\cite{BW-parameters, LHCb-PAPER-2022-050}, while the $\Lstar_{J=1/2}$ resonances are modelled using a normalised polynomial function $\mathcal{P}$, studied with simulation samples: 
\begin{align}
    \text{PDF}\left(m(p\Km)\right) = {f_{3/2}}  \left|\text{BW}_\text{rel}\left(m (p\Km)\right) \right|^2 + \left(1-{f_{3/2}}\right)\mathcal{P} \left(m(p\Km)\right),
\label{eq:totalMassPDF}
\end{align}
where $m(p\Km)$ is the invariant mass of the $p\Km$ final states. 

The presence of so far unknown strong phase differences between the \Lstar resonances affects the angular distributions. The necessity of including the interferences was derived from the study of simulated samples that reproduce nontrivial interference effects between the different resonances. The underlying models are described in Ref.~\cite{Beck:2022spd}. Studying various strong phase-difference patterns with the above mentioned samples reveals that the interferences impact the $\cos\theta_p$ distribution by shifting and scaling the distribution, while leaving the $\cos\theta_\ell$ and $m(p\Km)$ shapes unchanged. 
Combining the Eqs.~\ref{eq:angobs_simpel_norm} and \ref{eq:totalMassPDF} including a parametrisation of the interferences allows the angular PDF to be written as  
\begin{align}
   \text{PDF} (\theta_\ell, \theta_p)
    &= f_{3/2} \Bigg[ 
        \left(1 - \frac{1}{2} S_{1cc} \right) \left(1 - \cos^2\theta_\ell\right) 
        + S_{1cc} \cos^2\theta_\ell 
        + \frac{4}{3} A_{\mathrm{FB},3/2}^\ell \cos\theta_\ell 
    \Bigg] \nonumber\\
    & \times \left( \frac{1}{4} + \frac{3}{4} \cos^2\theta_p \right) \notag \\
    & + \left(1 - f_{3/2}\right) \Bigg[ 
        \frac{1}{2} \left(1 - K_{1cc}\right) \left(1 - \cos^2\theta_\ell\right) 
        + K_{1cc} \cos^2\theta_\ell 
        + \frac{2}{3} A_{\textrm{FB},1/2}^\ell \cos\theta_\ell 
    \Bigg] \notag \\
    & \times \left( 
        \frac{3 - i_2}{3} + i_1 \cos\theta_p + i_2 \cos^2\theta_p 
    \right),
    \label{eq:PDF_ang_int12_synthese}
\end{align}
where the parameters $i_1$ and $i_2$ 
are introduced to accommodate potential interferences between the various \Lstar resonances. The derived angular fit model is validated on those same samples by demonstrating that the obtained parameter values from the fit correspond to the generated values, which is shown in Appendix~\ref{supl:Interferences}. 
The main observables of interest are $A_\text{FB,\,3/2}^\ell$ and $S_{1cc}$, due to their sensitivity to physics beyond the SM. 
The remaining parameters, \ie $K_{1cc}$, $A^\ell_\text{FB,\,1/2}$, $i_1$ and $i_2$, are treated as nuisance parameters in the absence of any theoretical modelling. 

%-------%----------------%--------------- ^ ^ ---------------%
\section{Detector and simulation} 
\label{sec:detector}
The LHCb detector~\cite{LHCb-DP-2022-002,LHCb-DP-2008-001,LHCb-DP-2014-002} is a single-arm forward spectrometer covering the pseudorapidity range $2 < \eta < 5$, designed for the study of particles containing \bquark\ or \cquark\ quarks. 
The detector elements that are particularly relevant for this analysis are: 
a silicon-strip vertex detector surrounding the $pp$ interaction region that allows \cquark\ and \bquark\ hadrons to be identified from their characteristically long flight distance; a tracking system that provides a measurement of the momentum, $p$, of charged particles; 
two ring-imaging Cherenkov detectors that are able to discriminate between different species of charged hadrons; 
an electromagnetic calorimeter to reconstruct electron and photon energy deposits; and a muon system to identify muons. Events are selected online using a trigger~\cite{LHCb-DP-2012-004,LHCb-DP-2019-001} that consists of a hardware and a software stage. The trigger requirements follow the strategy employed in Ref.~\cite{LHCb-PAPER-2022-050}.

Simulated samples are used to determine the efficiency of the candidate selection as well as the distributions of specific sources of background. 
In the simulation, $pp$ collisions are generated using \pythia~\cite{Sjostrand:2007gs} with a specific \lhcb configuration~\cite{LHCb-PROC-2010-056}.
Decays of unstable particles are described by \evtgen~\cite{Lange:2001uf}, using \photos~\cite{davidson2015photos} to generate final-state radiation from the muons, protons and kaons. The \decay{\Lb}{\Lres\mumu} and control mode \decay{\Lb}{p\Km \jpsi} decays, with \jpsi\to \mumu, are simulated using a uniform phase-space model. The intermediate resonant $p\Km$ structure in the \decay{\Lb}{p\Km \jpsi} decay is weighted according to the amplitude model described in Ref.~\cite{LHCb-PAPER-2015-029}, the accuracy of which is sufficient for this analysis. The \Lb lifetime is weighted in the simulation to match its known value~\cite{PDG2022}.

All simulated \Lb decays are corrected using \decay{\Lb}{p\Km \jpsi} decays in data to account for known imperfections in modelling of the detector occupancy and the distribution of the \Lb momentum component transverse to the beam.
Similarly, the particle identification and trigger response are also corrected in simulation to give a reliable description of data. The analysis uses weights to take into account the relative integrated luminosities in the various data-taking periods.

%%======

\section{Candidate selection}
\label{sec:selection}
The selection of  \decay{\Lb}{\Lz (1520)\mumu} decays follows closely that described in Ref.\cite{LHCb-PAPER-2022-050}.
The \Lb candidate is reconstructed by combining a $\Lres$ candidate with two oppositely charged particle tracks identified as muons.
The \Lres candidate consists of two oppositely charged particle tracks, identified as a proton and a kaon, that originate from a common vertex and have an invariant mass in the interval $1470< m(p\Km)< 1570$\,\mevcc. 
Background is further suppressed by requirements on the quality of the \Lb vertex, the compatibility of the \Lb candidate with having originated from the primary \pp interaction vertex (PV), the flight distance significance of the \Lb candidate from the PV, and the separation of the final-state particles from the PV. 
The tree-level transitions $\Lb \to p\Km\jpsi$ serve as control modes. The \qsq region where the electroweak penguin diagram dominates is referred to as ``rare mode''. The peaking backgrounds vetoed in the control and rare mode include the $\phi$ mass veto applied on the $\proton_{\to \Kp}\Km$ invariant mass, where the proton is misidentified as a kaon. In addition, the backgrounds of $\Bp \to \Kp\mumu$ are vetoed in the $\Kp\mumu$ and $p_{\to \Kp}\mumu$ invariant mass, where in the latter the kaon is misidentified as a proton. In the rare mode only the \Dz and \jpsi contributions are vetoed in the $\Km \mup$ and $\Km_{\to \mun}\mup$ invariant mass. Semileptonic and hadronic decays such as $\Lb \to \Lc X$ are not vetoed, since this disturbs the angular distribution of $\cos\theta_\ell$. A systematic uncertainty is evaluated to account for this background.

Finally, to increase the signal purity, a multivariate boosted decision tree classifier (BDT) is used \cite{TMVA4} against combinatorial background, which is formed by random combinations of tracks. The BDT is trained using a dedicated sample of simulated signal events, while candidates in the upper $p\Km\mumu$ invariant-mass sideband, $m(p\Km\mumu) \in [5700, 5800]$\,\mevcc serve as a proxy for modelling the combinatorial background. The variables used in the training include kinematic and topological properties of the \Lb, the intermediate-state and final-state particles.

%%======
\section{Angular efficiency}
\label{sec:angular_eff}
The geometrical acceptance of the LHCb detector, the trigger selection, candidate reconstruction and selection may distort the angular distributions of the selected candidates. These effects are accounted for using simulation corrected as explained in Sec.~\ref{sec:detector}. For signal candidates, the acceptance is parametrised using Legendre polynomials: even orders up to four for $\cos\theta_\ell$ and all orders up to six for $\cos\theta_p$, which are found empirically to be significant while describing the acceptance across the whole phase space. The angular PDF does not depend on the angle $\phi$ as shown in Eq.~\ref{eq:PDF_ang_int12_synthese}, therefore no angular acceptance is needed to describe this angle. Within a given \qsq bin, studies of the linear correlations between $\cos\theta_\ell$ and $\cos\theta_p$ demonstrated that the angular acceptance can be factorised. 
The total angular PDF is constructed as the product of the angular PDF, shown in Eq.~\ref{eq:PDF_ang_int12_synthese}, and the angular acceptance. 
The projections of the angular acceptance can be found in Appendix~\ref{supl:AngAcc}. 

\section{Mass and angular distributions}
\label{sec:mass_angular_fits}
\begin{figure}[tbp]
	\begin{center}
		\includegraphics[width=0.45 \linewidth]{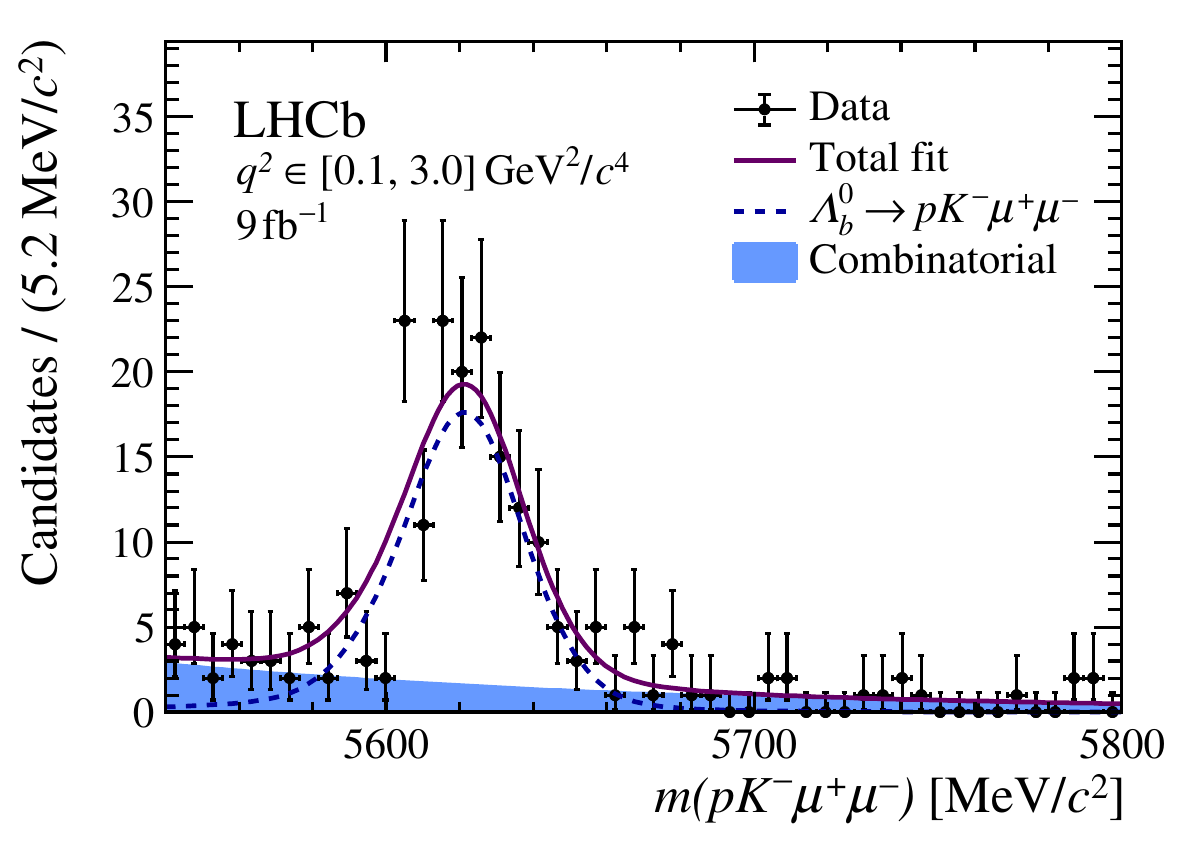}
		\includegraphics[width=0.45 \linewidth]{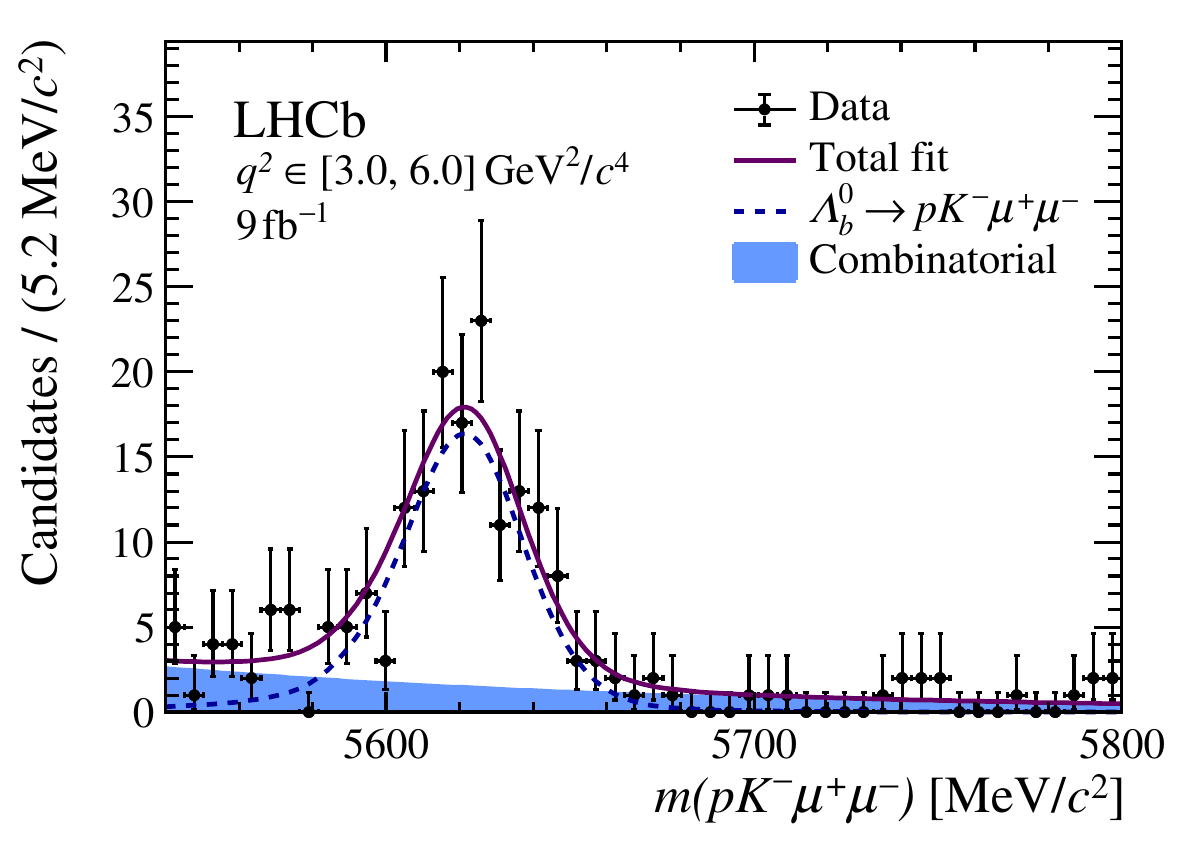}
		\includegraphics[width=0.45 \linewidth]{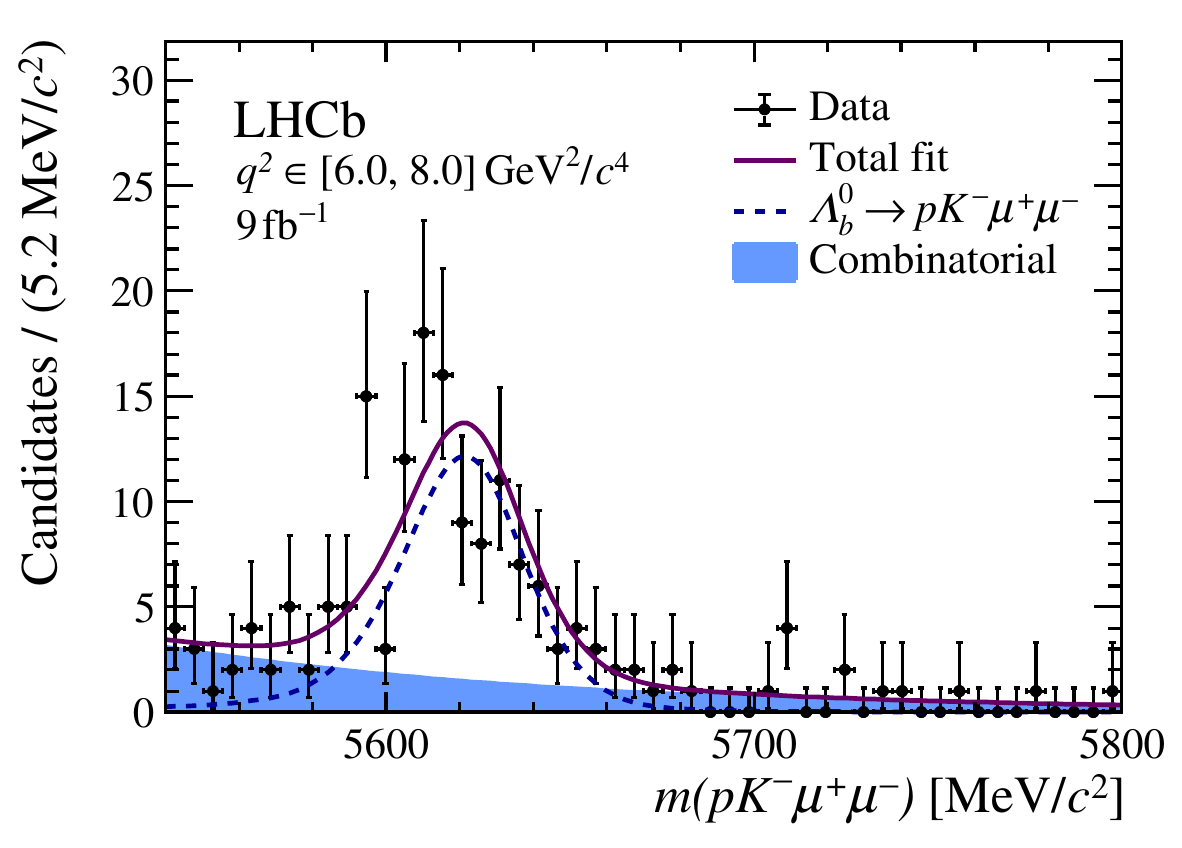}
		\includegraphics[width=0.45 \linewidth]{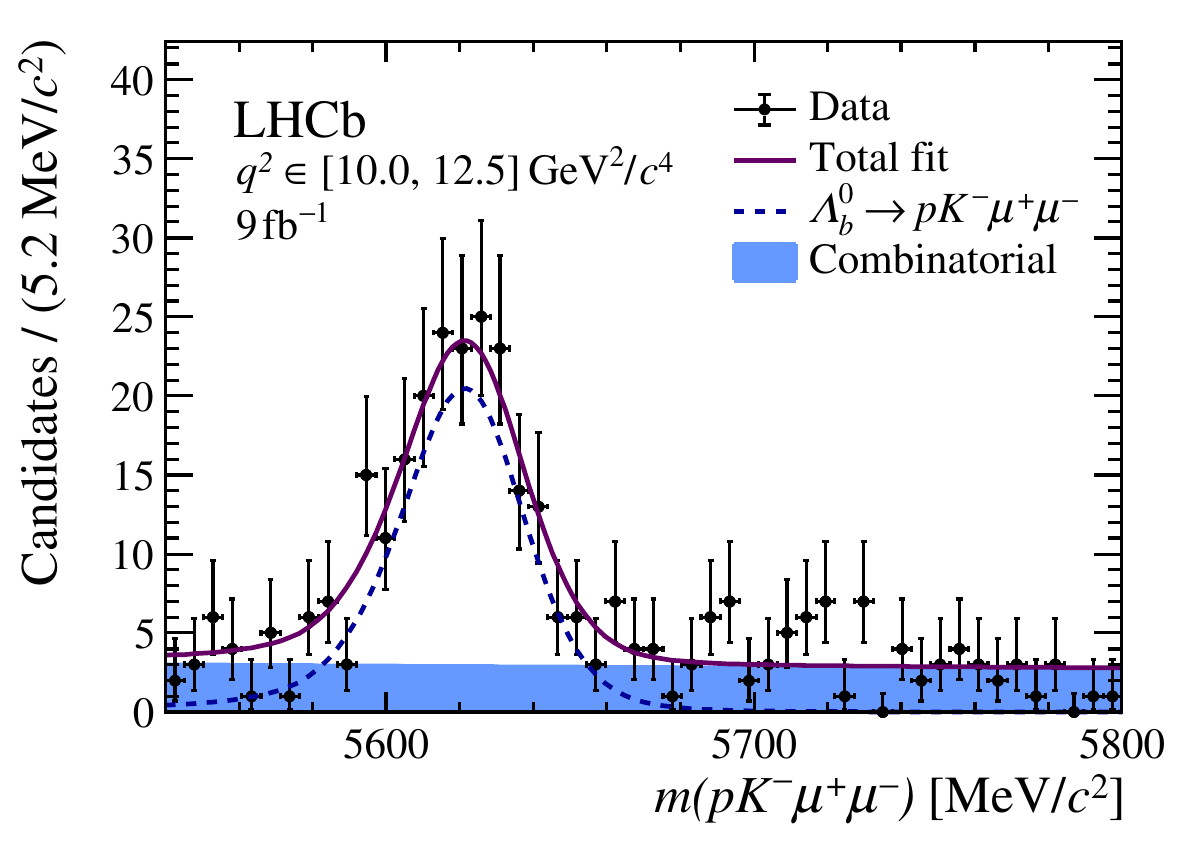}
		\includegraphics[width=0.45 \linewidth]{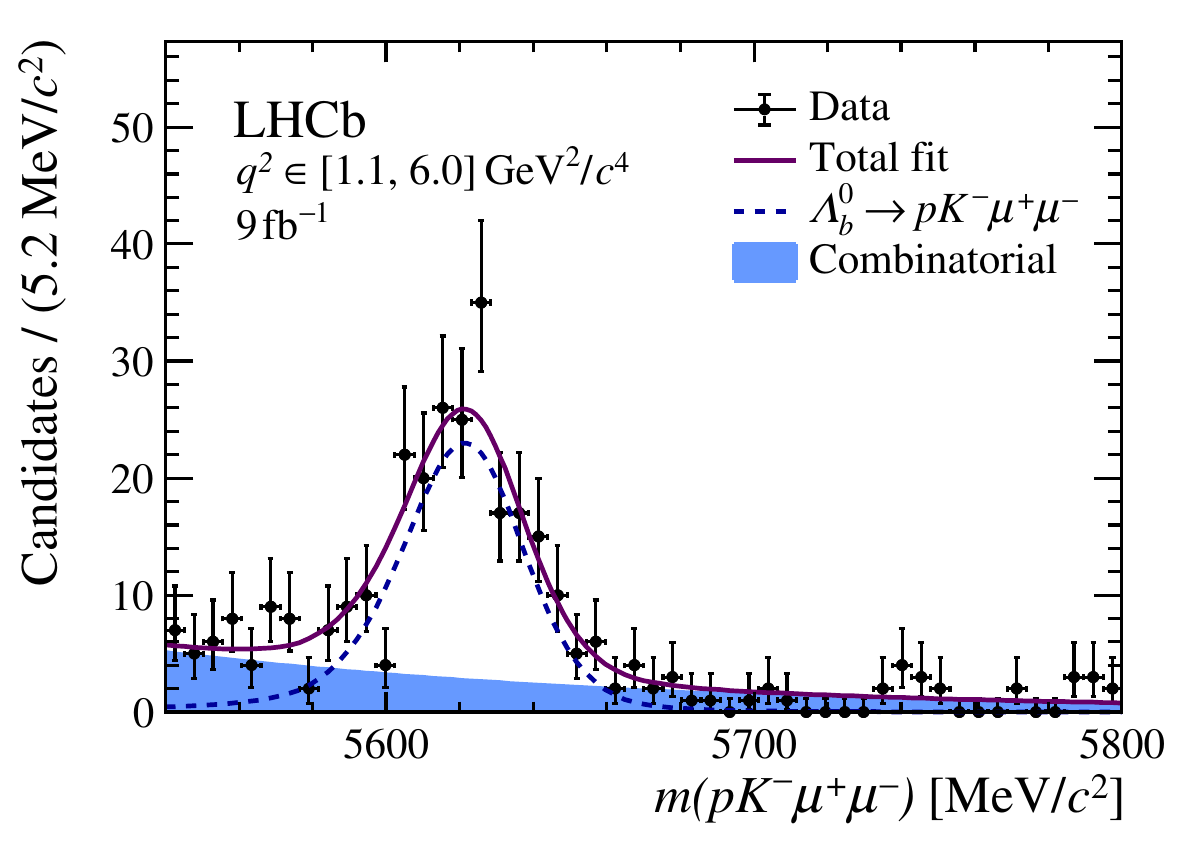}
		\caption{Invariant-mass distributions $m(pK^{-}\mu^+\mu^-)$ in each of the five $q^2$ regions considered, with fits shown for signal and combinatorial background components.}
		\label{fig:Datafits_rare}
	\end{center}
\end{figure}

The signal yield in each of the \qsq bins is estimated by performing an extended unbinned maximum-likelihood fit to the $pK^{-}\mumu$ invariant-mass distribution, which is fitted in each \qsq bin independently. 
The signal invariant-mass distribution is modelled using a Hypatia function \cite{MartinezSantos:2013ltf}, where the tails are determined from a fit to simulation. The mass of the \Lb baryon is fixed to the value obtained from a fit to the high-yield control mode in data. A scale factor is derived from data and simulation in the $\jpsi$ control mode and applied to the \Lb resolution obtained from simulation in the rare mode, in order to obtain the expected resolution in data.  
The remaining combinatorial background is modelled by an exponential distribution. Figure~\ref{fig:Datafits_rare} shows the distributions of $m(p\Km\mumu)$ for selected candidates, together with the fits described above.

 The fraction of combinatorial background is determined from the \Lb mass fit. In the next step, the $p\Km$ lineshape is fitted using Eq.~\ref{eq:totalMassPDF}, which takes into account the description of the \Lres and the $\Lstar_{J=1/2}$ resonances; the combinatorial background is modelled using a polynomial function of order one. This fit allows the fraction of the \Lres component, $f_{3/2}$, to be determined. 
 
\begin{figure}[htbp]
	\begin{center}
		\includegraphics[width=0.32 \linewidth]{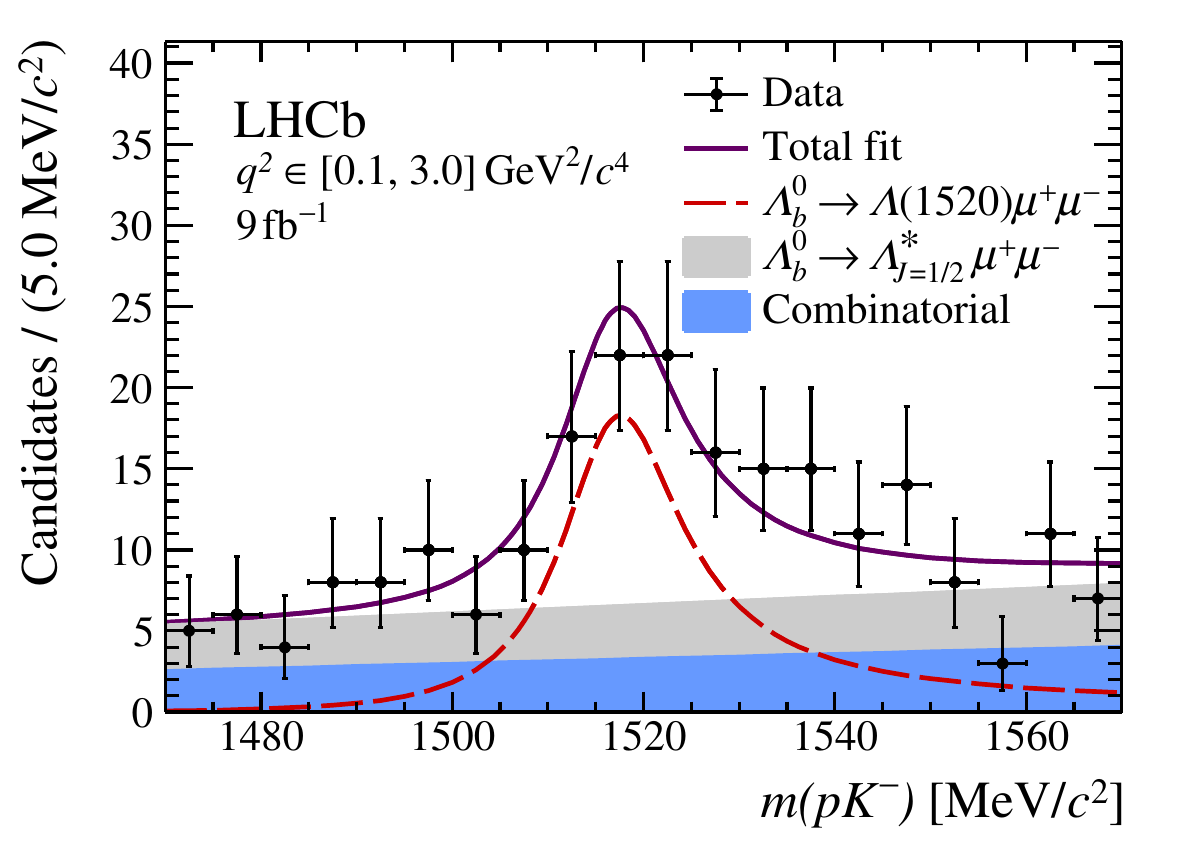}
		\includegraphics[width=0.32 \linewidth]{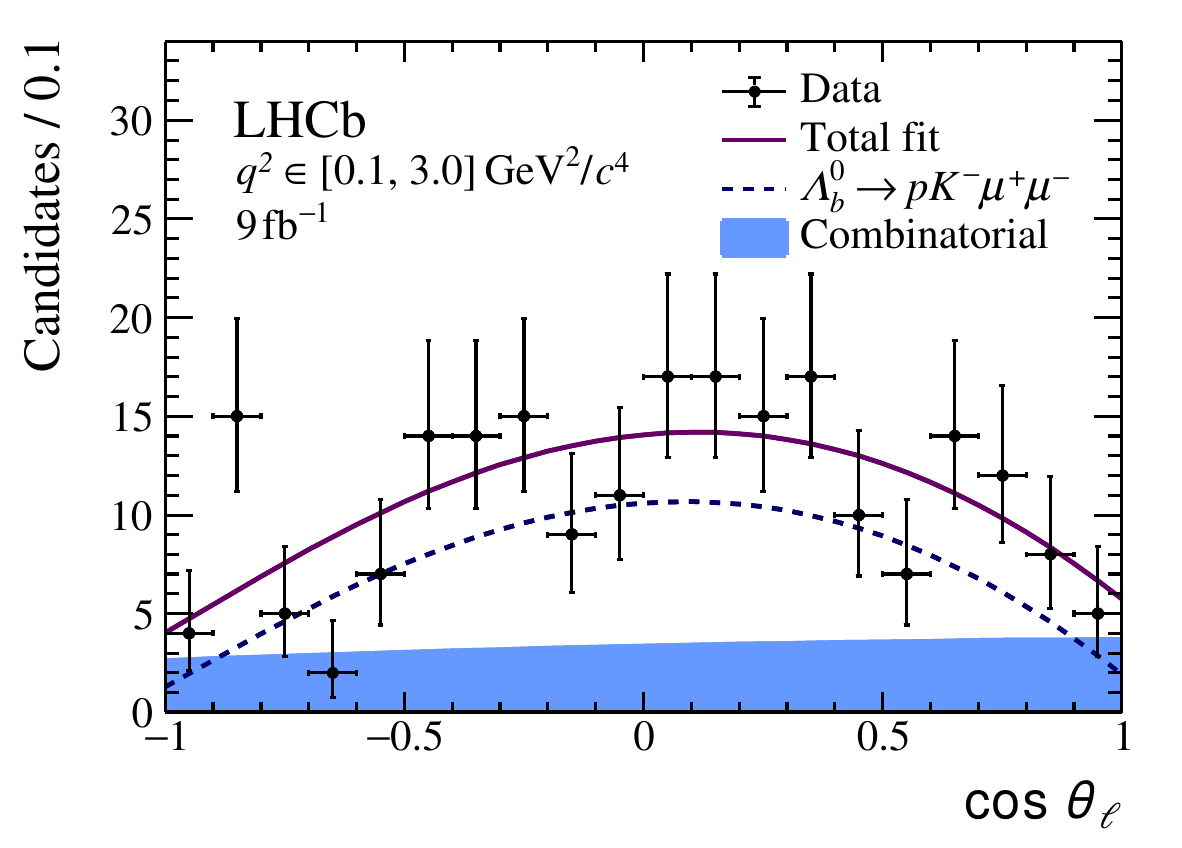}
		\includegraphics[width=0.32 \linewidth]{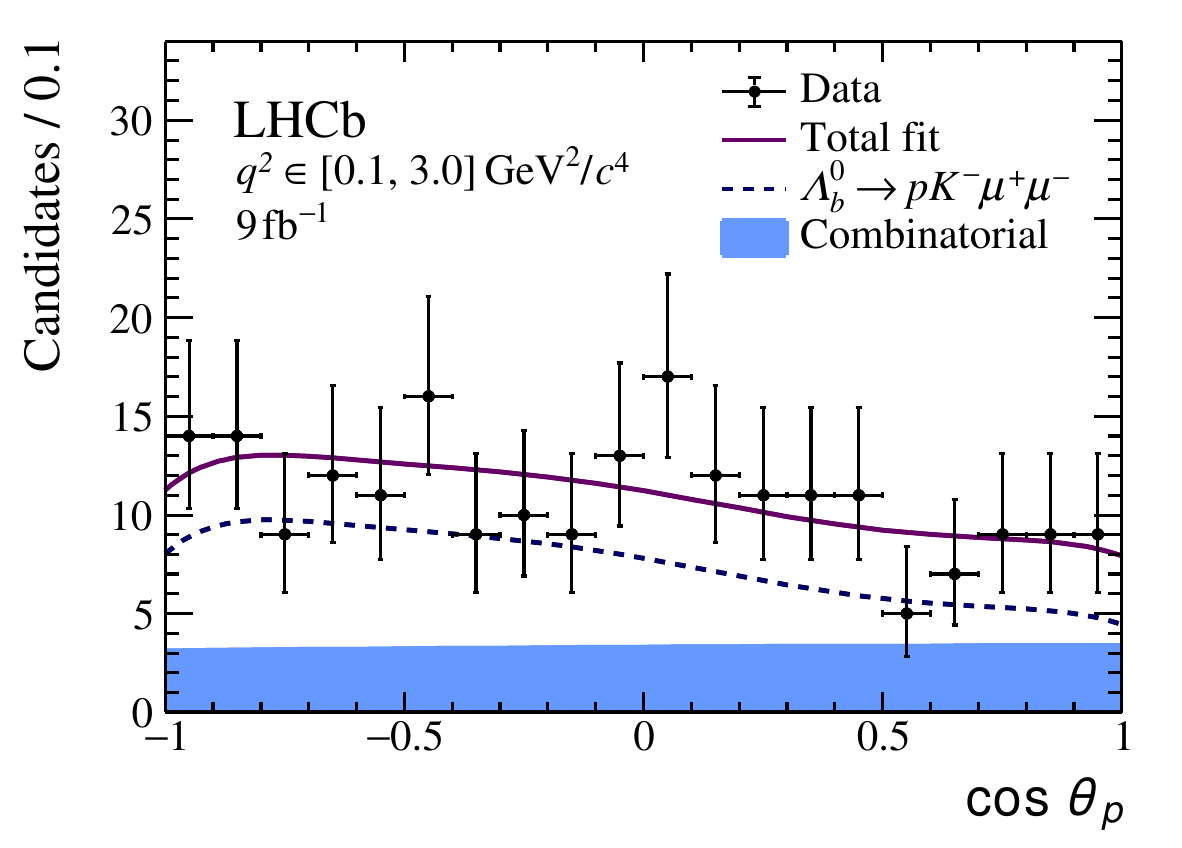}
		\includegraphics[width=0.32 \linewidth]{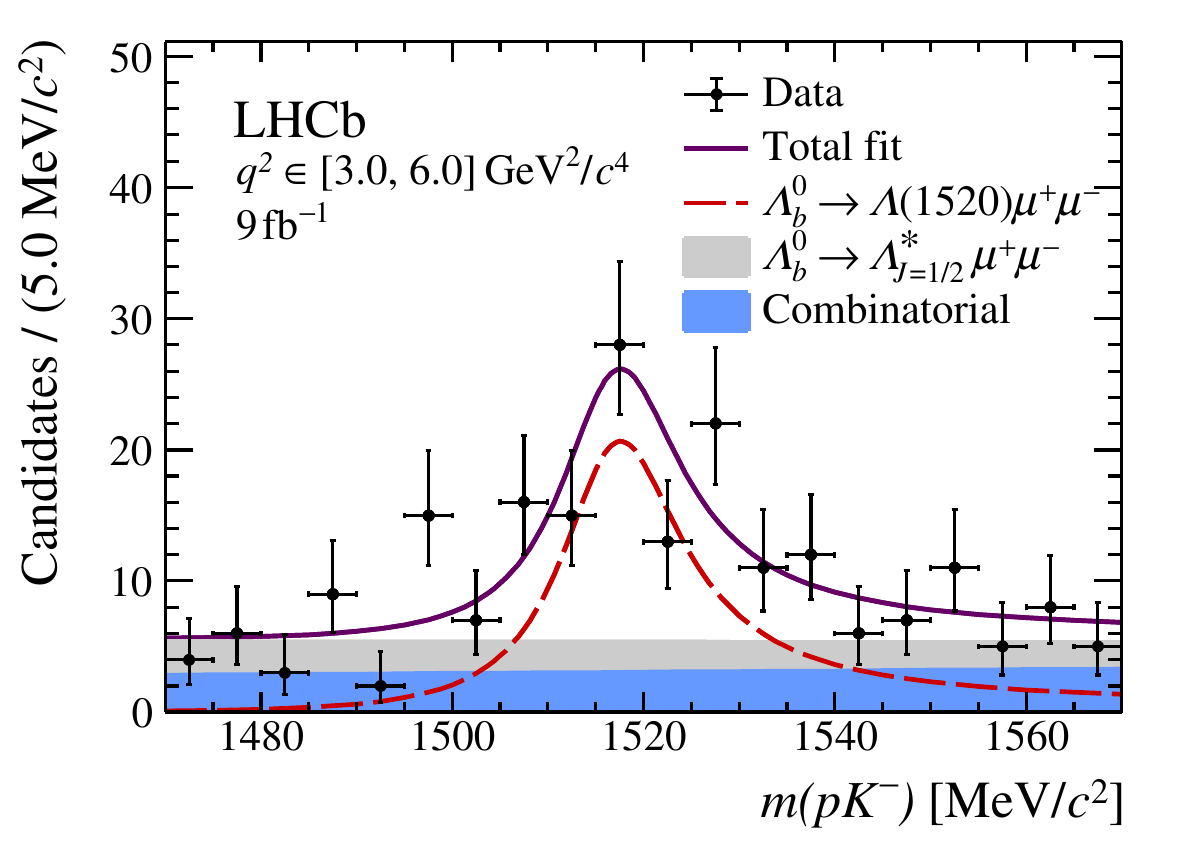}
		\includegraphics[width=0.32 \linewidth]{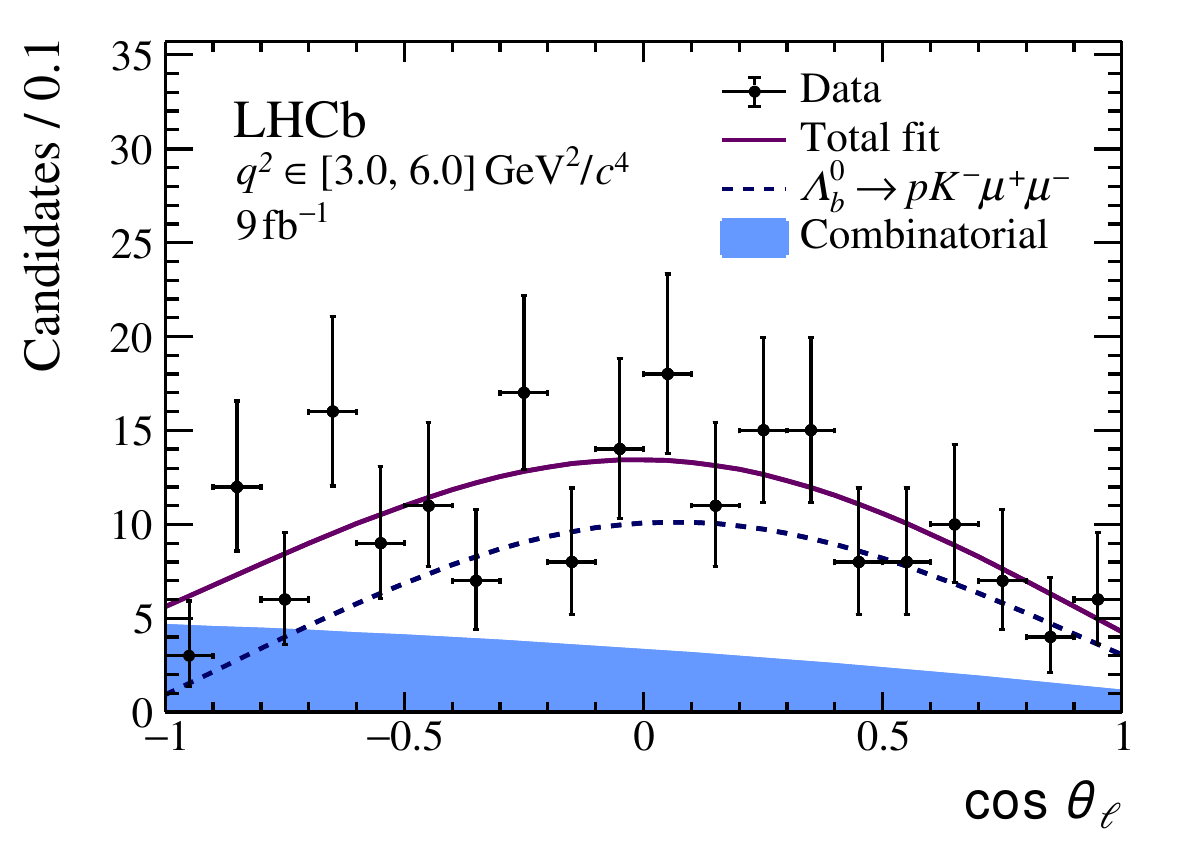}
		\includegraphics[width=0.32 \linewidth]{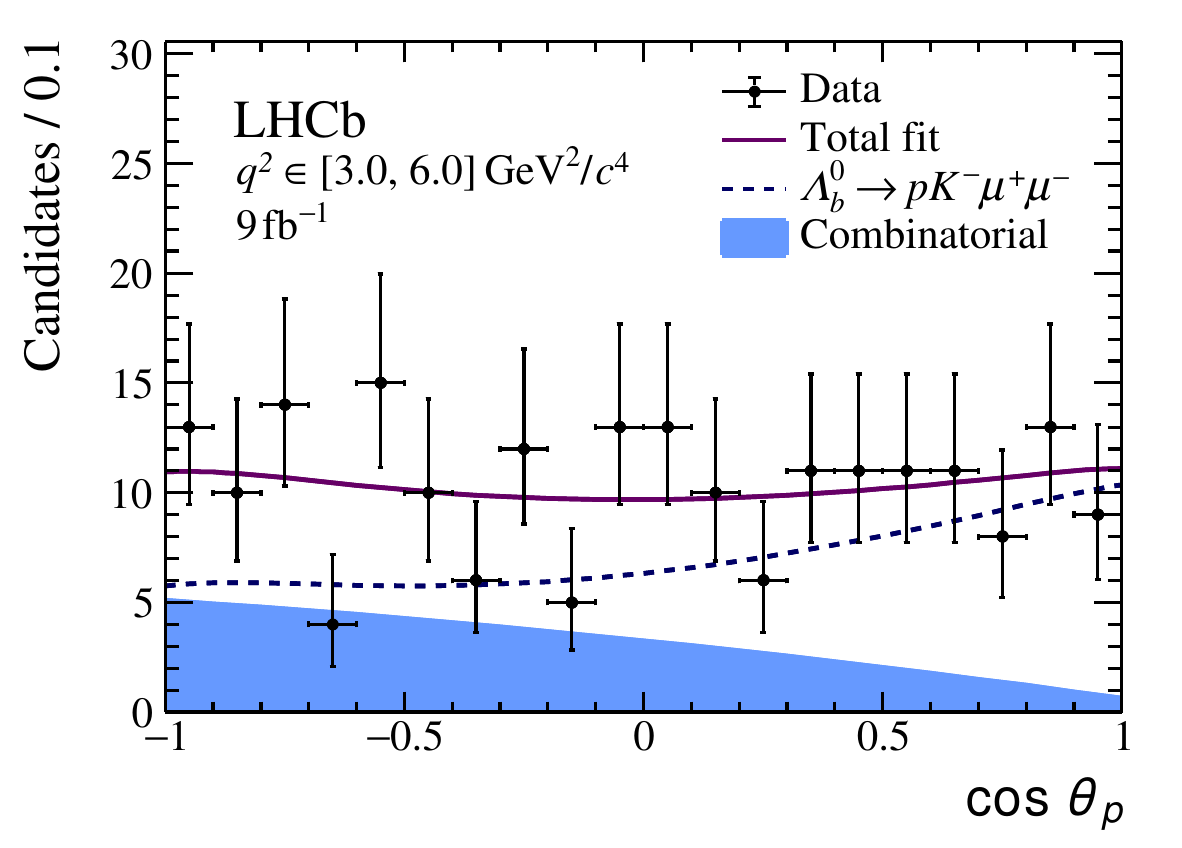}
		\includegraphics[width=0.32 \linewidth]{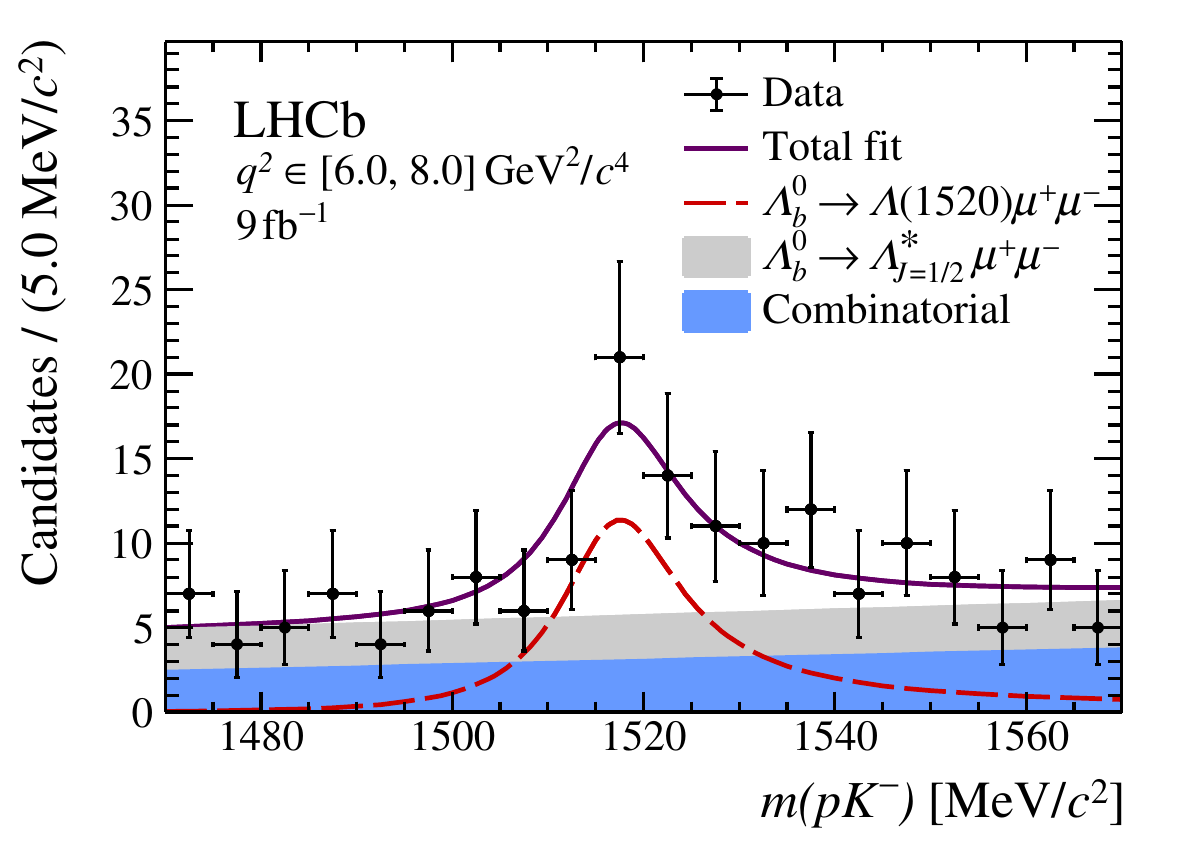}
		\includegraphics[width=0.32 \linewidth]{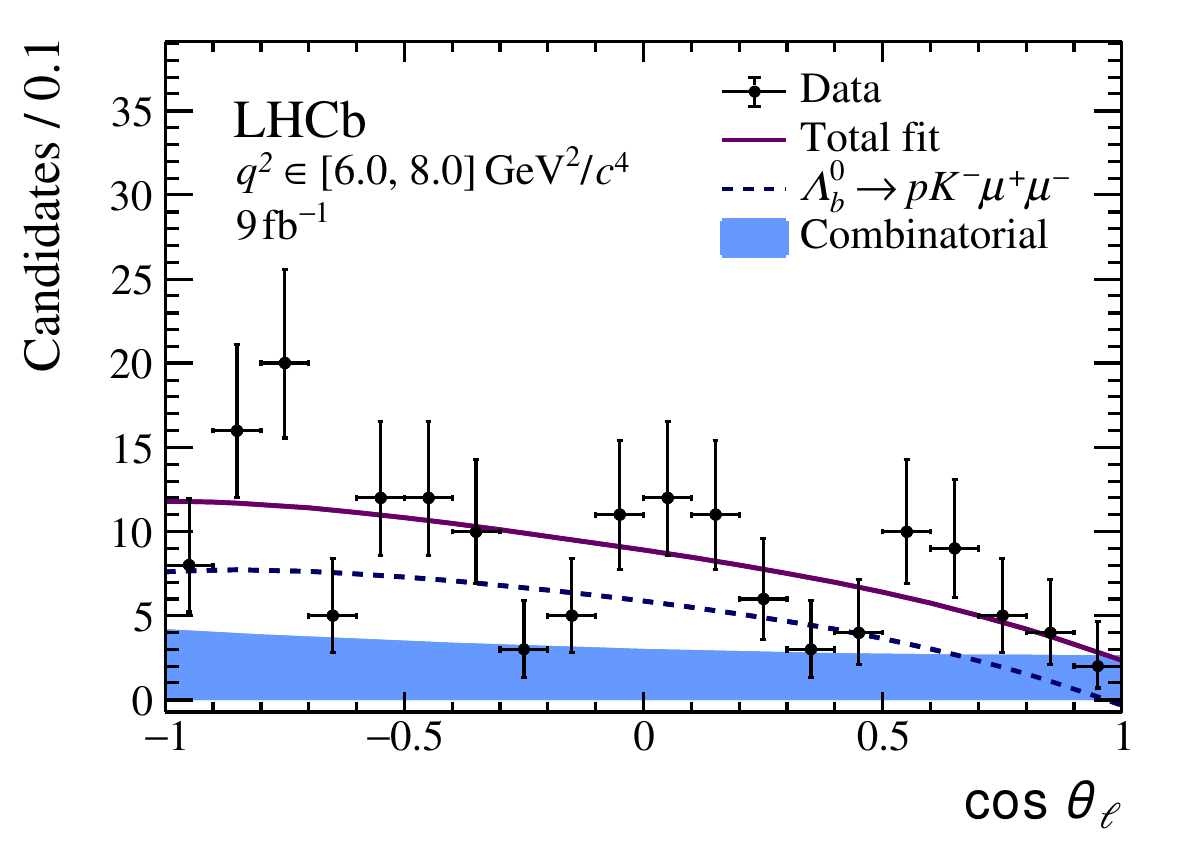}
		\includegraphics[width=0.32 \linewidth]{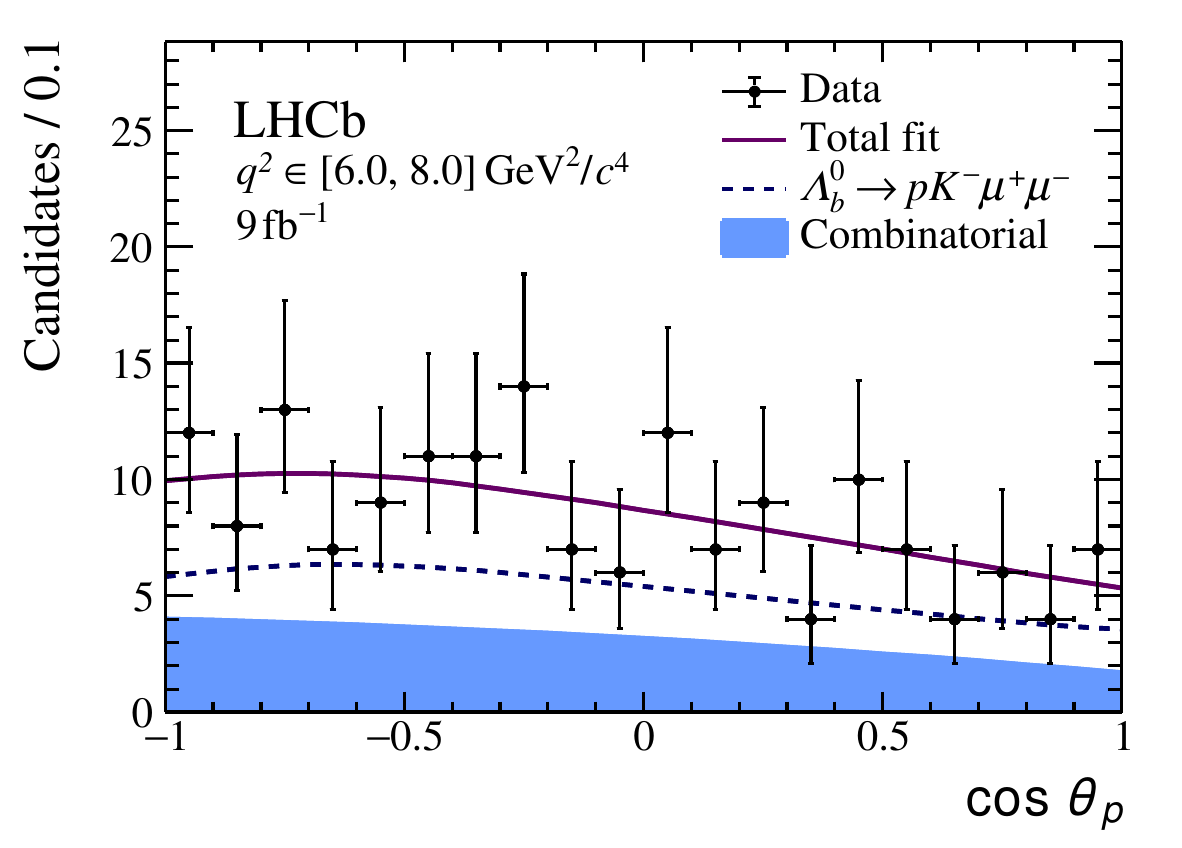}
		\includegraphics[width=0.32 \linewidth]{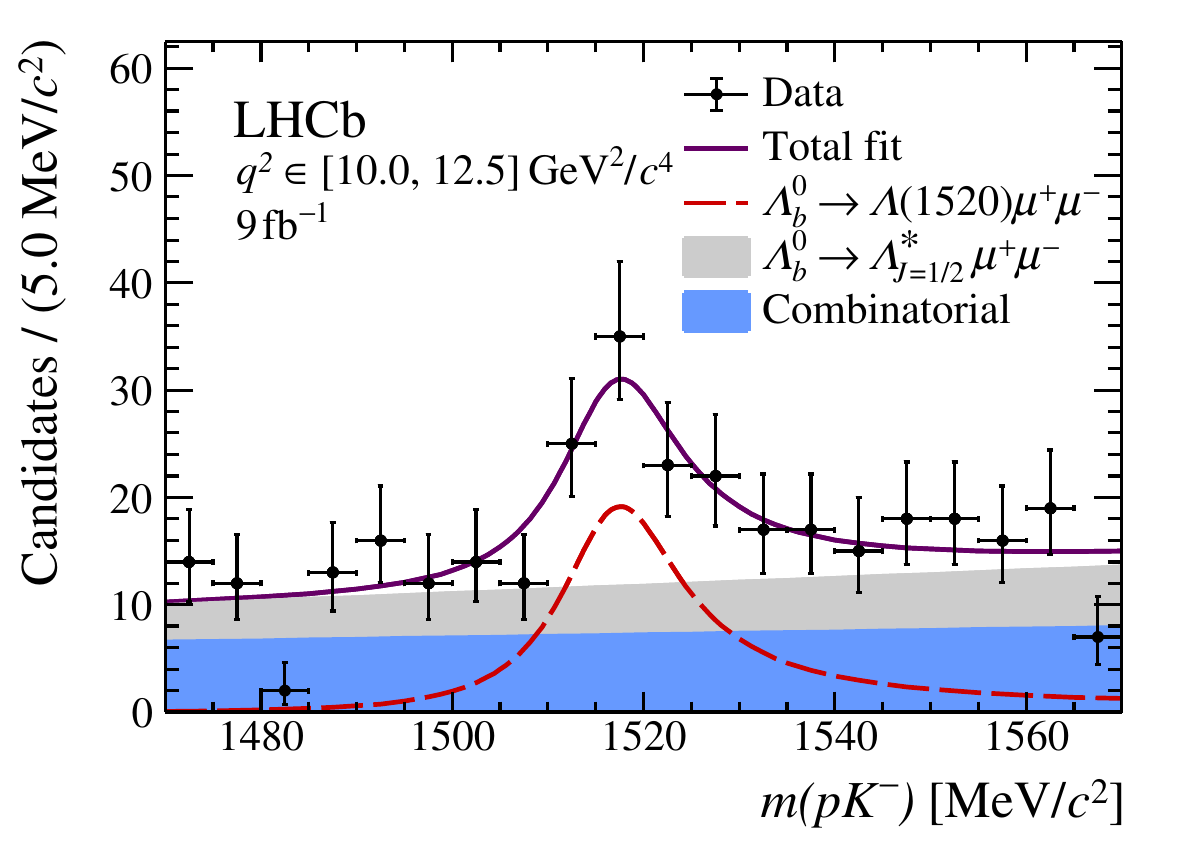}
		\includegraphics[width=0.32 \linewidth]{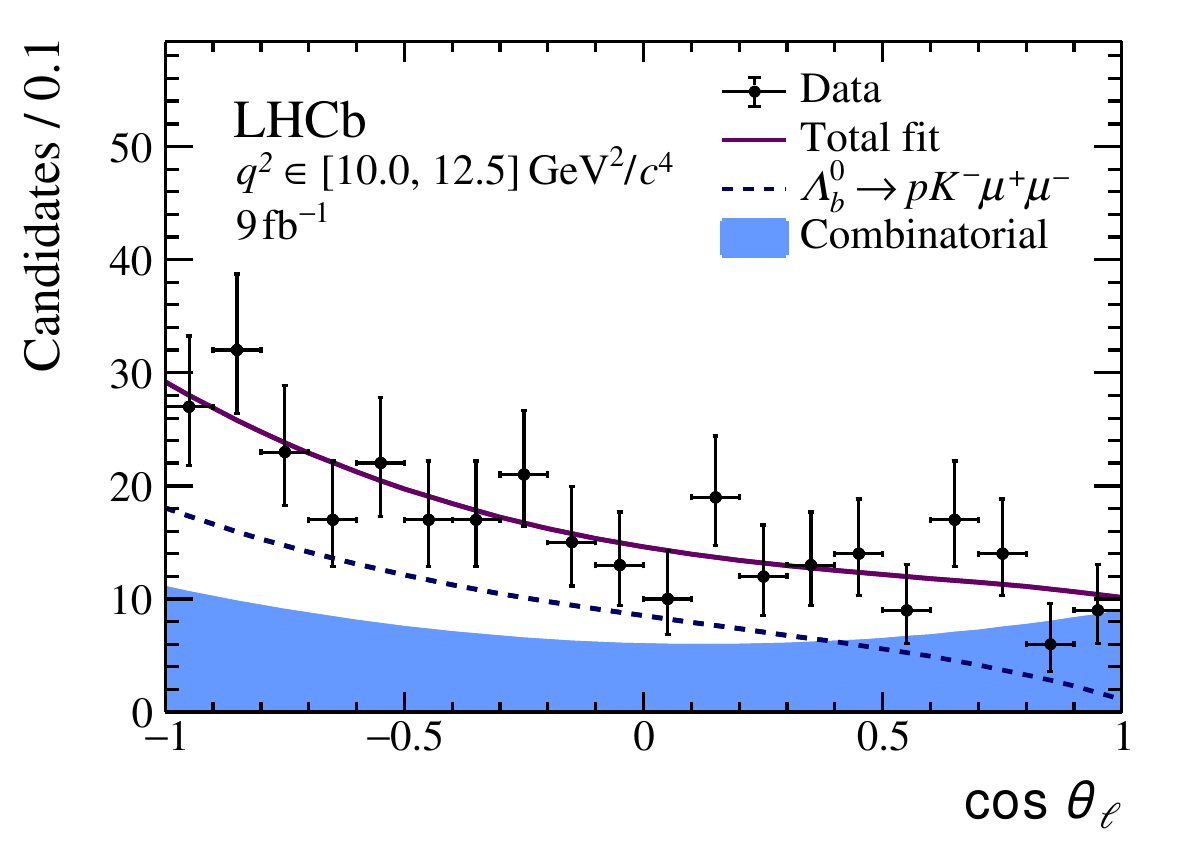}
		\includegraphics[width=0.32 \linewidth]{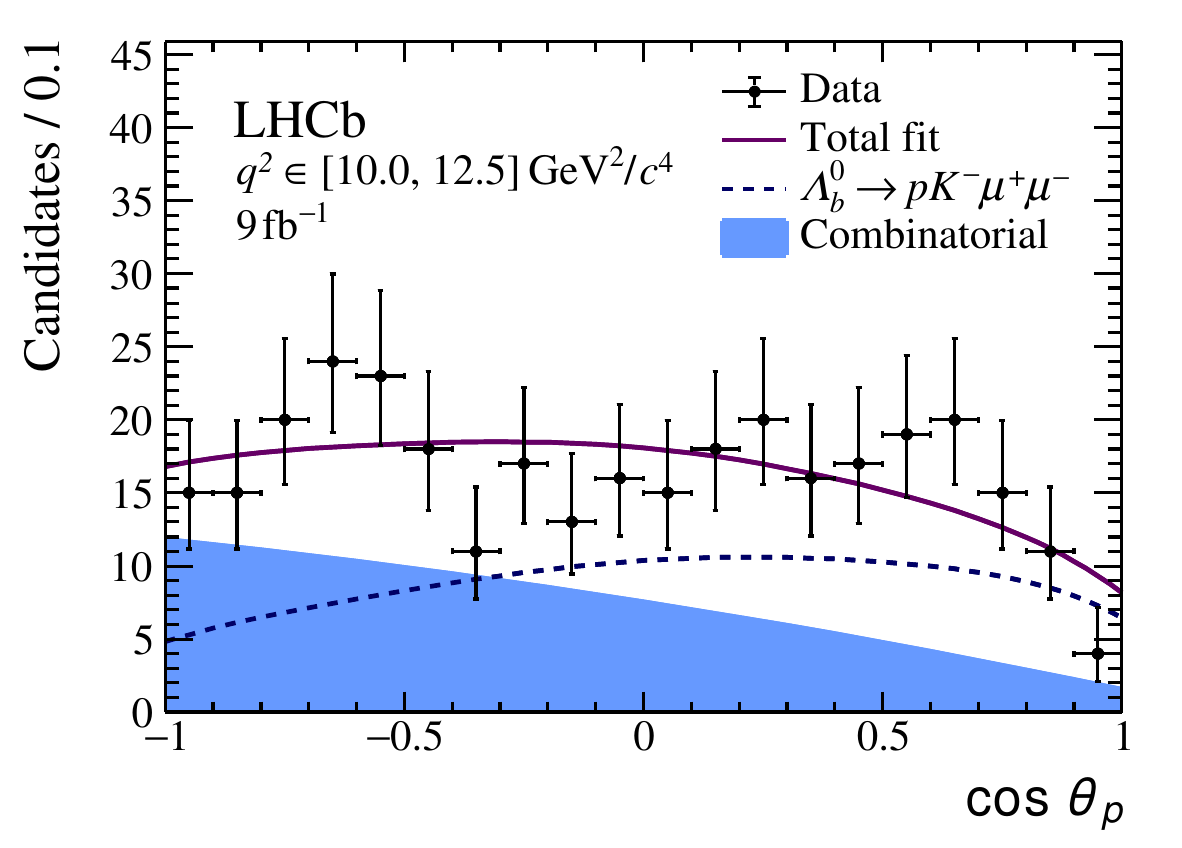}
		\includegraphics[width=0.32 \linewidth]{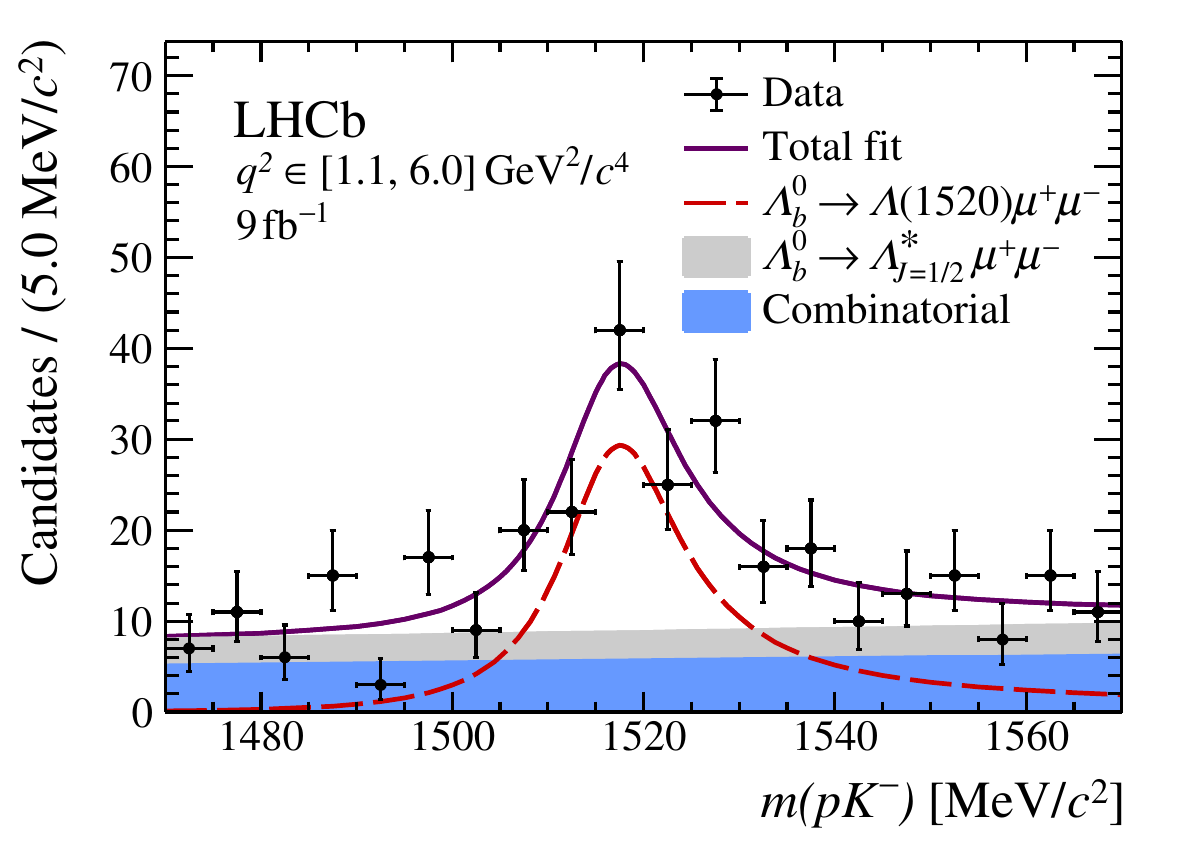}
		\includegraphics[width=0.32 \linewidth]{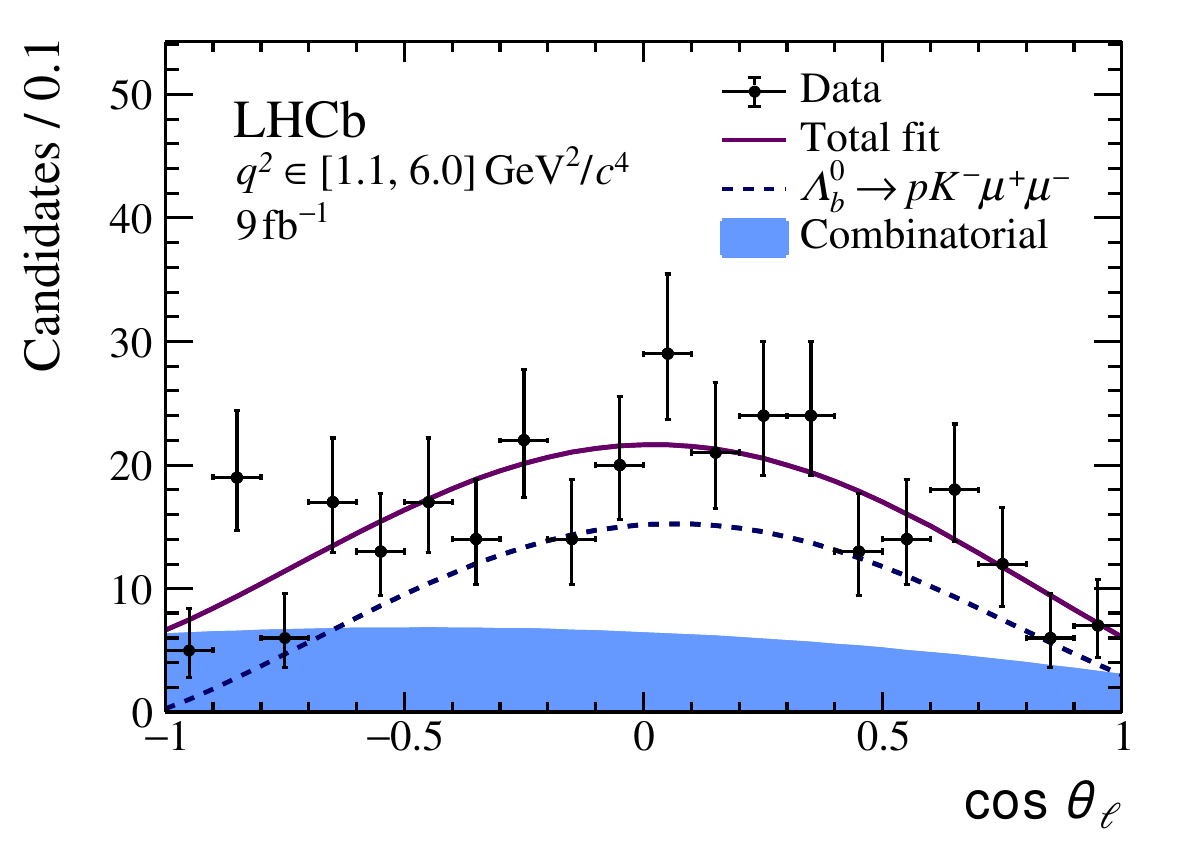}
		\includegraphics[width=0.32 \linewidth]{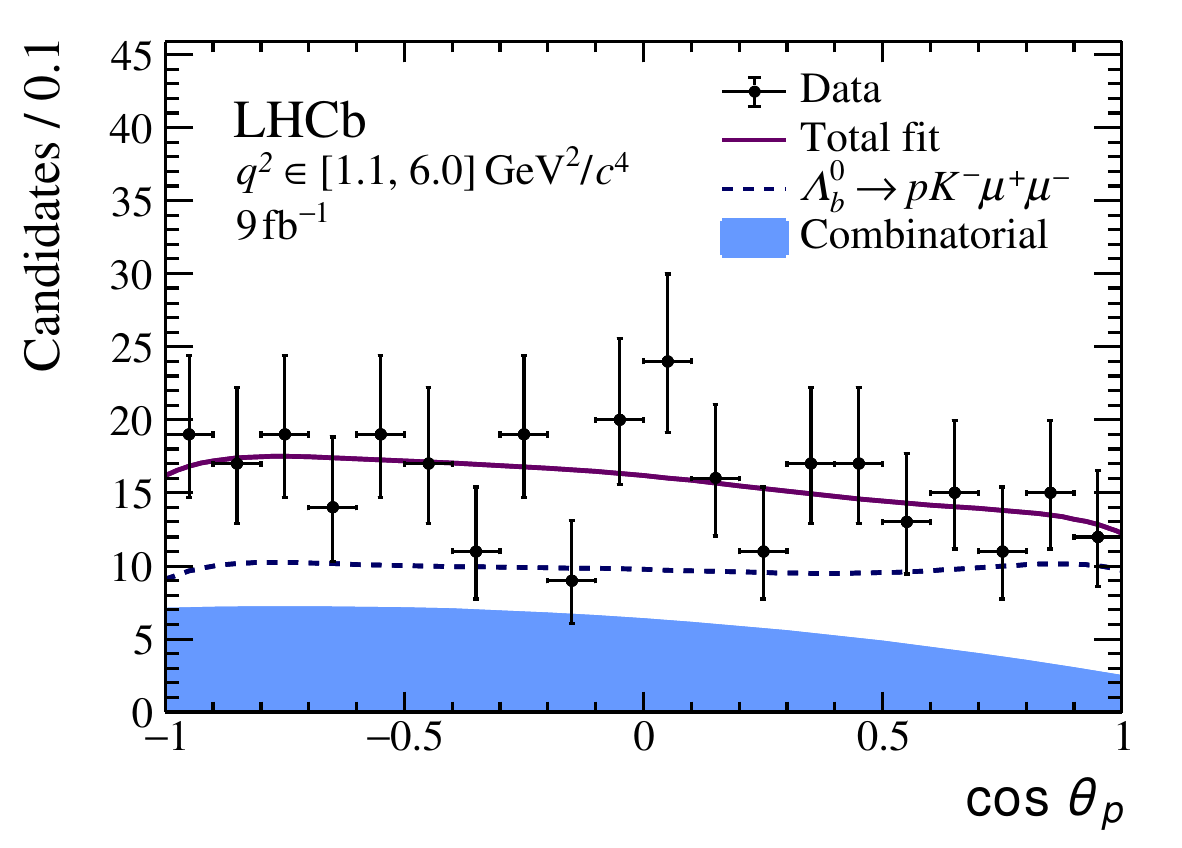}
    \end{center}
	\caption{(Left) Invariant-mass distribution in $pK^{-}$ and angular distributions (middle) $\cos\theta_\ell$ and (right) $\cos\theta_p$. The fit projections are overlaid. The five $q^2$ bins are shown from top to bottom. The ${\it \Lambda}_{\it b}^{0} \to {\it \Lambda}(1520)\mu^+\mu^-$ and ${\it \Lambda}_{\it b}^{0} \to {\it \Lambda}^{*}_{J=1/2}\mu^+\mu^-$ contributions are drawn separately in the $m(pK^-)$ projections.}
	\label{fig:SignalModeFit_wCombi1}
\end{figure}

 Finally, the $\cos\theta_\ell$ and $\cos\theta_p$ distributions are fitted using Eq.~\ref{eq:PDF_ang_int12_synthese} to describe the \Lres and $\Lstar_{J=1/2}$ components, which are shown in Fig.~\ref{fig:SignalModeFit_wCombi1}. The high-mass sideband of the $p\Km \mumu$ invariant mass, above 5700\,\mevcc, in data is used to model the angular distribution of the combinatorial background assuming factorisation, where a linear correlation does not exceed 3\%. For the extraction of its shape, the BDT working point is loosened in order to increase the sample size. Its shape is described using second-order Chebyshev polynomial functions. At different BDT working points, the angular distribution is checked to be consistent in the higher and lower $m(p\Km\mumu)$ sidebands. The angular parameters that are allowed to vary in the angular fits are $A^\ell_\mathrm{FB,\,3/2}$, $S_{1cc}$ for the \Lres contribution and $A^{\ell}_\mathrm{FB,\,1/2}$, $K_{1cc}$ for the  $\Lstar_{J=1/2}$  contribution, while $i_1$, $i_2$ correspond to the interference terms.

The fitting procedure for the $m(p\Km)$ and angular PDFs is validated extensively using samples of differing size and content. First, pseudoexperiments are generated according to the values obtained from the rare-mode fit to data, each with a size equal to the yields obtained in data. The convergence rate is observed to vary between 70\% and 88\%. This exercise is repeated with samples that are a factor of 100 larger in size, which improves the fit convergence to be within 85\% and 100\%. In both configurations, the pull distributions of the pseudoexperiments show biases. The generation of data-sized samples far from the physical boundaries demonstrates that these biases disappear, which shows that the statistical behaviour of the fit is well understood.

For the $\qsq \in [0.1,3.0]$ and $[1.1, 6.0]$\,\gevgevcccc intervals, pseudoexperiments are generated with parameter values obtained from the $m(p\Km)$ and angular fits to data. When the results of these pseudoexperiments show biases or the coverage is under- or overestimated, corrections to central values or the statistical uncertainties are derived. 
The size of the biases has at most a value of 18\% of the statistical uncertainty in $S_{1cc}$ and 17\% in $A_\text{FB,\,3/2}^\ell$. 
For the other \qsq bins, the angular PDF reported in Eq.~\ref{eq:PDF_ang_int12_synthese} is slightly negative in some regions of the ($\cos\theta_\ell$, $\cos\theta_p$) space, as visible for the $\cos\theta_\ell$ projection for the \mbox{[6.0, 8.0]}\,\gevgevcccc interval in Fig.~\ref{fig:SignalModeFit_wCombi1}. A dedicated procedure is adopted for these intervals to evaluate corrections on the central values and the statistical uncertainties. It consists of searching for a point in the 6-dimensional parameter space, \ie all the six parameters appearing in Eq.~\ref{eq:PDF_ang_int12_synthese}, which is the closest to the fit result, but where the PDF in the ($\cos\theta_\ell$, $\cos\theta_p$) space is fully positive. This prevents technical issues in the generation of pseudoexperiments. 

Finally, data in the dimuon mass region around the \jpsi meson, $\qsq \in [8.0, 10.0]$\,\gevgevcccc, where the value of  $A^{\ell}_\text{FB,\,3/2}$ is expected to vanish, are also used in two configurations: once using the dataset that contains about 43\,000 signal candidates\cite{LHCb-PAPER-2022-050} and once bootstrapping the dataset to a size corresponding to the rare mode yields \ie about 120 signal candidates. 
In the first configuration, the value of $A^{\ell}_\text{FB,\,3/2}$ is found to be compatible with zero as expected. All the pseudoexperiments converged. The Gaussian fit to the pulls of the pseudoexperiments demonstrates that the fits on data are unbiased and the uncertainties are well estimated. In the second configuration, fits are performed to 1000 bootstrapped data samples. The mean of the Gaussian fit to the pulls of these fits is not biased and the uncertainty is well estimated. However, in this case, the convergence rate decreased from 100\% to a similar rate as observed in the rare-mode fit. This test is performed to disentangle the impact of the small sample size and the location of the observable values with respect to the physical boundary.

\section{Systematic uncertainties}
\label{sec:systematics}
Potential sources of systematic uncertainty are identified as those that may modify the mass or angular distribution of the signal or background components in the analysis. Each source is evaluated by comparing the baseline analysis with alternative variations, and a corresponding systematic uncertainty is assigned whenever a significant effect is observed. These are estimated as the difference between the baseline and alternative configurations of the fit. A full breakdown of the systematic uncertainties for both $A^{\ell}_\mathrm{FB,\,3/2}$ and $S_{1cc}$ in each region of $\qsq$ is presented in Tables~\ref{tab:SystematicsSummary-AFB} and \ref{tab:SystematicsSummary-S1cc}.
\begin{table}[tb]
    \centering
    \caption{Absolute values of the considered systematic uncertainties in \qsq intervals, which have the units \gevgevcccc, for $A^\ell_\text{FB,\,3/2}$. Systematic uncertainties that do not apply are indicated by ``--''.}
    
   \resizebox{\textwidth}{!}{\begin{tabular}{l l | rrrrr }

    \textbf{Source} & \textbf{Variation} &  \textbf{[0.1,\,3.0]}\hspace{-2mm} & \textbf{[3.0,\,6.0]}\hspace{-2mm} & \textbf{[6.0,\,8.0]}\hspace{-2mm} & \textbf{[10.0,\,12.5]}\hspace{-4mm} & \textbf{[1.1,\,6.0]}\hspace{-1mm}\\
\toprule
    \multirow{5}{*}{\Lb mass fit} & \Lb mass shape &
    $<0.001$ & $0.012$ & $0.002$ & $0.014$ & $0.005$\; \\
     & $M_\Lb$ and $\Gamma_{\!\Lb}$ &
    $<0.001$ & $<0.001$ & $<0.001$ & $0.002$ & $<0.001$\; \\
     & Comb. shape &
     $<0.001$ & $0.035$ & $0.001$ & $0.001$ & $0.002$\; \\
     & Constrain $f_\text{comb}$ &
     $0.001$ & $0.002$ & $0.001$ & $0.002$ & $<0.001$\;\\
     & Correct bias in $f_\text{comb}$ &
     $<0.001$ & $0.004$ & $0.001$ & --\hspace{4mm} & $<0.001$\; \\
\midrule
    \multirow{3}{*}{Background} & Increased comb. sample & 
     $0.068$ & $0.051$ & $0.007$ & $0.049$ & $0.018$\; \\
     & Reduced comb. sample & 
     $0.013$ & $0.087$ & $0.003$ & $0.070$ & $0.045$\; \\
     & \Lc veto & 0.083 & 0.108 & 0.083 & 0.018 & 0.036\; \\
\midrule
    \multirow{6}{*}{$m(p\Km)$ fit} & Constrain $f_{3/2}$ & 
    $0.001$ & $<0.001$ & $0.002$ & $0.003$ & $<0.001$\; \\
    ~ & Correct bias in $f_{3/2}$ & 
    --\hspace{4mm} & $0.002$ & --\hspace{4mm} & --\hspace{4mm} & --\hspace{4mm}\; \\
     ~ & $M_{\Lres}$ \& $\Gamma_{\!\Lres}$  & 
     $<0.001$  &$0.001$ & $0.002$ & $0.002$ & $<0.001$\; \\
     ~ & $r_{\Lres}$ value   & 
     $<0.001$ & $<0.001$ & $<0.001$ & $0.001$ & $<0.001$\; \\  
     & $\Lstar_{J=1/2}$ model & 
     $<0.001$ & $<0.001$ & $<0.001$ & $0.015$ &  $<0.001$\;  \\
     & Comb. shape & 
     $<0.001$ & $<0.001$ & $<0.001$ & $<0.001$ &  $<0.001$\;  \\
\midrule
    \multirow{3}{*}{Angular acc.} & Simulation corr. & $0.013$ & $0.003$ & $0.001$ & $0.011$ & $0.003$\;  \\
     & Legendre order & $0.004$ & $0.005$ & $0.003$ & $0.011$ &  $<0.001$\; \\
     & Simulation size & $<0.001$ & $<0.001$ & $0.003$ & $0.001$ &  $<0.001$\;\\
\midrule
    \multirow{2}{*}{Angular fit} & {Comb. shape} & $0.023$ & $0.003$ & $0.001$ & $0.018$ & $0.004$\; \\
     & Fit bias  & $0.007$ & $0.009$ & $0.008$ & $0.008$ & $0.007$\; \\
    \toprule
    \multicolumn{2}{l|}{Total systematic uncertainty} & $0.111$ & $0.153$ & $0.085$ & $0.093$ & $0.061$\; \\
    \bottomrule
    \end{tabular}}
    \label{tab:SystematicsSummary-AFB}
\end{table}

\begin{table}[tb]
    \centering
    \small  
    \caption{Absolute values of the considered systematic uncertainties in \qsq intervals for $S_{1cc}$ in the units \gevgevcccc. Systematic uncertainties that do not apply are indicated by ``--''.}
    
   \resizebox{\textwidth}{!}{\begin{tabular}{l l|rrrrr }
 \textbf{Source} & \textbf{Variation} &  \textbf{[0.1,\,3.0]}\hspace{-2mm} & \textbf{[3.0,\,6.0]}\hspace{-2mm} & \textbf{[6.0,\,8.0]}\hspace{-2mm} & \textbf{[10.0,\,12.5]}\hspace{-4mm} & \textbf{[1.1,\,6.0]}\hspace{-1mm}\\
\toprule
    \multirow{5}{*}{$\Lb$ mass fit} & \Lb mass shape &
    $0.010$ & $0.014$ & $0.041$ & $0.019$ & $0.016$\; \\
    & $M_\Lb$ and $\Gamma_{\!\Lb}$ &
    $0.002$ & $0.001$ & $0.005$ & $0.003$ & $<0.001$\; \\
     & Comb. shape & 
     $0.001$ & $0.065$ & $0.012$ & $0.001$ & $0.011$\; \\
     & Constrain $f_\text{comb}$ &
     $<0.001$ & $0.006$ & $0.003$ & $0.005$ & $0.001$\; \\
     & Correct $f_\text{comb}$ &
     $0.006$ & $0.008$ & $0.011$ &  --\hspace{4mm} & $0.002$\; \\
\midrule
     \multirow{3}{*}{Background}& Increased comb. sample 
     & $0.129$ & $0.007$ & $0.104$ & $0.044$ & $0.101$\; \\
     & Reduced comb. sample & 
     $0.004$ & $0.219$ & $0.022$ & $0.152$ & $0.076$\;  \\ 
     & \Lc veto & $0.167$ & $0.166$ & $0.122$ & $0.058$ & $0.120$\; \\
\midrule
    \multirow{6}{*}{$m(p\Km)$  fit} & Constrain $f_{3/2}$ & 
    $0.003$ & $0.001$ & $0.013$ & $<0.001$ & $<0.001$\;  \\
     & Correct bias in $f_{3/2}$ & 
      --\hspace{4mm} & $0.007$ &  --\hspace{4mm} &  --\hspace{4mm} &  --\hspace{4mm}\; \\
     & $M_{\Lres}$ \& $\Gamma_{\!\Lres}$ & 
     $<0.001$      & $<0.001$ & $0.012$ & $<0.001$ & $<0.001$\; \\
     ~ & $r_{\Lres}$ value   & 
     $<0.001$ & $0.002$ & $0.006$ & $<0.001$ & $<0.001$\; \\ 
     & $\Lstar_{J=1/2}$ model & 
     $0.004$  & $<0.001$ & $0.004$ & $0.004$ & $<0.001$\;  \\
     & Comb. shape & 
     $<0.001$ & $<0.001$ & $<0.001$ & $<0.001$ &  $<0.001$\;   \\
\midrule
    \multirow{3}{*}{Angular acc.} & Simulation corr. & $0.060$ & $0.009$ &  $<0.001$ & $0.001$ & $0.016$\; \\
     & Legendre order & $0.002$ & $<0.001$ & $<0.001$ & $0.004$ & $0.003$\;  \\
     & Simulation  size & $0.009$ & $0.007$ & $0.003$ & $0.016$ & $0.003$\; \\
\midrule
    \multirow{2}{*}{Angular fit} & {Comb. shape} & $0.007$ & $0.013$ & $0.016$ & $0.049$ & $<0.001$\; \\
     & Fit bias  & $0.013$ & $0.016$ & $0.019$ & $0.013$ & $0.013$\; \\   
\toprule
    \multicolumn{2}{l|}{Total systematic uncertainty} & $0.221$ & $0.284$ & $0.170$ & $0.178$ & $0.177$\; \\
    \bottomrule 
    \end{tabular}}
    \label{tab:SystematicsSummary-S1cc}
\end{table}

%--------------------
Effects related to the parametrisation of the $m(p\Km\mu^+\mu^-)$ distribution are investigated by changing the baseline model for the signal component from a Hypatia function to a double-sided Crystal Ball \cite{Skwarnicki:1986xj}.
In order to increase the overall fit stability, the \Lb mass and width parameters are fixed in the baseline fit to the values determined using the high-yield \decay{\Lb}{p\Km \jpsi} control-mode data. In the alternative fit, these parameters are Gaussian constrained. 
The sensitivity to the background component model is tested by changing the baseline choice 
from an exponential function to a first-order polynomial.
The fraction of combinatorial background is constrained using the fits to $m(p\Km)$ data rather than being fixed. Pseudoexperiments testing the fit of the \Lb mass lineshape indicated the presence of a relatively small bias on the fraction of the combinatorial background. A systematic uncertainty is assigned to take this bias into account. 

%--------------------

The yield in the high-mass sideband range of the combinatorial background sample is increased by loosening the nominal BDT requirement further and a systematic uncertainty is evaluated. 
Another systematic uncertainty is evaluated by removing all background candidates used both for the extraction of the combinatorial background shape describing the $m(p\Km)$ and the angular distribution and for the fit to extract the angular observables. 
An additional systematic uncertainty is assigned to account for the potential background contamination from $\Lb \to \Lc (p\Km \pip) \mun \bar{\nu}_{\mu}$ and $\Lb \to \Lc (p\Km \pip) \pim $ decays with pions misidentified as muons. A veto to an interval of 30\,\mevcc around the \Lc mass is applied in the $p\Km\mup$ invariant mass and a systematic uncertainty evaluated with this alternative dataset.

%--------------------

In each $q^2$ interval, the fraction $f_{3/2}$ of the $\Lres$ resonance is determined from a fit to the $m(p\Km)$ lineshape and fixed to its central value in the subsequent angular fit. As an alternative, $f_{3/2}$ is treated as a constrained parameter, with a Gaussian constraint centred on the value obtained from the lineshape fit. When a bias is observed on $f_{3/2}$, which is only the case for the $\qsq \in [3.0,6.0]$~\gevgevcccc bin, a systematic uncertainty on the angular observables is computed by correcting for it.

The mass and the width of the \Lres resonance are allowed to vary with Gaussian constraints around the values with the associated uncertainties obtained from the fit to data of the \jpsi control mode. The value of the interaction radius of the \Lres BW, $r_{\Lres}$, is varied from 0 to 3\,$\gev^{-1}$ according to Ref.~\cite{Ryd:2005zz}.

The description of the $\Lstar_{J=1/2}$ component is changed from the nominal model of a polynomial of order one to a sum of two exponential functions, while for the combinatorial background the shape parameters are constrained instead of being fixed. 

%--------------------

Uncertainties are considered on the corrections applied to the simulated sample.
Effects arising from modelling the efficiency in $\cos\theta_\ell$ and $\cos\theta_p$ are evaluated by increasing the order of the corresponding polynomial by one even order in each of the observables separately.
The limited size of the simulation sample may also have an impact on the description of the angular acceptance. Variations have been obtained by bootstrapping the simulation sample to its original size. 

%--------------------
Extensive pseudoexperiment studies are performed by varying the sample size and using different parameter and observable values. The pull distributions when generating pseudoexperiments based on the fit result values indicate the presence of biases due to the proximity of the obtained fitted values to physical boundaries. The uncertainty on the bias is accounted for as part of the systematic uncertainties. 

The dominant systematic uncertainties arise from the background contribution, in particular the combinatorial background and the hadronic and semileptonic decays of the \Lb baryon via a \Lc baryon.

\section{Results}
\label{sec:results}
An angular analysis of the $\Lb\to\Lres\mu^+\mu^-$ decays is reported, using a dataset collected with the LHCb detector between 2011 and 2018, corresponding to an integrated luminosity of 9\invfb of proton-proton collisions. The analysis is performed in five bins of \qsq. The angular observables $A^\ell_\text{FB,\,3/2}$ and $S_{1cc}$ are measured for the first time in $\Lb\to\Lres\mu^+\mu^-$ decays. The angular fit model had been applied for the first time, including the dominant and narrow \Lres resonance, as well as the contributions of spin $ J = 1/2$ $\Lambdares^{*}$ resonances in the $\Lres$ mass window. The interferences between the different $\Lambdares^{*}$ resonance states in the \Lres region are modelled in the angular fit and have been found, as anticipated from the studies described in Appendix~\ref{supl:Interferences}, to be non-negligible. 

\begin{figure} [t]
    \centering
    \includegraphics[width=0.48\linewidth]{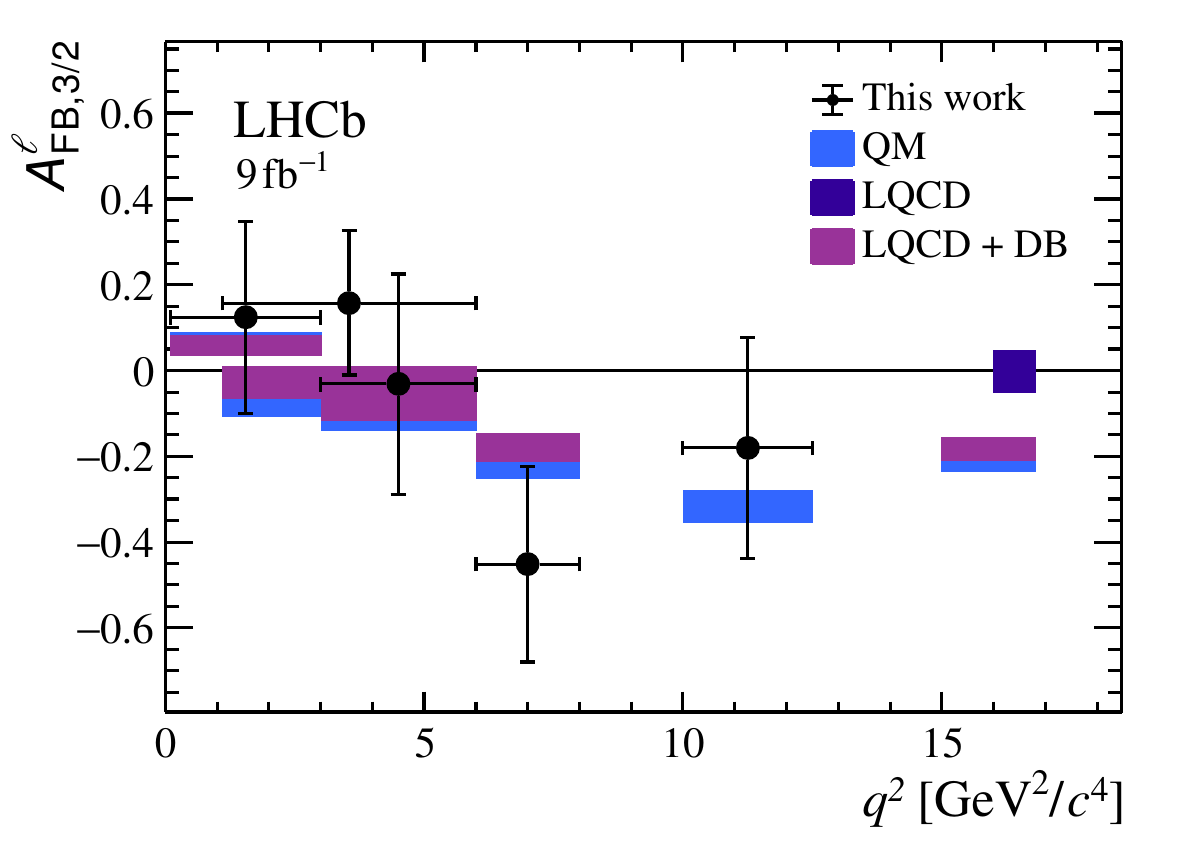}
    \includegraphics[width=0.48\linewidth]{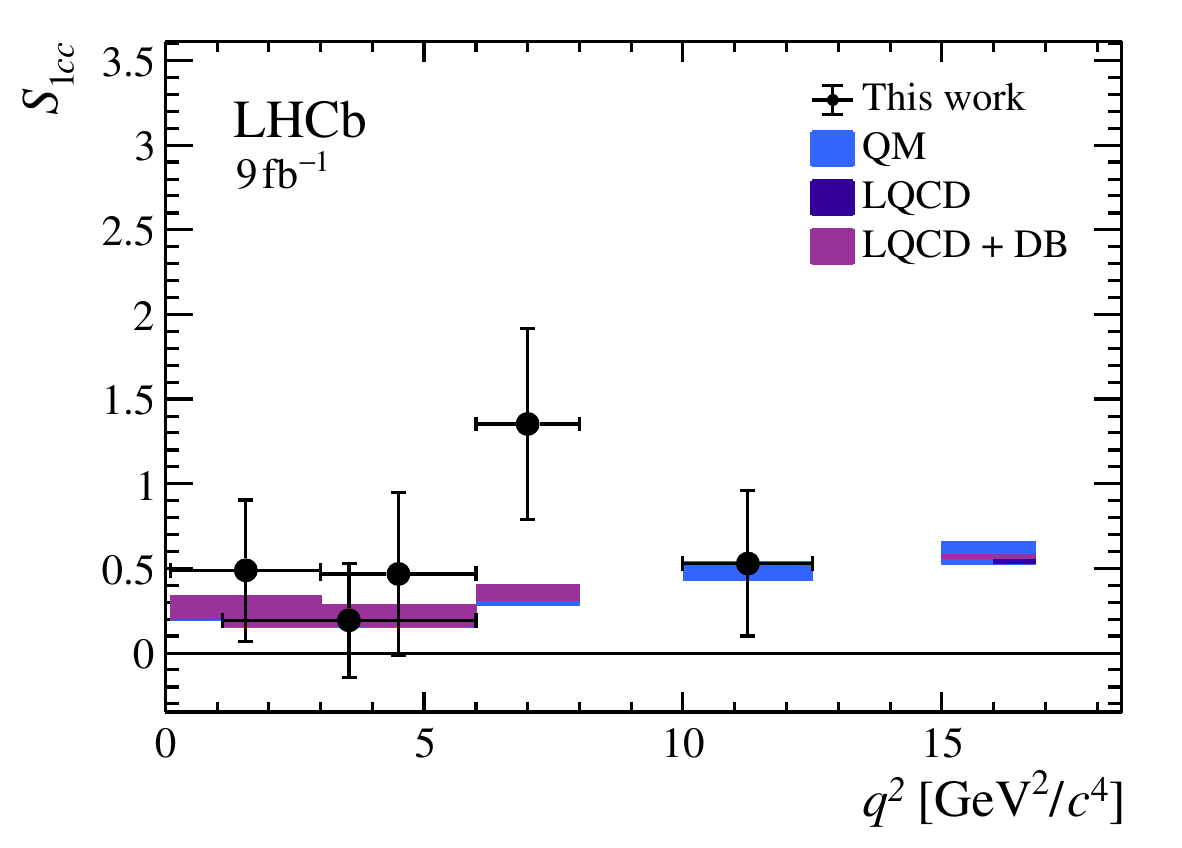}

    \caption{Angular observables (left) $A^\ell_\text{FB,\,3/2}$ and (right) $S_{1cc}$ in bins of $q^2$, compared with theoretical predictions~\cite{Descotes-Genon:2019dbw} based on the form factors from Lattice QCD~\cite{Meinel:2020owd, Meinel:2021mdj}, the joint Lattice and dispersive bound~\cite{Amhis:2022vcd}, plus the Quark Model (QM)~\cite{Descotes-Genon:2019dbw, Mott:2011cx}. The error bars show the total uncertainties. }   
    \label{fig:our-favorite-projections}
\end{figure}

The results of the angular observables measured in the five \qsq bins displayed in Table~\ref{tab:angobs_fitvals_rareMode_unblinded} are compared to SM predictions in Fig.~\ref{fig:our-favorite-projections}. The measured observable values are compatible with the SM predictions of Refs.~\cite{Descotes-Genon:2019dbw, Meinel:2020owd, Meinel:2021mdj, Amhis:2022vcd}. 
The measurement is statistically dominated. 
The statistical correlations between the two angular observables are provided in Appendix~\ref{supl:Correlations}.

Further improvement in precision can be expected from the analysis of larger data samples collected by the LHCb experiment in the future. This will allow the extension of the complexity of the models employed as well as provide potential discrimination of the various theoretical predictions.

\begin{table}[tb]
    \centering
    \small
    \caption{Angular observable values in each of the rare \qsq bins with their associated corrected (first) statistical and (second) systematic uncertainty. }
    \begin{tabular}{c|cc}
		\toprule
        $ \qsq [\gevgevcccc]$ & $A^\ell_\text{FB,\,3/2}$ & $S_\text{1cc}$ \\
        \midrule
		$[0.1, 3.0]$ 
        & $\phm0.13 \pm 0.22 \pm 0.11$ 
        & $0.49 \pm 0.40 \pm 0.22$  \\
        
		$[3.0,6.0]$    
        & $-0.03 \pm 0.21 \pm 0.15$ 
        & $0.47 \pm 0.39 \pm 0.28$  \\
        
		$[6.0, 8.0]$  
        & $-0.45 \pm 0.21 \pm 0.09$ 
        & $1.35 \pm 0.54 \pm 0.17$  \\
        
		$[10.0, 12.5]$
        & $-0.18 \pm 0.24 \pm 0.09$ 
        & $0.53 \pm 0.39 \pm 0.18$  \\
        
		$[1.1, 6.0]$ 
        & $\phm0.16 \pm 0.15 \pm 0.06$ 
        & $0.19 \pm 0.30 \pm 0.18$  \\
        \bottomrule
    \end{tabular}
    \label{tab:angobs_fitvals_rareMode_unblinded}
\end{table}

\section*{Acknowledgements}
%
% These Acknowledgements valid from 16/06/2026
%
\noindent We express our gratitude to our colleagues in the CERN
accelerator departments for the excellent performance of the LHC. We
thank the technical and administrative staff at the LHCb
institutes.
We acknowledge support from CERN and from the national agencies:
ARC (Australia);
CAPES, CNPq, FAPERJ and FINEP (Brazil); 
MOST and NSFC (China); 
CNRS/IN2P3 and CEA (France);  % added CEA 26/02/2026
BMFTR, DFG and MPG (Germany);
NKFIH (Hungary);              % added 16/06/2026
INFN (Italy); 
NWO (Netherlands); 
MNiSW and NCN (Poland); 
MEC/IFA (Romania); 
%MSHE (Russia); 
MICIU and AEI (Spain);
SNSF and SER (Switzerland); 
NASU (Ukraine); 
STFC (United Kingdom); 
DOE NP and NSF (USA).
%%%%%%%%%%%%%%%%%%%%%%%%%%%%%%%%%%%%%%%%%%%%%
We acknowledge the computing resources that are provided by ARDC (Australia), 
CBPF (Brazil),
CERN, 
IHEP and LZU (China),
IN2P3 (France), 
KIT and DESY (Germany), 
INFN (Italy), 
SURF (Netherlands),
Polish WLCG (Poland),
IFIN-HH (Romania), % http://dx.doi.org/10.13039/100019931,"Institutul National de Cercetare-Dezvoltare pentru Fizica si Inginerie Nucleara 'Horia Hulubei'"
%RRCKI and Yandex LLC (Russia), 
PIC (Spain), CSCS (Switzerland), 
GridPP (United Kingdom),
and NSF (USA).  % added Feb2026
%%%%%%%%%%%%%%%%%%%%%%%%%%%%%%%%%%%%%%%%%% 
We are indebted to the communities behind the multiple open-source
software packages on which we depend.
%%%%%%%%%%%%%%%%%%%%%%%%%%%%%%%%%%%%%%%%%%
Individual groups or members have received support from
% ARC and ARDC (Australia); % moved to national 16/01/2025
RTP (Australia), % added 06/03/2026
FWO Odysseus grant G0ASD25N (Belgium), % added 20/4/2026
Key Research Program of Frontier Sciences of CAS, CAS PIFI, CAS CCEPP (China); 
%Fundamental Research Funds for the Central Universities,  and Sci.\
%\& Tech.\ Program of Guangzhou (China); Removed 24/11/25
Minciencias (Colombia);
EPLANET, Marie Sk\l{}odowska-Curie Actions, ERC and NextGenerationEU (European Union);
A*MIDEX, ANR, IPhU and Labex P2IO, and R\'{e}gion Auvergne-Rh\^{o}ne-Alpes (France);
%RFBR, RSF and Yandex LLC (Russia);
Alexander-von-Humboldt Foundation (Germany);
ICSC (Italy); 
%GVA, XuntaGal, GENCAT, Inditex, InTalent and Prog.~Atracci\'on Talento, CM (Spain);
Severo Ochoa and Mar\'ia de Maeztu Units of Excellence, GVA, XuntaGal, GENCAT, InTalent-Inditex and Prog.~Atracci\'on Talento CM (Spain);
%XuntaGal --> Xunta de Galicia 
% SRC (Sweden);  % removed 27/02/2026 - end of grant
the Leverhulme Trust, the Royal Society and UKRI (United Kingdom).

\newpage
{\noindent\normalfont\bfseries\Large Appendices}
\label{sec:appendix}

\appendix

\section{Interference hypotheses}
\label{supl:Interferences}

Dedicated simulation samples are generated following Ref.~\cite{Beck:2022spd}, where random combinations of the strong phase differences $e^{\pm i(\varphi_{\Lambdares(1520)} - \varphi_{\Lambdares(X)})}$ are generated to test the different interference hypotheses. Table~\ref{tab:Interferences} summarises the values of the different phase differences $\Delta \varphi_{X} \equiv \varphi_{\Lambdares(1520)} - \varphi_{\Lambdares(X)}$, which are picked in a random manner for the $\Lambdares(1405)$ and $\Lambdares(1600)$ resonance. Each of the samples has a size of $10^6$ events. 
\begin{table}[b]
	\begin{center}
		\caption{Random values of phase combinations of the $\Lambdares(1405)$ and $\Lambdares(1600)$ with respect to the \Lres resonance. }
        \begin{tabular}{ccc}
			\toprule
			Phase combination & $\Delta \varphi_{1405}$ & $\Delta \varphi_{1600}$ \\
			\midrule
			0 & $0.00\pi$ & $0.00\pi$ \\
			1 & $1.38\pi$ & $1.93\pi$ \\
			2 & $1.10\pi$ & $1.61\pi$ \\
			3 & $0.43\pi$ & $0.62\pi$ \\
			4 & $0.06\pi$ & $1.38\pi$ \\
			5 & $1.41\pi$ & $0.70\pi$ \\
			\bottomrule
		\end{tabular}

		\label{tab:Interferences}
	\end{center}
\end{table}

\begin{figure}[htbp]
	\begin{center}
		\includegraphics[width=0.4 \linewidth]{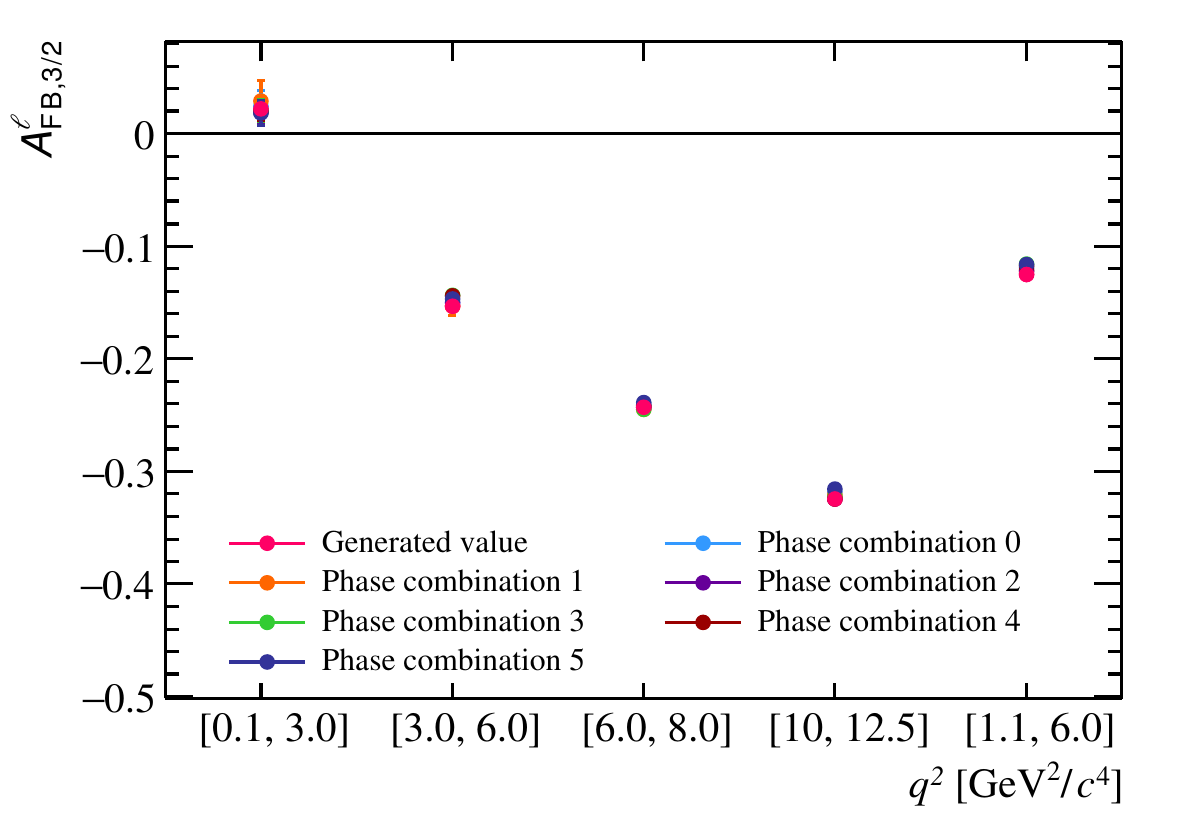}
		\includegraphics[width=0.4 \linewidth]{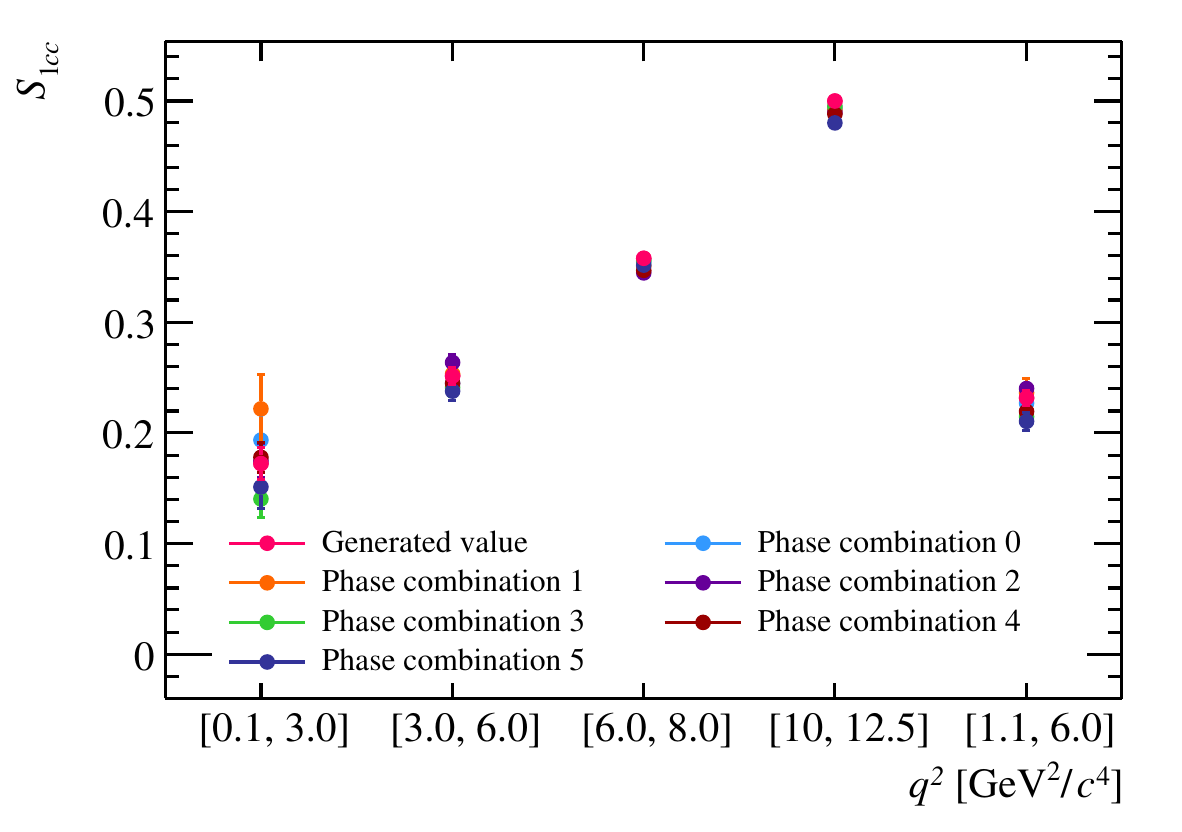}
		\includegraphics[width=0.4 \linewidth]{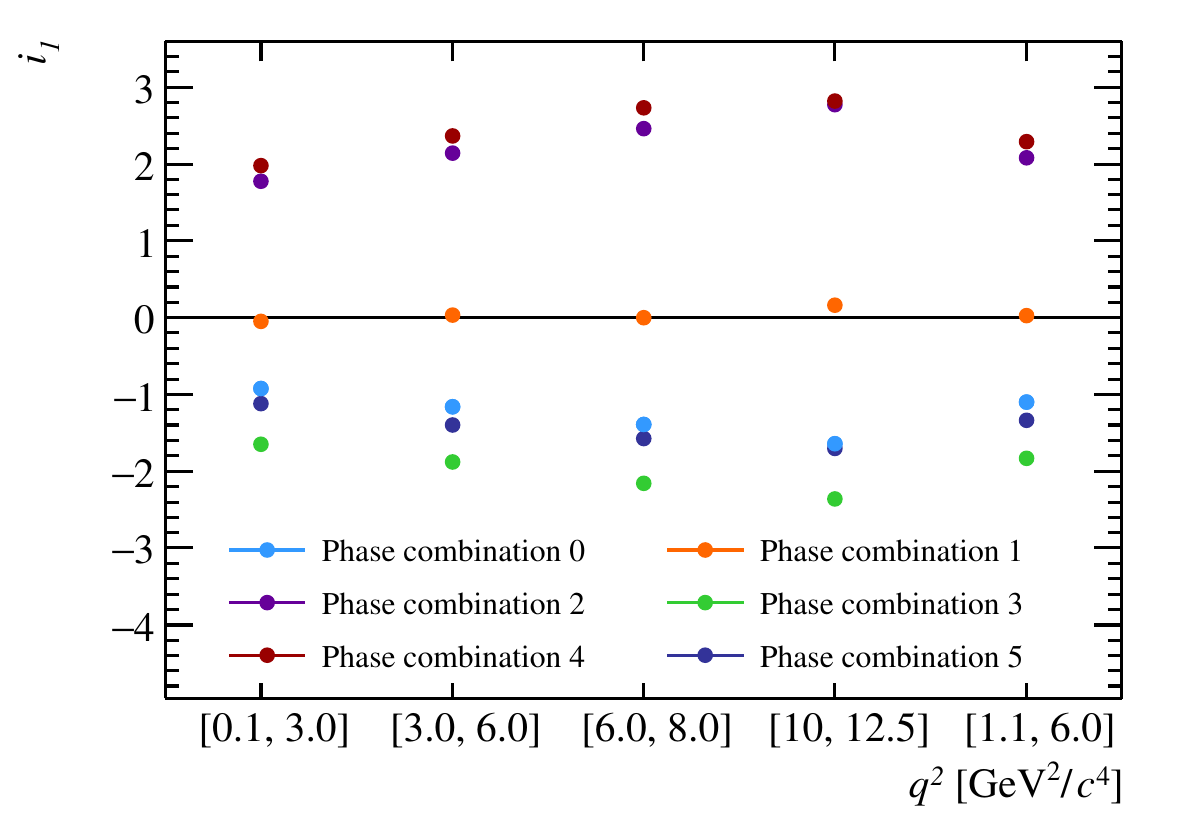}
		\includegraphics[width=0.4 \linewidth]{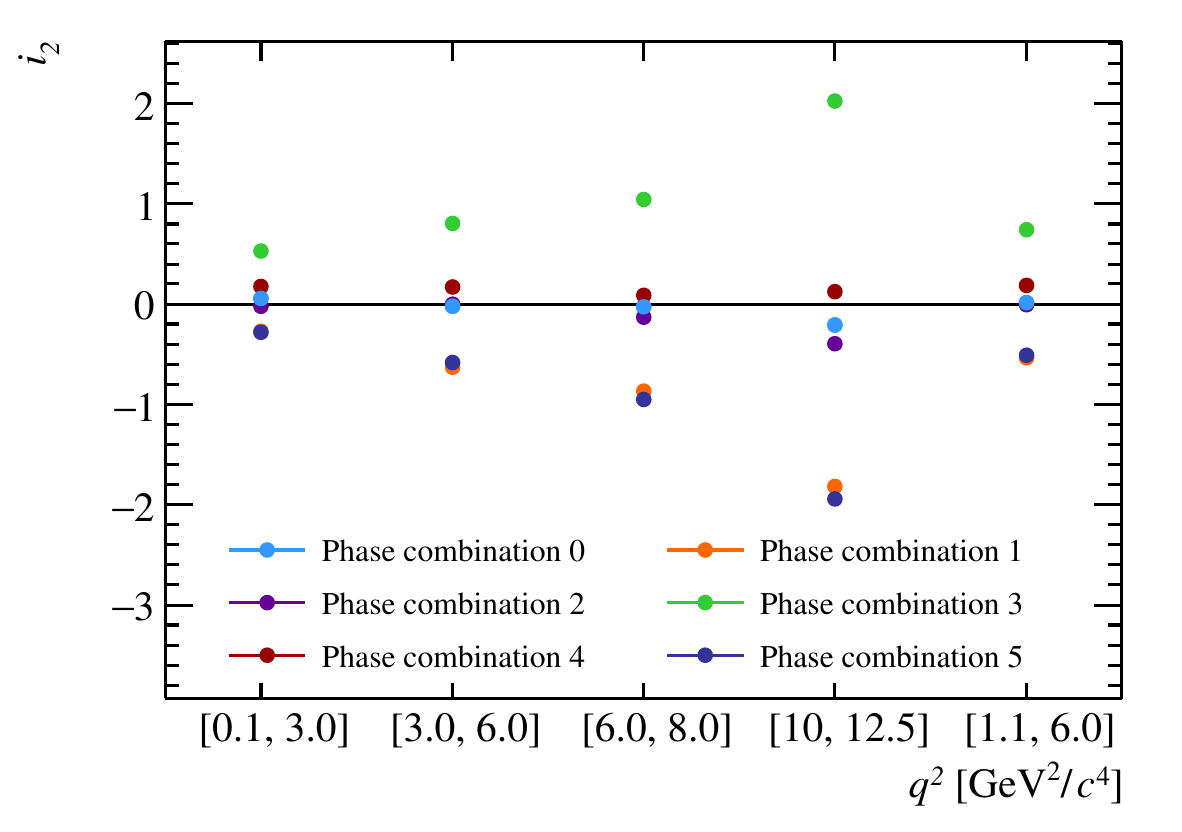}
	\end{center}
	\caption{Resulting parameter values with their associated uncertainties obtained by fitting the dedicated simulated samples with the angular fit model. }
	\label{fig:Summary_FitParams_tot}	
\end{figure}

The angular PDF is validated with these dedicated simulated samples and the resulting fit parameter values of the parameters of interests and the interference parameters are presented in Fig.~\ref{fig:Summary_FitParams_tot}. The associated uncertainty corresponds to the uncertainty on the fit parameters associated to the sample size in each of the \qsq bins. 
\newpage

\section{Angular acceptance plots}
\label{supl:AngAcc}

\begin{figure}[htbp]
	\begin{center}
		\includegraphics[width=0.32 \linewidth]{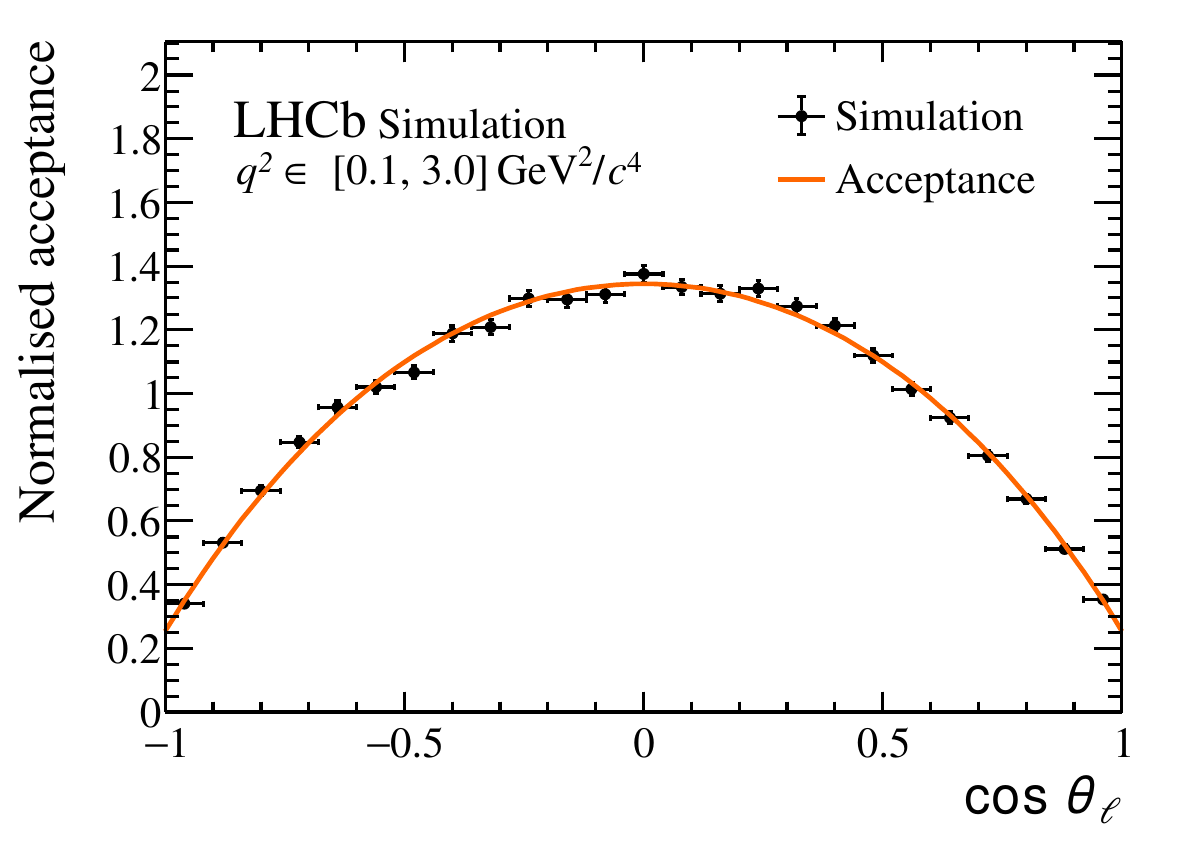}
		\includegraphics[width=0.32 \linewidth]{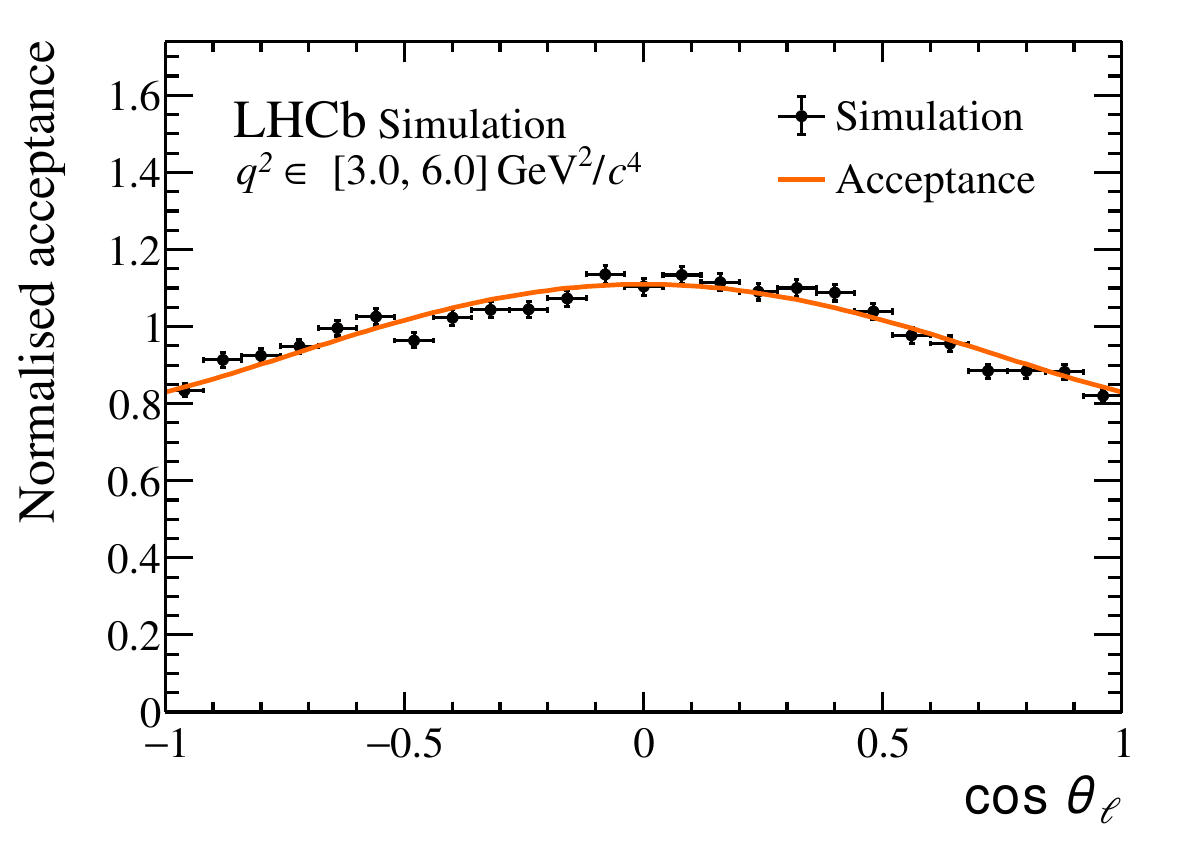}
		\includegraphics[width=0.32 \linewidth]{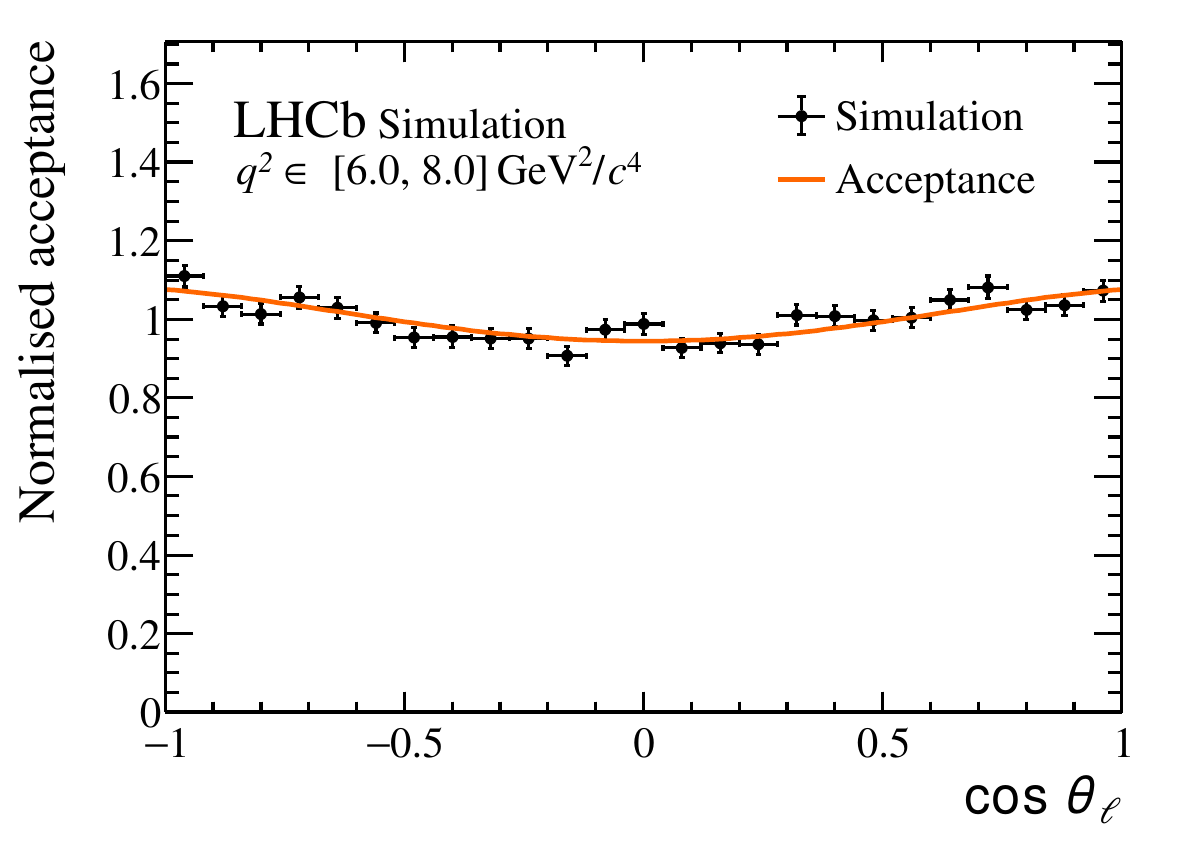}
		\includegraphics[width=0.32 \linewidth]{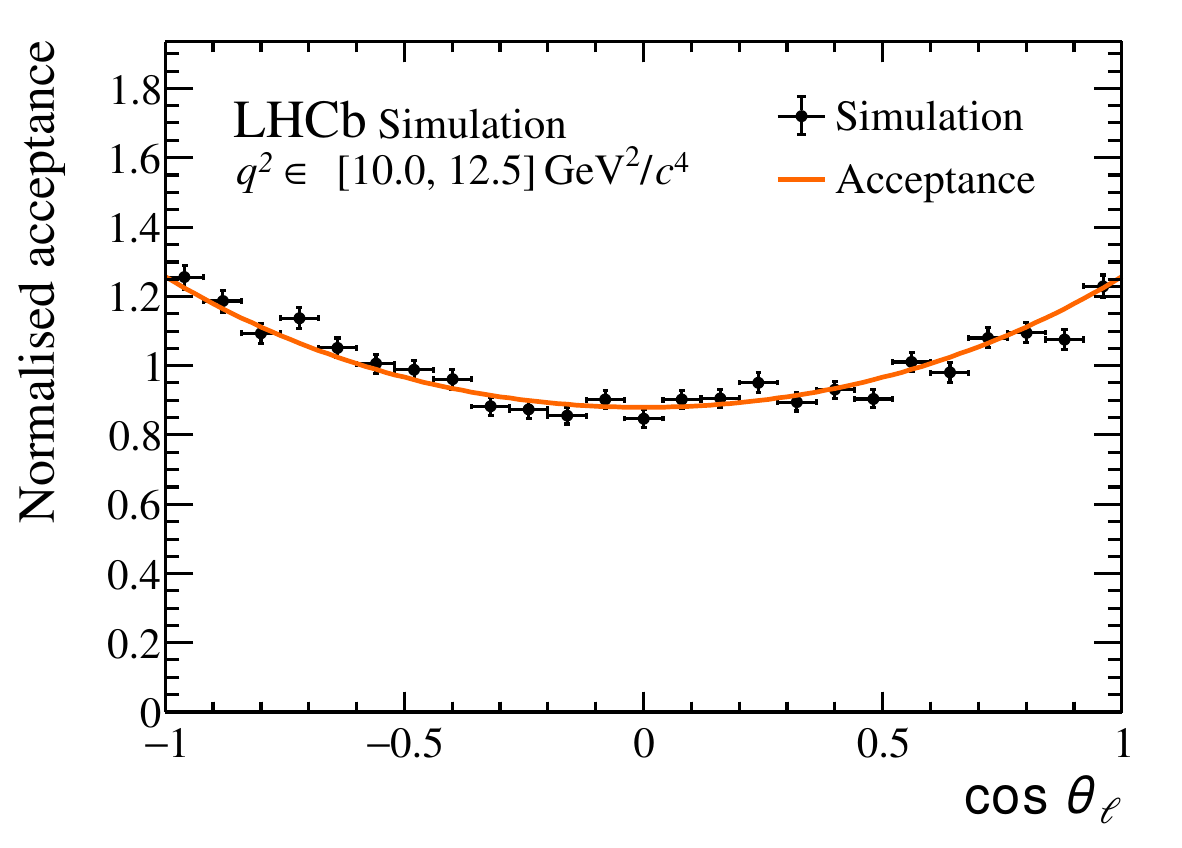}
		\includegraphics[width=0.32 \linewidth]{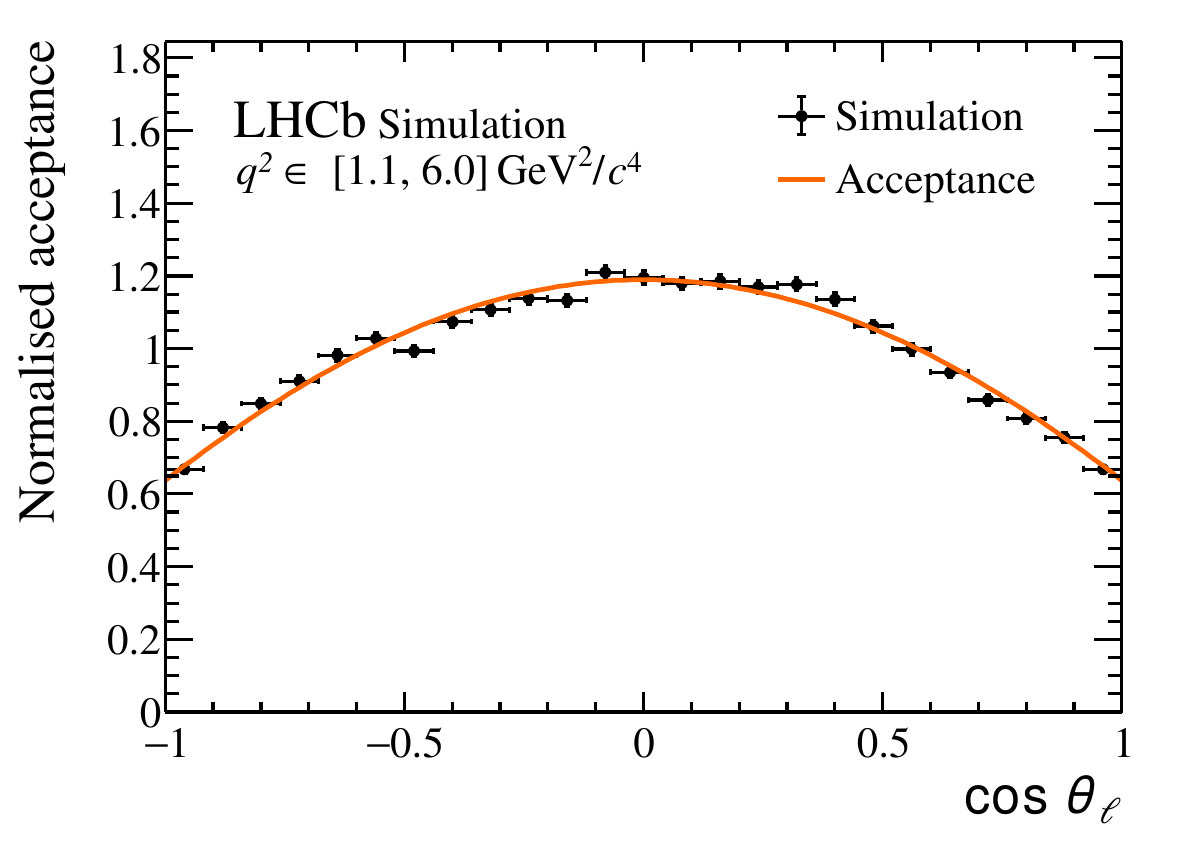}
		\caption{Simulated distribution of $\cos\theta_\ell$ in the ${\it \Lambda}_{\it b}^0 \to {\it \Lambda}(1520)\mu^{+}\mu^{-}$ decay for the five $q^2$ bins. Phase-space simulated samples after full selection and the corrections are shown as data points, projections of the angular acceptance model are represented by orange lines.}
		\label{fig:AngAcc_cosTl}
	\end{center}
\end{figure}

\begin{figure}[htbp]
	\begin{center}
		\includegraphics[width=0.32 \linewidth]{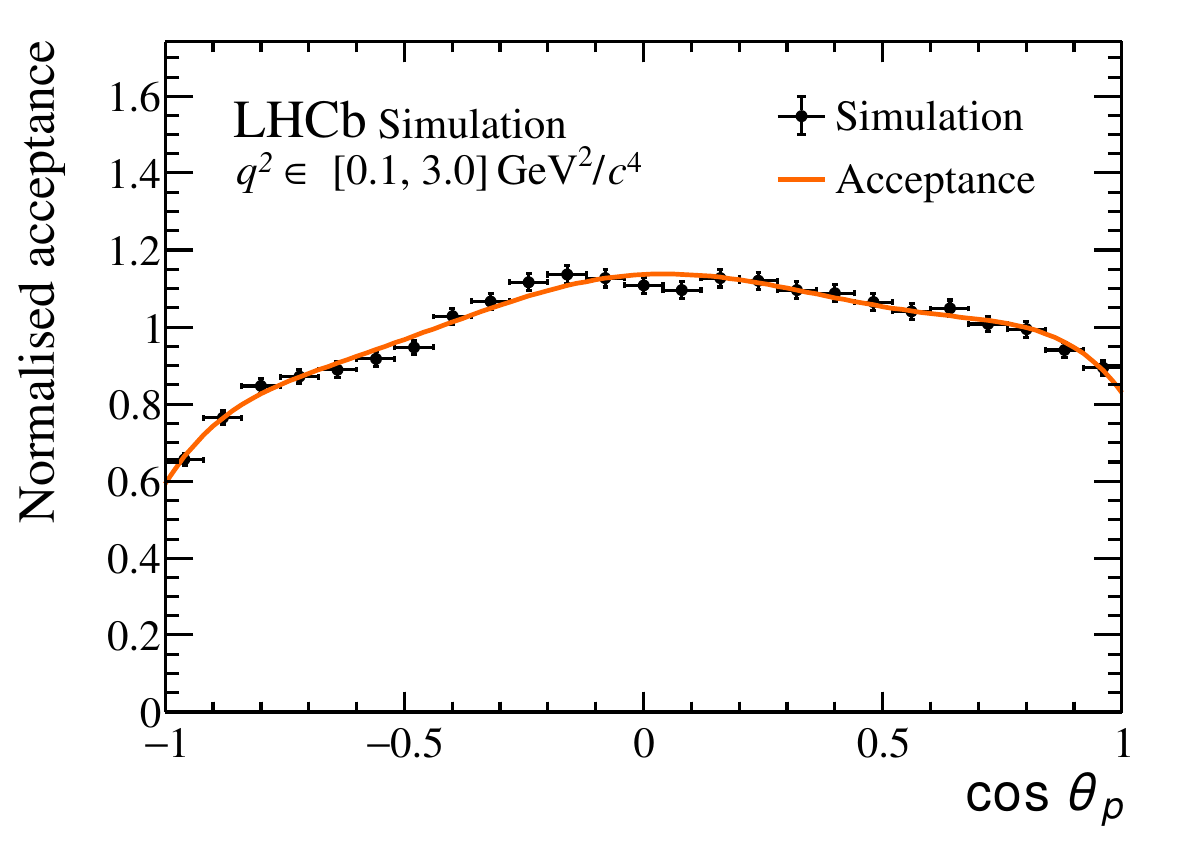}
		\includegraphics[width=0.32 \linewidth]{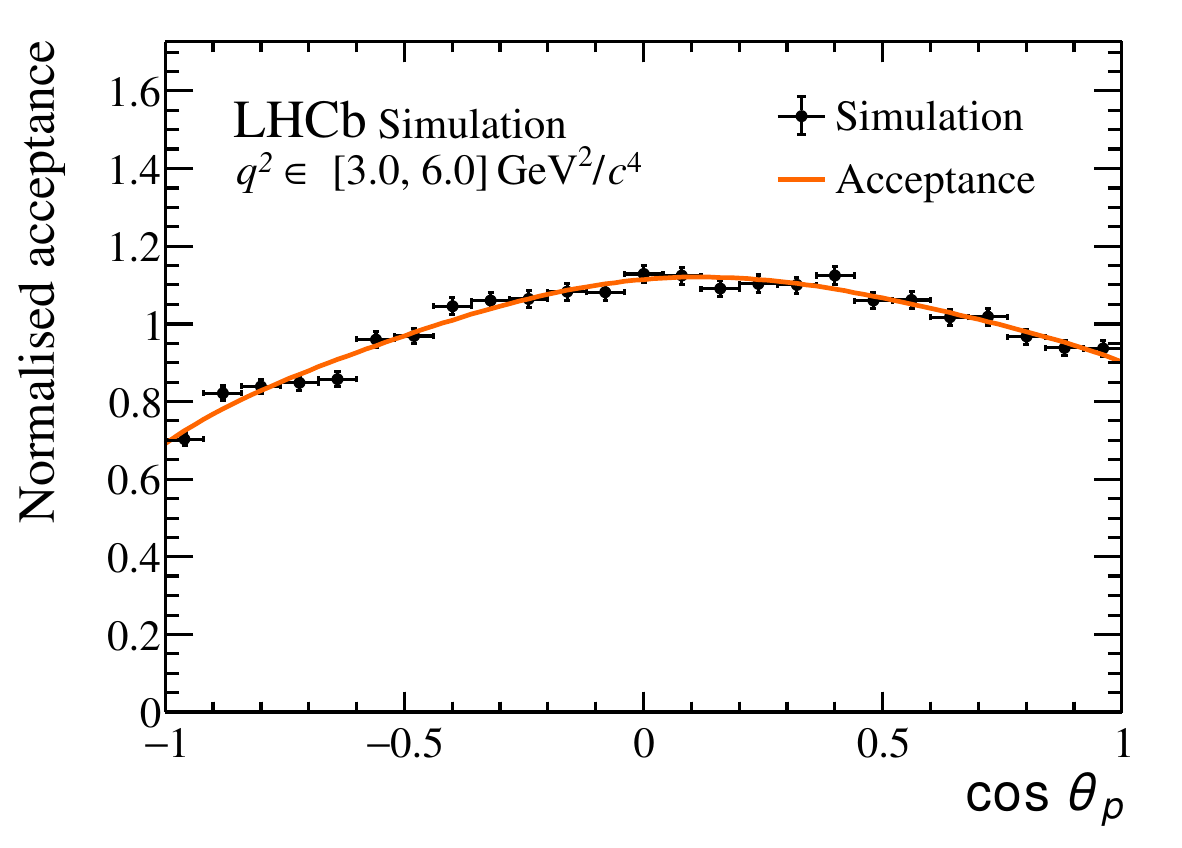}
		\includegraphics[width=0.32 \linewidth]{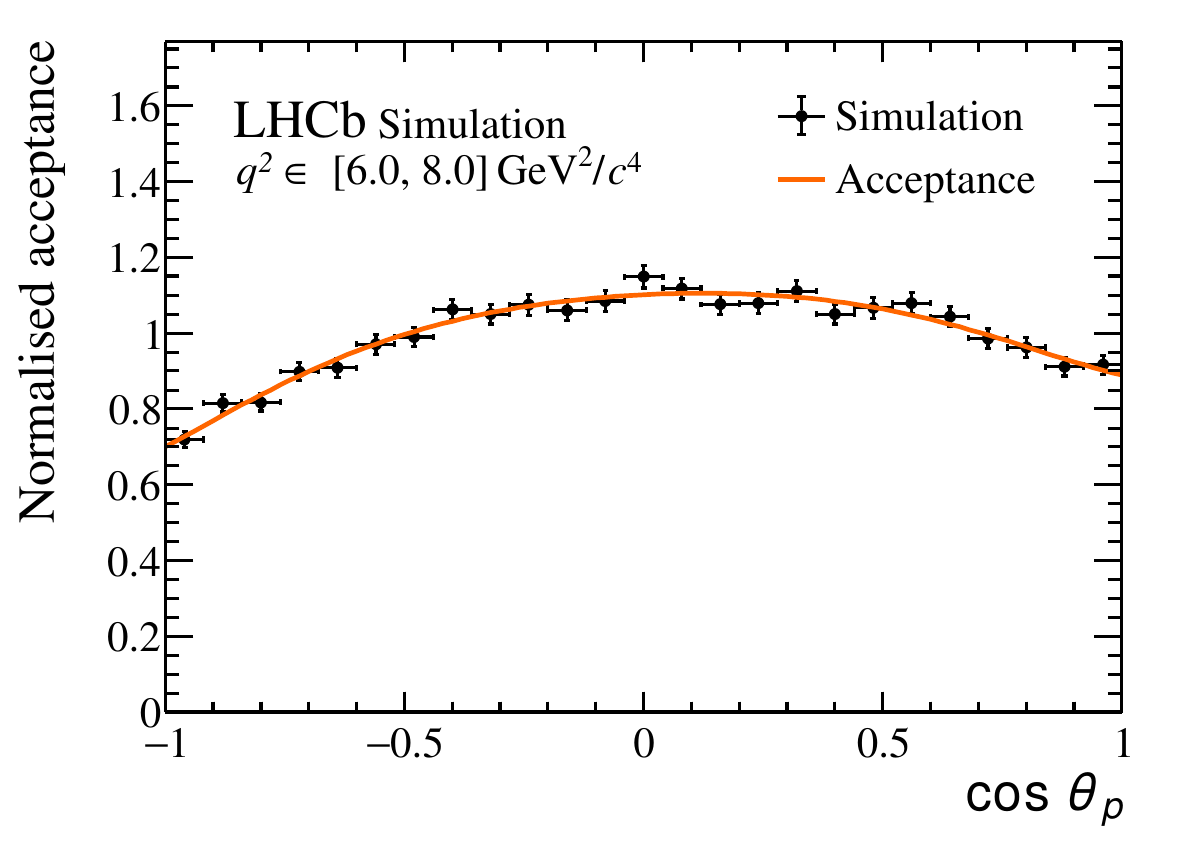}
		\includegraphics[width=0.32 \linewidth]{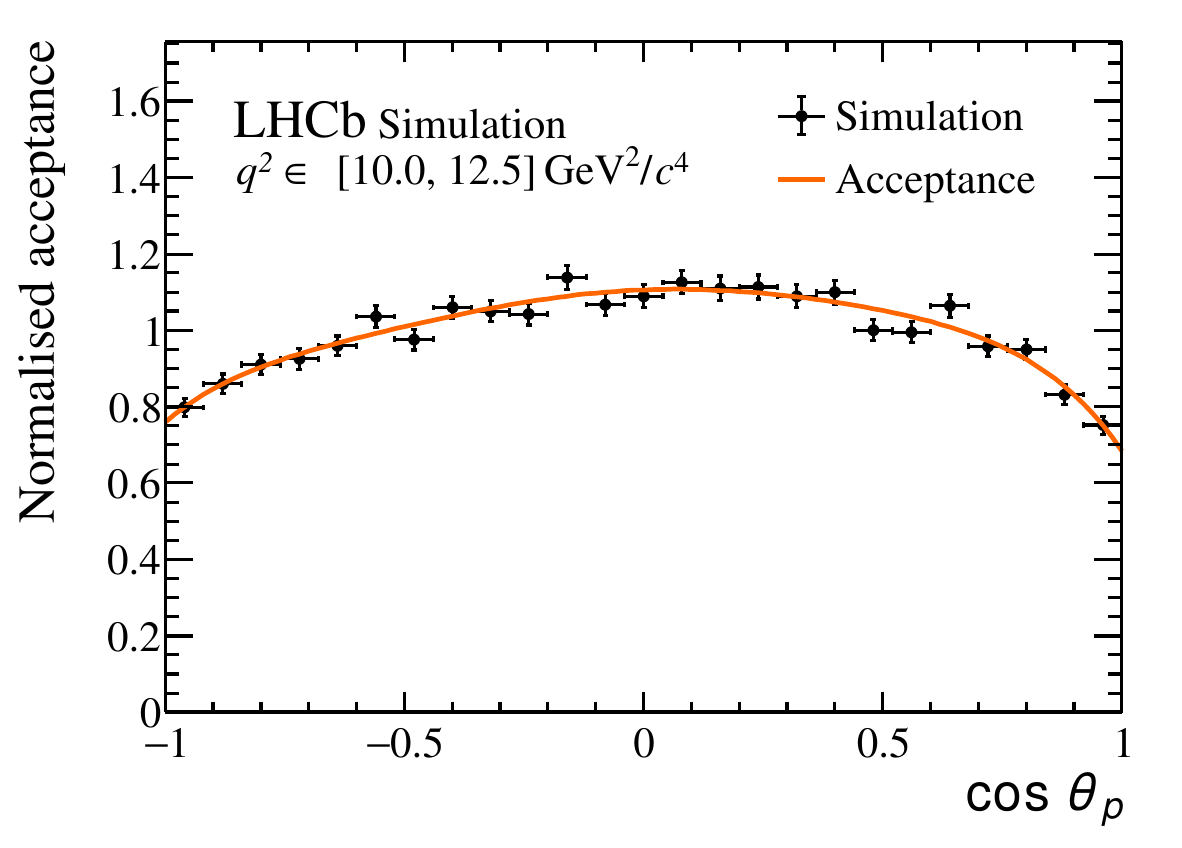}
		\includegraphics[width=0.32 \linewidth]{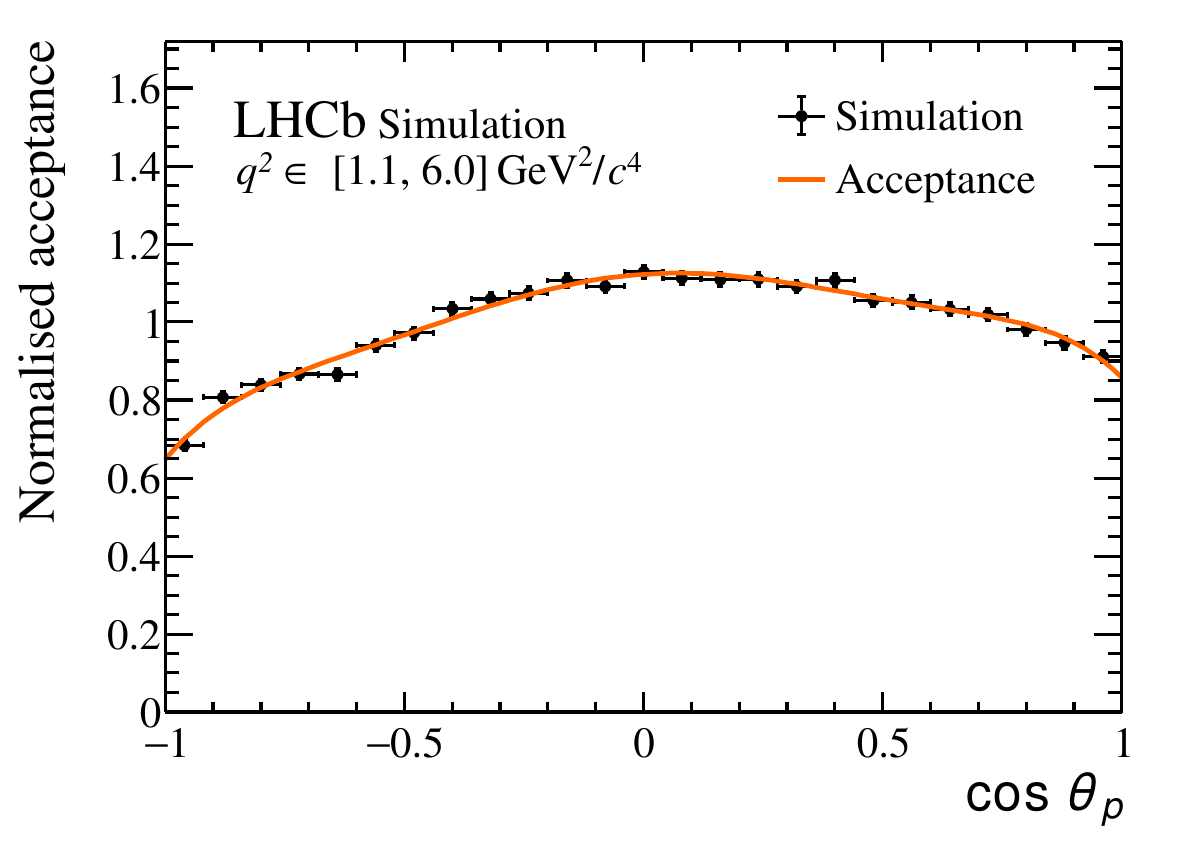}
		\caption{Simulated distributions of $\cos\theta_p$ in the ${\it \Lambda}_{\it b}^0 \to {\it \Lambda}(1520)\mu^{+}\mu^{-}$ decay for the five \qsq bins. Phase-space simulated samples after full selection and the corrections are shown as data points, projections of the angular acceptance model are represented by orange lines. }
		\label{fig:AngAcc_cosTp}
	\end{center}
\end{figure}

The $\cos\theta_\ell$ and $\cos\theta_p$ projections of the angular acceptance are presented in Fig.~\ref{fig:AngAcc_cosTl} and \ref{fig:AngAcc_cosTp}. 

\section{Correlation matrices for the parameters of interest}
\label{supl:Correlations}
\begin{figure}[htbp]
	\begin{center}
		\includegraphics[width=0.32 \linewidth]{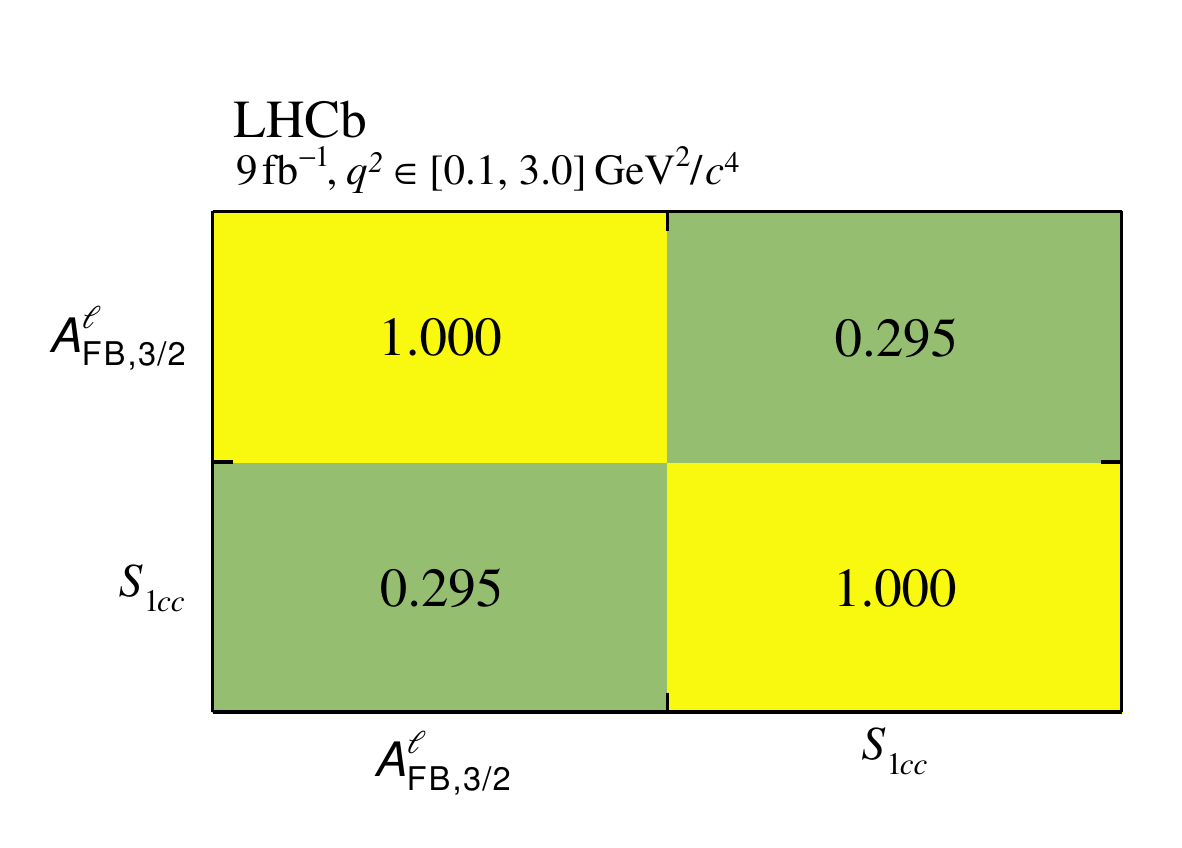}
		\includegraphics[width=0.32 \linewidth]{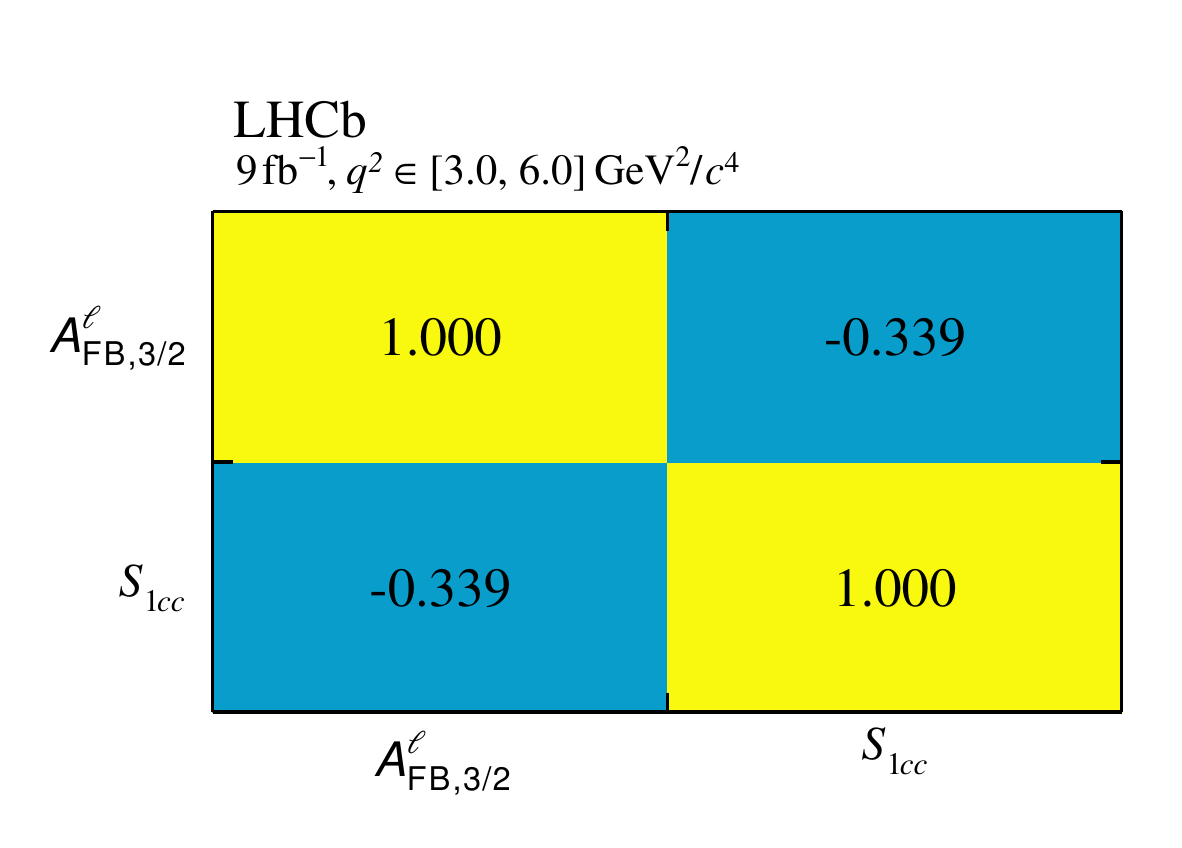}
		\includegraphics[width=0.32\linewidth]{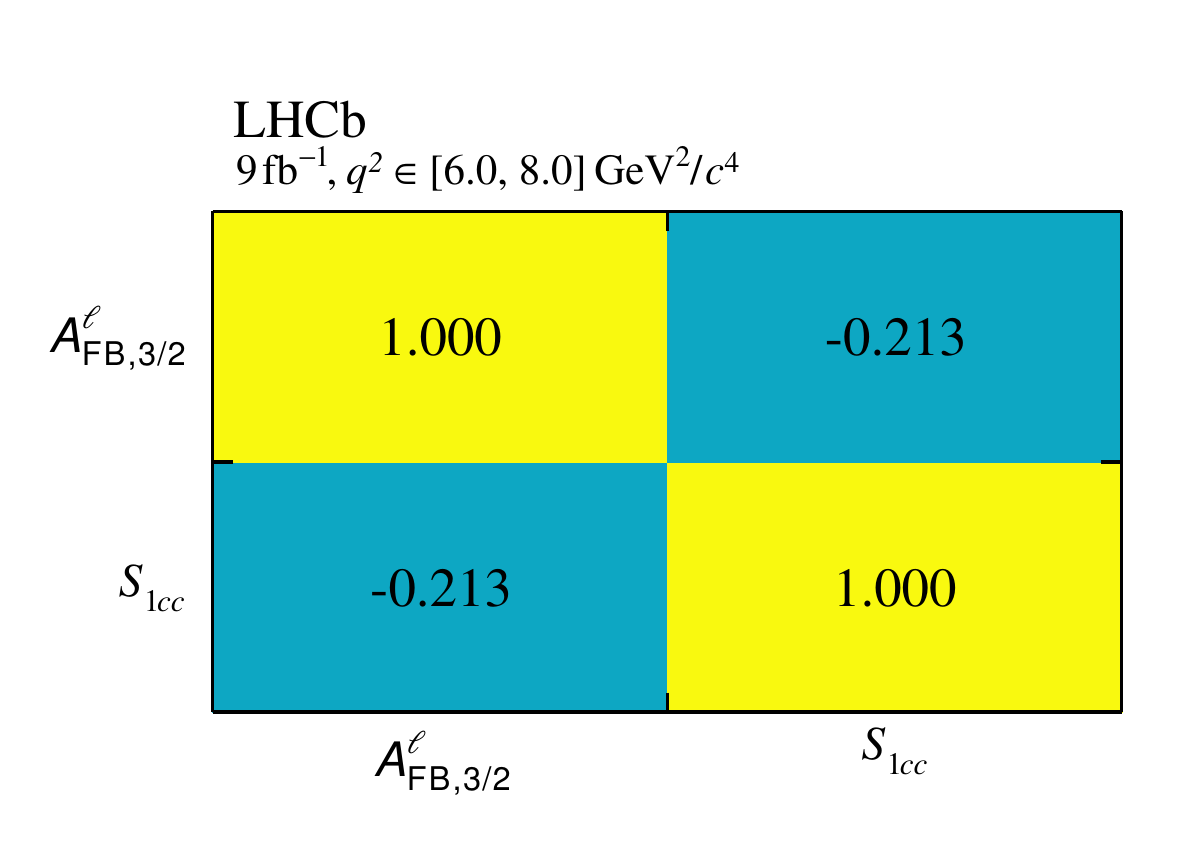}
		\includegraphics[width=0.32 \linewidth]{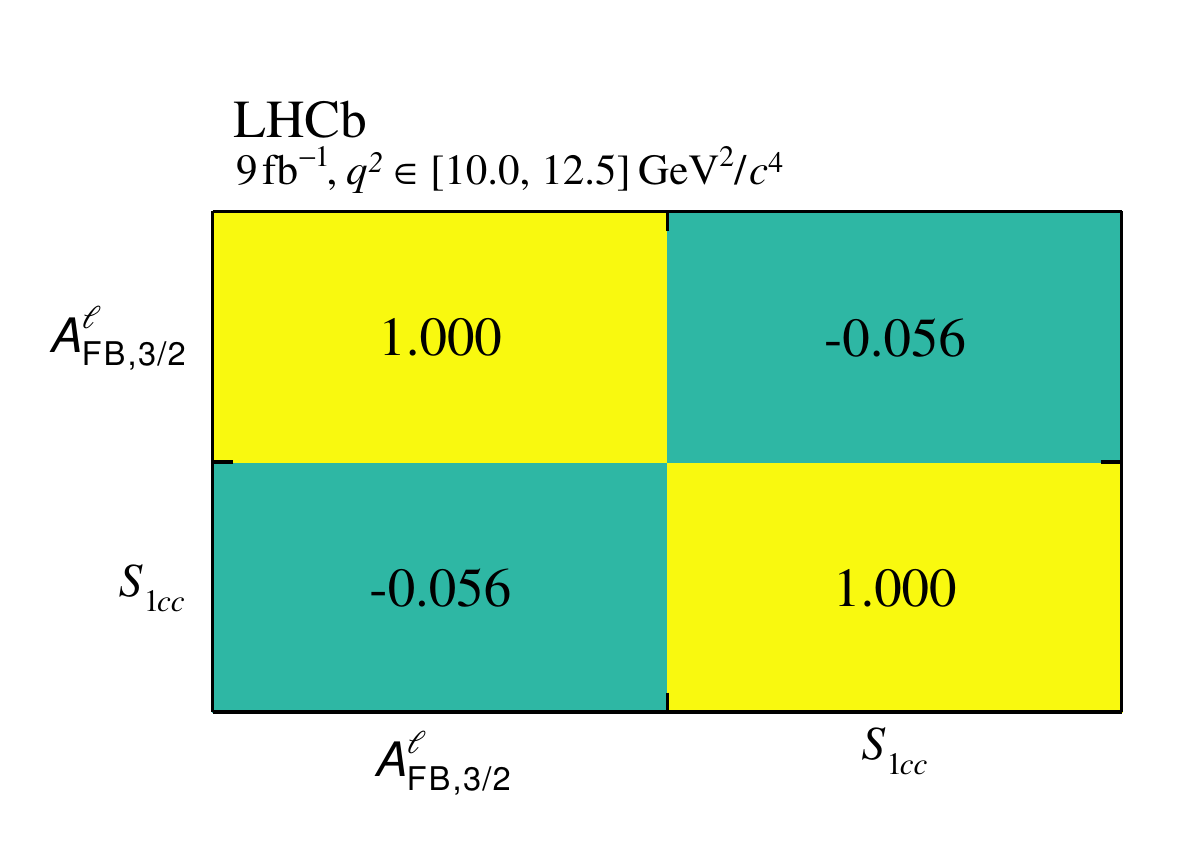}
		\includegraphics[width=0.32 \linewidth]{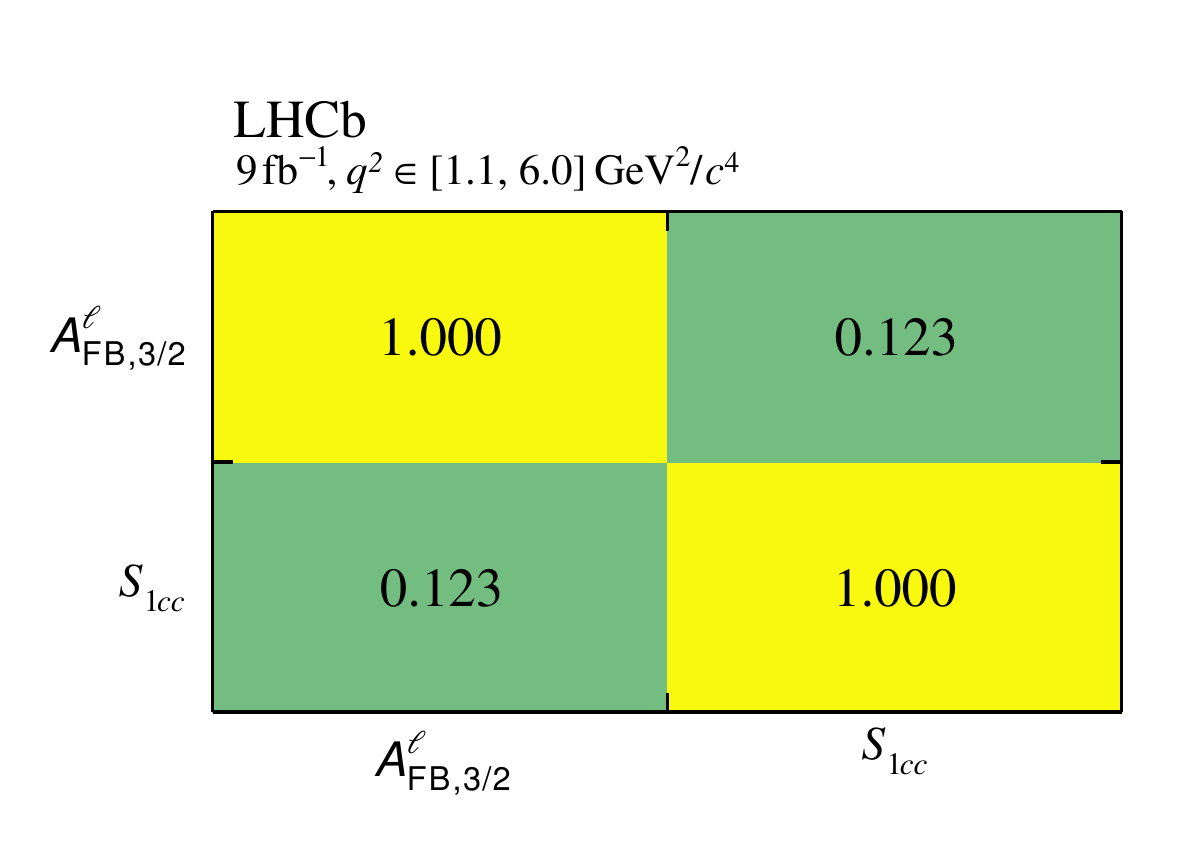}
	\end{center}
    \caption{Statistical correlation matrices for the parameters of interest.}
    \label{fig:Correlations}
\end{figure}

The statistical correlation matrices for the parameters of interest are shown in Fig.~\ref{fig:Correlations}.

\newpage

\addcontentsline{toc}{section}{References}
%\setboolean{inbibliography}{true}
\bibliographystyle{LHCb}
\bibliography{main,standard,LHCb-PAPER,LHCb-CONF,LHCb-DP,LHCb-TDR}

\ifx\mcitethebibliography\mciteundefinedmacro
\PackageError{LHCb.bst}{mciteplus.sty has not been loaded}
{This bibstyle requires the use of the mciteplus package.}\fi
\providecommand{\href}[2]{#2}
\begin{mcitethebibliography}{10}
\mciteSetBstSublistMode{n}
\mciteSetBstMaxWidthForm{subitem}{\alph{mcitesubitemcount})}
\mciteSetBstSublistLabelBeginEnd{\mcitemaxwidthsubitemform\space}
{\relax}{\relax}

\bibitem{LHCb-PAPER-2014-006}
LHCb collaboration, R.~Aaij {\em et~al.}, \ifthenelse{\boolean{articletitles}}{\emph{{Differential branching fractions and isospin asymmetries of \mbox{\decay{\B}{K^{(*)}\mumu}} decays}}, }{}\href{https://doi.org/10.1007/JHEP06(2014)133}{JHEP \textbf{06} (2014) 133}, \href{http://arxiv.org/abs/1403.8044}{{\normalfont\ttfamily arXiv:1403.8044}}\relax
\mciteBstWouldAddEndPuncttrue
\mciteSetBstMidEndSepPunct{\mcitedefaultmidpunct}
{\mcitedefaultendpunct}{\mcitedefaultseppunct}\relax
\EndOfBibitem
\bibitem{CMS:2024syx}
CMS collaboration, A.~Hayrapetyan {\em et~al.}, \ifthenelse{\boolean{articletitles}}{\emph{{Test of lepton flavor universality in \mbox{$\Bpm\to \Kpm\mup\mun$} and $\Bpm\to \Kpm\ep\en$ decays in proton-proton collisions at \mbox{$\sqrt{s} = 13$}\,TeV}}, }{}\href{https://doi.org/10.1088/1361-6633/ad4e65}{Rept.\ Prog.\ Phys.\  \textbf{87} (2024) 077802}, \href{http://arxiv.org/abs/2401.07090}{{\normalfont\ttfamily arXiv:2401.07090}}\relax
\mciteBstWouldAddEndPuncttrue
\mciteSetBstMidEndSepPunct{\mcitedefaultmidpunct}
{\mcitedefaultendpunct}{\mcitedefaultseppunct}\relax
\EndOfBibitem
\bibitem{Belle:2019xld}
Belle collaboration, S.~Choudhury {\em et~al.}, \ifthenelse{\boolean{articletitles}}{\emph{{Test of lepton flavor universality and search for lepton flavor violation in $B \rightarrow K\ell \ell$ decays}}, }{}\href{https://doi.org/10.1007/JHEP03(2021)105}{JHEP \textbf{03} (2021) 105}, \href{http://arxiv.org/abs/1908.01848}{{\normalfont\ttfamily arXiv:1908.01848}}\relax
\mciteBstWouldAddEndPuncttrue
\mciteSetBstMidEndSepPunct{\mcitedefaultmidpunct}
{\mcitedefaultendpunct}{\mcitedefaultseppunct}\relax
\EndOfBibitem
\bibitem{LHCb-PAPER-2021-014}
LHCb collaboration, R.~Aaij {\em et~al.}, \ifthenelse{\boolean{articletitles}}{\emph{{Branching fraction measurements of the rare $\Bs \to \phiz \mup\mun$ and $\Bs \to f_2^\prime(1525) \mup\mun$ decays}}, }{}\href{https://doi.org/10.1103/PhysRevLett.127.151801}{Phys.\ ~Rev.\ ~Lett.\  \textbf{127} (2021) 151801}, \href{http://arxiv.org/abs/2105.14007}{{\normalfont\ttfamily arXiv:2105.14007}}\relax
\mciteBstWouldAddEndPuncttrue
\mciteSetBstMidEndSepPunct{\mcitedefaultmidpunct}
{\mcitedefaultendpunct}{\mcitedefaultseppunct}\relax
\EndOfBibitem
\bibitem{LHCb-PAPER-2020-002}
LHCb collaboration, R.~Aaij {\em et~al.}, \ifthenelse{\boolean{articletitles}}{\emph{{Measurement of \CP-averaged observables in the \mbox{\decay{\Bz}{\Kstarz\mumu}} decay}}, }{}\href{https://doi.org/10.1103/PhysRevLett.125.011802}{Phys.\ ~Rev.\ ~Lett.\  \textbf{125} (2020) 011802}, \href{http://arxiv.org/abs/2003.04831}{{\normalfont\ttfamily arXiv:2003.04831}}\relax
\mciteBstWouldAddEndPuncttrue
\mciteSetBstMidEndSepPunct{\mcitedefaultmidpunct}
{\mcitedefaultendpunct}{\mcitedefaultseppunct}\relax
\EndOfBibitem
\bibitem{LHCb-PAPER-2024-011}
LHCb collaboration, R.~Aaij {\em et~al.}, \ifthenelse{\boolean{articletitles}}{\emph{{Comprehensive analysis of local and nonlocal amplitudes in the $\Bd\to\Kstarz\mumu$ decay}}, }{}\href{https://doi.org/10.1007/JHEP09(2024)026}{{JHEP} \textbf{09} (2024) 026}, Erratum \href{https://doi.org/10.1007/JHEP05(2025)208}{ibid.\   \textbf{05} (2025) 208}, \href{http://arxiv.org/abs/2405.17347}{{\normalfont\ttfamily arXiv:2405.17347}}\relax
\mciteBstWouldAddEndPuncttrue
\mciteSetBstMidEndSepPunct{\mcitedefaultmidpunct}
{\mcitedefaultendpunct}{\mcitedefaultseppunct}\relax
\EndOfBibitem
\bibitem{LHCb-PAPER-2023-032}
LHCb collaboration, R.~Aaij {\em et~al.}, \ifthenelse{\boolean{articletitles}}{\emph{{Determination of short- and long-distance contributions in $\Bz \to \Kstarz\mup\mun$ decays}}, }{}\href{https://doi.org/10.1103/PhysRevD.109.052009}{Phys.\ ~Rev.\  \textbf{D109} (2024) 052009}, \href{http://arxiv.org/abs/2312.09102}{{\normalfont\ttfamily arXiv:2312.09102}}\relax
\mciteBstWouldAddEndPuncttrue
\mciteSetBstMidEndSepPunct{\mcitedefaultmidpunct}
{\mcitedefaultendpunct}{\mcitedefaultseppunct}\relax
\EndOfBibitem
\bibitem{LHCb-PAPER-2023-033}
LHCb collaboration, R.~Aaij {\em et~al.}, \ifthenelse{\boolean{articletitles}}{\emph{{Amplitude analysis of the $\Bz \to \Kstarz\mup\mun$ decay}}, }{}\href{https://doi.org/10.1103/PhysRevLett.132.131801}{Phys.\ ~Rev.\ ~Lett.\  \textbf{132} (2024) 131801}, \href{http://arxiv.org/abs/2312.09115}{{\normalfont\ttfamily arXiv:2312.09115}}\relax
\mciteBstWouldAddEndPuncttrue
\mciteSetBstMidEndSepPunct{\mcitedefaultmidpunct}
{\mcitedefaultendpunct}{\mcitedefaultseppunct}\relax
\EndOfBibitem
\bibitem{LHCb-PAPER-2025-041}
LHCb collaboration, R.~Aaij {\em et~al.}, \ifthenelse{\boolean{articletitles}}{\emph{{Comprehensive analysis of the $\Bz\to\Kstarz\mup\mun$ decay}}, }{}\href{https://doi.org/10.1103/24g9-yn9d}{{Phys.\ ~Rev.\ ~Lett.\ } \textbf{137} (2026) 021802}, \href{http://arxiv.org/abs/2512.18053}{{\normalfont\ttfamily arXiv:2512.18053}}\relax
\mciteBstWouldAddEndPuncttrue
\mciteSetBstMidEndSepPunct{\mcitedefaultmidpunct}
{\mcitedefaultendpunct}{\mcitedefaultseppunct}\relax
\EndOfBibitem
\bibitem{LHCb-PAPER-2026-023}
LHCb collaboration, R.~Aaij {\em et~al.}, \ifthenelse{\boolean{articletitles}}{\emph{{Model-independent measurement of the transversity amplitudes of the $ \Bz\to \Kstarz\mup\mun$ decay}}, }{}\href{http://arxiv.org/abs/2608.12215}{{\normalfont\ttfamily arXiv:2608.12215}}, {Submitted to JHEP}\relax
\mciteBstWouldAddEndPuncttrue
\mciteSetBstMidEndSepPunct{\mcitedefaultmidpunct}
{\mcitedefaultendpunct}{\mcitedefaultseppunct}\relax
\EndOfBibitem
\bibitem{CMS:2024atz}
CMS collaboration, A.~Hayrapetyan {\em et~al.}, \ifthenelse{\boolean{articletitles}}{\emph{{Angular analysis of the $B^0\to K^*(892)^0\mu^+\mu^-$ decay in proton-proton collisions at $\sqrt{s}$ = 13 TeV}}, }{}\href{https://doi.org/10.1016/j.physletb.2025.139406}{Phys.\ Lett.\  \textbf{B864} (2025) 139406}, \href{http://arxiv.org/abs/2411.11820}{{\normalfont\ttfamily arXiv:2411.11820}}\relax
\mciteBstWouldAddEndPuncttrue
\mciteSetBstMidEndSepPunct{\mcitedefaultmidpunct}
{\mcitedefaultendpunct}{\mcitedefaultseppunct}\relax
\EndOfBibitem
\bibitem{ATLAS:2018gqc}
ATLAS collaboration, M.~Aaboud {\em et~al.}, \ifthenelse{\boolean{articletitles}}{\emph{{Angular analysis of $B^0_d \rightarrow K^{*}\mu^+\mu^-$ decays in $pp$ collisions at $\sqrt{s}= 8$\,TeV with the ATLAS detector}}, }{}\href{https://doi.org/10.1007/JHEP10(2018)047}{JHEP \textbf{10} (2018) 047}, \href{http://arxiv.org/abs/1805.04000}{{\normalfont\ttfamily arXiv:1805.04000}}\relax
\mciteBstWouldAddEndPuncttrue
\mciteSetBstMidEndSepPunct{\mcitedefaultmidpunct}
{\mcitedefaultendpunct}{\mcitedefaultseppunct}\relax
\EndOfBibitem
\bibitem{Belle:2019oag}
Belle collaboration, A.~Abdesselam {\em et~al.}, \ifthenelse{\boolean{articletitles}}{\emph{{Test of lepton-flavor universality in ${B\to K^\ast\ell^+\ell^-}$ decays at Belle}}, }{}\href{https://doi.org/10.1103/PhysRevLett.126.161801}{Phys.\ Rev.\ Lett.\  \textbf{126} (2021) 161801}, \href{http://arxiv.org/abs/1904.02440}{{\normalfont\ttfamily arXiv:1904.02440}}\relax
\mciteBstWouldAddEndPuncttrue
\mciteSetBstMidEndSepPunct{\mcitedefaultmidpunct}
{\mcitedefaultendpunct}{\mcitedefaultseppunct}\relax
\EndOfBibitem
\bibitem{Belle:2016fev}
Belle collaboration, S.~Wehle {\em et~al.}, \ifthenelse{\boolean{articletitles}}{\emph{{Lepton-flavor-dependent angular analysis of $B\to K^\ast \ell^+\ell^-$}}, }{}\href{https://doi.org/10.1103/PhysRevLett.118.111801}{Phys.\ Rev.\ Lett.\  \textbf{118} (2017) 111801}, \href{http://arxiv.org/abs/1612.05014}{{\normalfont\ttfamily arXiv:1612.05014}}\relax
\mciteBstWouldAddEndPuncttrue
\mciteSetBstMidEndSepPunct{\mcitedefaultmidpunct}
{\mcitedefaultendpunct}{\mcitedefaultseppunct}\relax
\EndOfBibitem
\bibitem{LHCb-PAPER-2019-040}
LHCb collaboration, R.~Aaij {\em et~al.}, \ifthenelse{\boolean{articletitles}}{\emph{{Test of lepton universality with \mbox{\decay{\Lb}{p\Km\ellell}} decays}}, }{}\href{https://doi.org/10.1007/JHEP05(2020)040}{JHEP \textbf{05} (2020) 040}, \href{http://arxiv.org/abs/1912.08139}{{\normalfont\ttfamily arXiv:1912.08139}}\relax
\mciteBstWouldAddEndPuncttrue
\mciteSetBstMidEndSepPunct{\mcitedefaultmidpunct}
{\mcitedefaultendpunct}{\mcitedefaultseppunct}\relax
\EndOfBibitem
\bibitem{LHCb-PAPER-2015-029}
LHCb collaboration, R.~Aaij {\em et~al.}, \ifthenelse{\boolean{articletitles}}{\emph{{Observation of $\jpsi\proton$ resonances consistent with pentaquark states in \mbox{\decay{\Lb}{\jpsi\proton\Km}} decays}}, }{}\href{https://doi.org/10.1103/PhysRevLett.115.072001}{Phys.\ ~Rev.\ ~Lett.\  \textbf{115} (2015) 072001}, \href{http://arxiv.org/abs/1507.03414}{{\normalfont\ttfamily arXiv:1507.03414}}\relax
\mciteBstWouldAddEndPuncttrue
\mciteSetBstMidEndSepPunct{\mcitedefaultmidpunct}
{\mcitedefaultendpunct}{\mcitedefaultseppunct}\relax
\EndOfBibitem
\bibitem{LHCb-PAPER-2023-036}
LHCb collaboration, R.~Aaij {\em et~al.}, \ifthenelse{\boolean{articletitles}}{\emph{{Amplitude analysis of the $\Lb \to \proton \Km \gamma$ decay}}, }{}\href{https://doi.org/10.1007/JHEP06(2024)098}{{JHEP} \textbf{06} (2024) 098}, \href{http://arxiv.org/abs/2403.03710}{{\normalfont\ttfamily arXiv:2403.03710}}\relax
\mciteBstWouldAddEndPuncttrue
\mciteSetBstMidEndSepPunct{\mcitedefaultmidpunct}
{\mcitedefaultendpunct}{\mcitedefaultseppunct}\relax
\EndOfBibitem
\bibitem{PDG2026}
Particle Data Group, F.~Takahashi {\em et~al.}, \ifthenelse{\boolean{articletitles}}{\emph{{\href{http://pdg.lbl.gov/}{Review of particle physics}}}, }{}Mod.\ Phys.\  \textbf{A41} (2026) 2630011, to be published\relax
\mciteBstWouldAddEndPuncttrue
\mciteSetBstMidEndSepPunct{\mcitedefaultmidpunct}
{\mcitedefaultendpunct}{\mcitedefaultseppunct}\relax
\EndOfBibitem
\bibitem{Meinel:2020owd}
S.~Meinel and G.~Rendon, \ifthenelse{\boolean{articletitles}}{\emph{{$\Lb \to \Lres\ell^+\ell^-$ form factors from lattice QCD}}, }{}\href{https://doi.org/10.1103/PhysRevD.103.074505}{Phys.\ Rev.\  \textbf{D103} (2021) 074505}, \href{http://arxiv.org/abs/2009.09313}{{\normalfont\ttfamily arXiv:2009.09313}}\relax
\mciteBstWouldAddEndPuncttrue
\mciteSetBstMidEndSepPunct{\mcitedefaultmidpunct}
{\mcitedefaultendpunct}{\mcitedefaultseppunct}\relax
\EndOfBibitem
\bibitem{Meinel:2021mdj}
S.~Meinel and G.~Rendon, \ifthenelse{\boolean{articletitles}}{\emph{{$\Lambdares_{c} \to \Lres$ form factors from lattice QCD and improved analysis of the $\Lb \to \Lres$ and $\Lb \to \Lambdares_{c}^{*}(2595,2625)$ form factors}}, }{}\href{https://doi.org/10.1103/PhysRevD.105.054511}{Phys.\ Rev.\  \textbf{D105} (2022) 054511}, \href{http://arxiv.org/abs/2107.13140}{{\normalfont\ttfamily arXiv:2107.13140}}\relax
\mciteBstWouldAddEndPuncttrue
\mciteSetBstMidEndSepPunct{\mcitedefaultmidpunct}
{\mcitedefaultendpunct}{\mcitedefaultseppunct}\relax
\EndOfBibitem
\bibitem{Amhis:2022vcd}
Y.~Amhis, M.~Bordone, and M.~Reboud, \ifthenelse{\boolean{articletitles}}{\emph{{Dispersive analysis of $\Lb \to \Lres$ local form factors}}, }{}\href{https://doi.org/10.1007/JHEP02(2023)010}{JHEP \textbf{02} (2023) 010}, \href{http://arxiv.org/abs/2208.08937}{{\normalfont\ttfamily arXiv:2208.08937}}\relax
\mciteBstWouldAddEndPuncttrue
\mciteSetBstMidEndSepPunct{\mcitedefaultmidpunct}
{\mcitedefaultendpunct}{\mcitedefaultseppunct}\relax
\EndOfBibitem
\bibitem{LHCb-PAPER-2022-050}
LHCb collaboration, R.~Aaij {\em et~al.}, \ifthenelse{\boolean{articletitles}}{\emph{{Measurement of the $\Lb \to \Lambdares(1520)\mup\mun$ differential branching fraction}}, }{}\href{https://doi.org/10.1103/PhysRevLett.131.151801}{Phys.\ ~Rev.\ ~Lett.\  \textbf{131} (2023) 151801}, \href{http://arxiv.org/abs/2302.08262}{{\normalfont\ttfamily arXiv:2302.08262}}\relax
\mciteBstWouldAddEndPuncttrue
\mciteSetBstMidEndSepPunct{\mcitedefaultmidpunct}
{\mcitedefaultendpunct}{\mcitedefaultseppunct}\relax
\EndOfBibitem
\bibitem{LHCb-PAPER-2024-024}
LHCb collaboration, R.~Aaij {\em et~al.}, \ifthenelse{\boolean{articletitles}}{\emph{{Analysis of $\Lb\to \proton\Km\mumu$ decays}}, }{}\href{https://doi.org/10.1007/JHEP12(2024)147}{{JHEP} \textbf{12} (2024) 147}, \href{http://arxiv.org/abs/2409.12629}{{\normalfont\ttfamily arXiv:2409.12629}}\relax
\mciteBstWouldAddEndPuncttrue
\mciteSetBstMidEndSepPunct{\mcitedefaultmidpunct}
{\mcitedefaultendpunct}{\mcitedefaultseppunct}\relax
\EndOfBibitem
\bibitem{Descotes-Genon:2019dbw}
S.~Descotes-Genon and M.~Novoa-Brunet, \ifthenelse{\boolean{articletitles}}{\emph{{Angular analysis of the rare decay $\Lb\to \Lres(\to NK)\ell^+\ell^-$}}, }{}\href{https://doi.org/10.1007/JHEP06(2019)136}{JHEP \textbf{06} (2019) 136}, Erratum \href{https://doi.org/10.1007/JHEP06(2020)102}{ibid.\   \textbf{06} (2020) 102}, \href{http://arxiv.org/abs/1903.00448}{{\normalfont\ttfamily arXiv:1903.00448}}\relax
\mciteBstWouldAddEndPuncttrue
\mciteSetBstMidEndSepPunct{\mcitedefaultmidpunct}
{\mcitedefaultendpunct}{\mcitedefaultseppunct}\relax
\EndOfBibitem
\bibitem{Das:2020cpv}
D.~Das and J.~Das, \ifthenelse{\boolean{articletitles}}{\emph{{The $\Lb\to\Lres(\to N\!\bar{K})\ell^+\ell^-$ decay at low-recoil in HQET}}, }{}\href{https://doi.org/10.1007/JHEP07(2020)002}{JHEP \textbf{07} (2020) 002}, \href{http://arxiv.org/abs/2003.08366}{{\normalfont\ttfamily arXiv:2003.08366}}\relax
\mciteBstWouldAddEndPuncttrue
\mciteSetBstMidEndSepPunct{\mcitedefaultmidpunct}
{\mcitedefaultendpunct}{\mcitedefaultseppunct}\relax
\EndOfBibitem
\bibitem{Blake:2017une}
T.~Blake and M.~Kreps, \ifthenelse{\boolean{articletitles}}{\emph{{Angular distribution of polarised $\Lb$ baryons decaying to $\Lambdares \ellell$}}, }{}\href{https://doi.org/10.1007/JHEP11(2017)138}{JHEP \textbf{11} (2017) 138}, \href{http://arxiv.org/abs/1710.00746}{{\normalfont\ttfamily arXiv:1710.00746}}\relax
\mciteBstWouldAddEndPuncttrue
\mciteSetBstMidEndSepPunct{\mcitedefaultmidpunct}
{\mcitedefaultendpunct}{\mcitedefaultseppunct}\relax
\EndOfBibitem
\bibitem{LHCb-PAPER-2015-009}
LHCb collaboration, R.~Aaij {\em et~al.}, \ifthenelse{\boolean{articletitles}}{\emph{{Differential branching fraction and angular analysis of \mbox{\decay{\Lb}{\Lz\mumu}} decays}}, }{}\href{https://doi.org/10.1007/JHEP06(2015)115}{JHEP \textbf{06} (2015) 115}, Erratum \href{https://doi.org/10.1007/JHEP09(2018)145}{ibid.\   \textbf{09} (2018) 145}, \href{http://arxiv.org/abs/1503.07138}{{\normalfont\ttfamily arXiv:1503.07138}}\relax
\mciteBstWouldAddEndPuncttrue
\mciteSetBstMidEndSepPunct{\mcitedefaultmidpunct}
{\mcitedefaultendpunct}{\mcitedefaultseppunct}\relax
\EndOfBibitem
\bibitem{LHCb-PAPER-2018-029}
LHCb collaboration, R.~Aaij {\em et~al.}, \ifthenelse{\boolean{articletitles}}{\emph{{Angular moments of the decay \mbox{\decay{\Lb}{\Lz\mumu}} at low hadronic recoil}}, }{}\href{https://doi.org/10.1007/JHEP09(2018)146}{JHEP \textbf{09} (2018) 146}, \href{http://arxiv.org/abs/1808.00264}{{\normalfont\ttfamily arXiv:1808.00264}}\relax
\mciteBstWouldAddEndPuncttrue
\mciteSetBstMidEndSepPunct{\mcitedefaultmidpunct}
{\mcitedefaultendpunct}{\mcitedefaultseppunct}\relax
\EndOfBibitem
\bibitem{Amhis:2020phx}
Y.~Amhis {\em et~al.}, \ifthenelse{\boolean{articletitles}}{\emph{{Prospects for New Physics searches with $\Lb \rightarrow \Lres {{\ell ^+} {\ell ^-}} $ decays}}, }{}\href{https://doi.org/10.1140/epjp/s13360-021-01194-5}{Eur.\ Phys.\ J.\ Plus \textbf{136} (2021) 614}, \href{http://arxiv.org/abs/2005.09602}{{\normalfont\ttfamily arXiv:2005.09602}}\relax
\mciteBstWouldAddEndPuncttrue
\mciteSetBstMidEndSepPunct{\mcitedefaultmidpunct}
{\mcitedefaultendpunct}{\mcitedefaultseppunct}\relax
\EndOfBibitem
\bibitem{Boer:2014kda}
P.~B\"oer, T.~Feldmann, and D.~van Dyk, \ifthenelse{\boolean{articletitles}}{\emph{{Angular analysis of the decay \mbox{$\Lb \to \Lambdares (\to N \pi) \ell^+\ell^-$}}}, }{}\href{https://doi.org/10.1007/JHEP01(2015)155}{JHEP \textbf{01} (2015) 155}, \href{http://arxiv.org/abs/1410.2115}{{\normalfont\ttfamily arXiv:1410.2115}}\relax
\mciteBstWouldAddEndPuncttrue
\mciteSetBstMidEndSepPunct{\mcitedefaultmidpunct}
{\mcitedefaultendpunct}{\mcitedefaultseppunct}\relax
\EndOfBibitem
\bibitem{BW-parameters}
A.~V. Sarantsev {\em et~al.}, \ifthenelse{\boolean{articletitles}}{\emph{{Hyperon II: Properties of excited hyperons}}, }{}\href{https://doi.org/10.1140/epja/i2019-12880-5}{Eur.\ Phys.\ J.\  \textbf{A55} (2019) 180}, \href{http://arxiv.org/abs/1907.13387}{{\normalfont\ttfamily arXiv:1907.13387}}\relax
\mciteBstWouldAddEndPuncttrue
\mciteSetBstMidEndSepPunct{\mcitedefaultmidpunct}
{\mcitedefaultendpunct}{\mcitedefaultseppunct}\relax
\EndOfBibitem
\bibitem{Beck:2022spd}
A.~Beck, T.~Blake, and M.~Kreps, \ifthenelse{\boolean{articletitles}}{\emph{{Angular distribution of $ \Lb\to p{K}^{-}{\ell}^{+}{\ell}^{-} $ decays comprising {\ensuremath{\Lambdares}} resonances with spin {\ensuremath{\leq}} 5/2}}, }{}\href{https://doi.org/10.1007/JHEP02(2023)189}{JHEP \textbf{02} (2023) 189}, \href{http://arxiv.org/abs/2210.09988}{{\normalfont\ttfamily arXiv:2210.09988}}\relax
\mciteBstWouldAddEndPuncttrue
\mciteSetBstMidEndSepPunct{\mcitedefaultmidpunct}
{\mcitedefaultendpunct}{\mcitedefaultseppunct}\relax
\EndOfBibitem
\bibitem{LHCb-DP-2022-002}
LHCb collaboration, R.~Aaij {\em et~al.}, \ifthenelse{\boolean{articletitles}}{\emph{{The LHCb Upgrade I}}, }{}\href{https://doi.org/10.1088/1748-0221/19/05/P05065}{{JINST} \textbf{19} (2024) P05065}, \href{http://arxiv.org/abs/2305.10515}{{\normalfont\ttfamily arXiv:2305.10515}}\relax
\mciteBstWouldAddEndPuncttrue
\mciteSetBstMidEndSepPunct{\mcitedefaultmidpunct}
{\mcitedefaultendpunct}{\mcitedefaultseppunct}\relax
\EndOfBibitem
\bibitem{LHCb-DP-2008-001}
LHCb collaboration, A.~A. Alves~Jr.\ {\em et~al.}, \ifthenelse{\boolean{articletitles}}{\emph{{The \lhcb detector at the LHC}}, }{}\href{https://doi.org/10.1088/1748-0221/3/08/S08005}{JINST \textbf{3} (2008) S08005}\relax
\mciteBstWouldAddEndPuncttrue
\mciteSetBstMidEndSepPunct{\mcitedefaultmidpunct}
{\mcitedefaultendpunct}{\mcitedefaultseppunct}\relax
\EndOfBibitem
\bibitem{LHCb-DP-2014-002}
LHCb collaboration, R.~Aaij {\em et~al.}, \ifthenelse{\boolean{articletitles}}{\emph{{LHCb detector performance}}, }{}\href{https://doi.org/10.1142/S0217751X15300227}{Int.\ J.\ Mod.\ Phys.\  \textbf{A30} (2015) 1530022}, \href{http://arxiv.org/abs/1412.6352}{{\normalfont\ttfamily arXiv:1412.6352}}\relax
\mciteBstWouldAddEndPuncttrue
\mciteSetBstMidEndSepPunct{\mcitedefaultmidpunct}
{\mcitedefaultendpunct}{\mcitedefaultseppunct}\relax
\EndOfBibitem
\bibitem{LHCb-DP-2012-004}
R.~Aaij {\em et~al.}, \ifthenelse{\boolean{articletitles}}{\emph{{The \lhcb trigger and its performance in 2011}}, }{}\href{https://doi.org/10.1088/1748-0221/8/04/P04022}{JINST \textbf{8} (2013) P04022}, \href{http://arxiv.org/abs/1211.3055}{{\normalfont\ttfamily arXiv:1211.3055}}\relax
\mciteBstWouldAddEndPuncttrue
\mciteSetBstMidEndSepPunct{\mcitedefaultmidpunct}
{\mcitedefaultendpunct}{\mcitedefaultseppunct}\relax
\EndOfBibitem
\bibitem{LHCb-DP-2019-001}
R.~Aaij {\em et~al.}, \ifthenelse{\boolean{articletitles}}{\emph{{Design and performance of the LHCb trigger and full real-time reconstruction in Run 2 of the LHC}}, }{}\href{https://doi.org/10.1088/1748-0221/14/04/P04013}{JINST \textbf{14} (2019) P04013}, \href{http://arxiv.org/abs/1812.10790}{{\normalfont\ttfamily arXiv:1812.10790}}\relax
\mciteBstWouldAddEndPuncttrue
\mciteSetBstMidEndSepPunct{\mcitedefaultmidpunct}
{\mcitedefaultendpunct}{\mcitedefaultseppunct}\relax
\EndOfBibitem
\bibitem{Sjostrand:2007gs}
T.~Sj\"{o}strand, S.~Mrenna, and P.~Skands, \ifthenelse{\boolean{articletitles}}{\emph{{A brief introduction to PYTHIA 8.1}}, }{}\href{https://doi.org/10.1016/j.cpc.2008.01.036}{Comput.\ Phys.\ Commun.\  \textbf{178} (2008) 852}, \href{http://arxiv.org/abs/0710.3820}{{\normalfont\ttfamily arXiv:0710.3820}}\relax
\mciteBstWouldAddEndPuncttrue
\mciteSetBstMidEndSepPunct{\mcitedefaultmidpunct}
{\mcitedefaultendpunct}{\mcitedefaultseppunct}\relax
\EndOfBibitem
\bibitem{LHCb-PROC-2010-056}
I.~Belyaev {\em et~al.}, \ifthenelse{\boolean{articletitles}}{\emph{{Handling of the generation of primary events in Gauss, the LHCb simulation framework}}, }{}\href{https://doi.org/10.1088/1742-6596/331/3/032047}{J.\ Phys.\ Conf.\ Ser.\  \textbf{331} (2011) 032047}\relax
\mciteBstWouldAddEndPuncttrue
\mciteSetBstMidEndSepPunct{\mcitedefaultmidpunct}
{\mcitedefaultendpunct}{\mcitedefaultseppunct}\relax
\EndOfBibitem
\bibitem{Lange:2001uf}
D.~J. Lange, \ifthenelse{\boolean{articletitles}}{\emph{{The EvtGen particle decay simulation package}}, }{}\href{https://doi.org/10.1016/S0168-9002(01)00089-4}{Nucl.\ Instrum.\ Meth.\  \textbf{A462} (2001) 152}\relax
\mciteBstWouldAddEndPuncttrue
\mciteSetBstMidEndSepPunct{\mcitedefaultmidpunct}
{\mcitedefaultendpunct}{\mcitedefaultseppunct}\relax
\EndOfBibitem
\bibitem{davidson2015photos}
N.~Davidson, T.~Przedzinski, and Z.~Was, \ifthenelse{\boolean{articletitles}}{\emph{{PHOTOS interface in C++: Technical and physics documentation}}, }{}\href{https://doi.org/https://doi.org/10.1016/j.cpc.2015.09.013}{Comput.\ Phys.\ Commun.\  \textbf{199} (2016) 86}, \href{http://arxiv.org/abs/1011.0937}{{\normalfont\ttfamily arXiv:1011.0937}}\relax
\mciteBstWouldAddEndPuncttrue
\mciteSetBstMidEndSepPunct{\mcitedefaultmidpunct}
{\mcitedefaultendpunct}{\mcitedefaultseppunct}\relax
\EndOfBibitem
\bibitem{PDG2022}
Particle Data Group, R.~L. Workman {\em et~al.}, \ifthenelse{\boolean{articletitles}}{\emph{{\href{http://pdg.lbl.gov/}{Review of particle physics}}}, }{}\href{https://doi.org/10.1093/ptep/ptac097}{Prog.\ Theor.\ Exp.\ Phys.\  \textbf{2022} (2022) 083C01}\relax
\mciteBstWouldAddEndPuncttrue
\mciteSetBstMidEndSepPunct{\mcitedefaultmidpunct}
{\mcitedefaultendpunct}{\mcitedefaultseppunct}\relax
\EndOfBibitem
\bibitem{TMVA4}
A.~Hoecker {\em et~al.}, \ifthenelse{\boolean{articletitles}}{\emph{{TMVA 4 - Toolkit for Multivariate Data Analysis with ROOT. Users Guide}}, }{}\href{http://arxiv.org/abs/physics/0703039}{{\normalfont\ttfamily arXiv:physics/0703039}}\relax
\mciteBstWouldAddEndPuncttrue
\mciteSetBstMidEndSepPunct{\mcitedefaultmidpunct}
{\mcitedefaultendpunct}{\mcitedefaultseppunct}\relax
\EndOfBibitem
\bibitem{MartinezSantos:2013ltf}
D.~Mart{\'\i}nez~Santos and F.~Dupertuis, \ifthenelse{\boolean{articletitles}}{\emph{{Mass distributions marginalized over per-event errors}}, }{}\href{https://doi.org/10.1016/j.nima.2014.06.081}{Nucl.\ Instrum.\ Meth.\  \textbf{A764} (2014) 150}, \href{http://arxiv.org/abs/1312.5000}{{\normalfont\ttfamily arXiv:1312.5000}}\relax
\mciteBstWouldAddEndPuncttrue
\mciteSetBstMidEndSepPunct{\mcitedefaultmidpunct}
{\mcitedefaultendpunct}{\mcitedefaultseppunct}\relax
\EndOfBibitem
\bibitem{Skwarnicki:1986xj}
T.~Skwarnicki, {\em {A study of the radiative cascade transitions between the Upsilon-prime and Upsilon resonances}}, PhD thesis, Institute of Nuclear Physics, Krakow, 1986, {\href{http://inspirehep.net/record/230779/}{DESY-F31-86-02}}\relax
\mciteBstWouldAddEndPuncttrue
\mciteSetBstMidEndSepPunct{\mcitedefaultmidpunct}
{\mcitedefaultendpunct}{\mcitedefaultseppunct}\relax
\EndOfBibitem
\bibitem{Ryd:2005zz}
A.~Ryd {\em et~al.}, \ifthenelse{\boolean{articletitles}}{\emph{{EvtGen: A Monte Carlo Generator for $B$-Physics}}, }{} 2005\relax
\mciteBstWouldAddEndPuncttrue
\mciteSetBstMidEndSepPunct{\mcitedefaultmidpunct}
{\mcitedefaultendpunct}{\mcitedefaultseppunct}\relax
\EndOfBibitem
\bibitem{Mott:2011cx}
L.~Mott and W.~Roberts, \ifthenelse{\boolean{articletitles}}{\emph{{Rare dileptonic decays of \Lb in a quark model}}, }{}\href{https://doi.org/10.1142/S0217751X12500169}{Int.\ J.\ Mod.\ Phys.\  \textbf{A27} (2012) 1250016}, \href{http://arxiv.org/abs/1108.6129}{{\normalfont\ttfamily arXiv:1108.6129}}\relax
\mciteBstWouldAddEndPuncttrue
\mciteSetBstMidEndSepPunct{\mcitedefaultmidpunct}
{\mcitedefaultendpunct}{\mcitedefaultseppunct}\relax
\EndOfBibitem
\end{mcitethebibliography}

\newpage
% LHCb collaboration author list
% Data extracted on August 19th, 2026 at 1:19pm for paper reference LHCb-PAPER-2026-037
\centerline
{\large\bf LHCb collaboration}
\begin
{flushleft}
\small
R.~Aaij$^{41}$\lhcborcid{0000-0003-0533-1952},
M.~Abdelfatah$^{73}$,
A.S.W.~Abdelmotteleb$^{61}$\lhcborcid{0000-0001-7905-0542},
C.~Abellan~Beteta$^{55}$\lhcborcid{0009-0009-0869-6798},
F.~Abudin\'en$^{63}$\lhcborcid{0000-0002-6737-3528},
T.~Ackernley$^{65}$\lhcborcid{0000-0002-5951-3498},
A.A.~Adefisoye$^{73}$\lhcborcid{0000-0003-2448-1550},
B.~Adeva$^{51}$\lhcborcid{0000-0001-9756-3712},
M.~Adinolfi$^{59}$\lhcborcid{0000-0002-1326-1264},
P.~Adlarson$^{89,46}$\lhcborcid{0000-0001-6280-3851}
% ...

{\footnotesize \it

% ...
$^{41}$Nikhef National Institute for Subatomic Physics, Amsterdam, Netherlands\\
% ...
$^{46}$National Center for Nuclear Research (NCBJ), Warsaw, Poland\\
% ...
$^{51}$Instituto Galego de F{\'\i}sica de Altas Enerx{\'\i}as (IGFAE), Universidade de Santiago de Compostela, Santiago de Compostela, Spain\\
% ...
$^{55}$Physik-Institut, Universit{\"a}t Z{\"u}rich, Z{\"u}rich, Switzerland\\
% ...
$^{59}$H.H. Wills Physics Laboratory, University of Bristol, Bristol, United Kingdom\\
% ...
$^{61}$Department of Physics, University of Warwick, Coventry, United Kingdom\\
% ...
$^{63}$School of Physics and Astronomy, University of Edinburgh, Edinburgh, United Kingdom\\
$^{64}$School of Physics and Astronomy, University of Glasgow, Glasgow, United Kingdom\\
$^{65}$Oliver Lodge Laboratory, University of Liverpool, Liverpool, United Kingdom\\
% ...
$^{73}$Syracuse University, Syracuse, NY, United States\\
% ...
$^{89}$Department of Physics and Astronomy, Uppsala University, Uppsala, Sweden, associated to $^{64}$\\
% ...
}
\end{flushleft}

\end{document}